\documentclass[11pt]{article}

\usepackage{amsthm,amsmath,amsfonts,amssymb,natbib,mathtools,mathrsfs,algorithm,framed,multirow}
\usepackage[noend]{algpseudocode}
\usepackage{ccaption}
\usepackage{longtable}

\usepackage{graphicx}
\usepackage{epstopdf}
\usepackage{epsfig}

\usepackage{xspace} 
\usepackage{xcolor}

\usepackage{enumerate}
\usepackage{verbatim}
\usepackage{subfigure}
\usepackage{bbm}
\usepackage{threeparttable}

\usepackage{JASA_manu}

\usepackage{xr}
\usepackage{authblk}

\newtheorem{definition}{Definition}[section]

\newcommand{\pr}{{\rm I}\kern-0.18em{\rm P}}

\newcommand\tab[1][.5cm]{\hspace*{#1}}
\newcommand{\tran}{^{\mkern-1.5mu\mathsf{T}}}
\newcommand{\norm}[1]{\left\lVert#1\right\rVert}
\newcommand{\R}{{\rm I}\kern-0.18em{\rm R}}
\newcommand{\N}{{\rm I}\kern-0.18em{\rm N}}
\newcommand{\Z}{{\rm Z}\kern-0.44em{\rm Z}}
\newcommand{\E}{{\rm I}\kern-0.18em{\rm E}}
\newcommand{\Prob}{{\rm I}\kern-0.18em{\rm P}}
\newcommand{\1}{{\rm 1}\kern-0.24em{\rm I}}
\newcommand{\cov}{\mathrm{cov}}
\newcommand{\var}{\mathrm{var}}

\newcommand{\bu}{\mathbf{u}}

\newcommand{\bmu}{\boldsymbol{\mu}}

\newcommand{\bbeta}{\boldsymbol{\beta}}
\newcommand{\bEta}{\boldsymbol{\eta}}

\newcommand{\bxi}{\boldsymbol{\xi}}

\newcommand{\bSigma}{\boldsymbol{\Sigma}}
\newcommand{\btheta}{\boldsymbol{\theta}}

\newcommand{\bOmega}{\boldsymbol{\Omega}}
\newcommand{\cB}{\mathcal{B}}
\newcommand{\cC}{\mathcal{C}}

\newcommand{\cG}{\mathcal{G}}

\newcommand{\cS}{\mathcal{S}}
\newcommand{\cU}{\mathcal{U}}

\newcommand{\cN}{\mathcal{N}}
\newcommand{\tod}{\overset{d}\longrightarrow}
\DeclareMathOperator*{\argmin}{arg\,min}
\DeclareMathOperator*{\argmax}{arg\,max}
\newtheorem{theorem}{Theorem}[section]
\newtheorem{lemma}[theorem]{Lemma}
\newtheorem{corollary}[theorem]{corollary}

\newtheorem{remark}{Remark}

\makeatletter
\def\BState{\State\hskip-\ALG@thistlm}
\makeatother

\makeatletter
\newcommand{\leqnomode}{\tagsleft@true}
\newcommand{\reqnomode}{\tagsleft@false}
\makeatother

\begin{document}

\title{Generalized Pearson correlation squares for capturing mixtures of bivariate linear dependences}

\author[1]{Jingyi Jessica Li}
\author[2, 4]{Xin Tong}
\author[3]{Peter J. Bickel}
\affil[1]{\footnotesize Department of Statistics, University of California, Los Angeles, CA}
\affil[2]{Department of Data Sciences and Operations, University of Southern California, Los Angeles, CA}
\affil[3]{Department of Statistics, University of California, Berkeley, CA}
\affil[4]{To whom correspondence should be addressed. Email: xint@marshall.usc.edu}

\date{}

\maketitle



\begin{abstract}

Motivated by the pressing needs for capturing complex but interpretable variable relationships in scientific research, here we generalize the squared Pearson correlation to capture a mixture of linear dependences between two real-valued random variables, with or without an index variable that specifies the line memberships. We construct generalized Pearson correlation squares by focusing on three aspects: the exchangeability of the two variables, the independence of parametric model assumptions, and the availability of population-level parameters. For the computation of the generalized Pearson correlation square from a sample without line-membership specification, we develop a $K$-lines clustering algorithm, where $K$, the number of lines, can be chosen in a data-adaptive way. With our defined population-level generalized Pearson correlation squares, we derive the asymptotic distributions of the sample-level statistics to enable efficient statistical inference.  Simulation studies verify the theoretical results and compare the generalized Pearson correlation squares with other widely-used association measures in terms of power. Gene expression data analysis demonstrates the effectiveness of the generalized Pearson correlation squares in capturing interpretable gene-gene relationships missed by other measures. We implement the estimation and inference procedures in an R package gR2.
	
	\vspace*{.15in}
	
\end{abstract}



\section{Introduction}\label{sec:intro}

\begin{figure}[htbp]
\includegraphics[width=\textwidth]{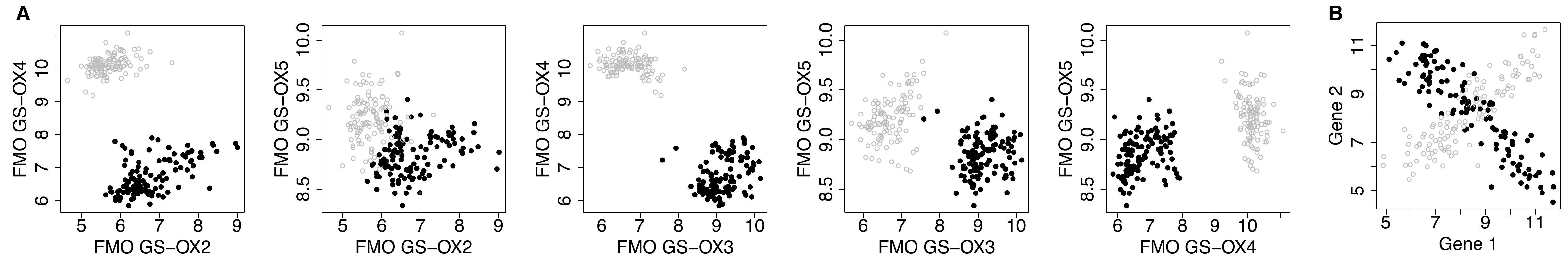}
\caption{\textbf{A}: Pairwise expression levels of \textit{Arabidopsis thaliana} genes. \textbf{B}: A simulated toy example. Grey circles and black dots indicate data from root and shoot tissues, respectively. \label{fig:motiv}}
\end{figure}

In scientific research, Pearson correlation and its rank-based variant Spearman correlation remain the most widely used association measures for describing the relationship between two scalar random variables. The reason underlying their popularity is two-fold: linear and monotone relationships\footnote{Monotone relationships becomes linear after values of each variable are transformed into ranks.} are widespread in nature and interpretable to human experts.
In many cases, though, the interesting relationship between two random variables often depends on another hidden categorical variable. Here is an example from gene expression data of  \textit{Arabidopsis thaliana}, a plant model organism, that many genes exhibit different linear dependences in root and shoot tissues \citep{li2008subclade, kim2012using}. Figure~\ref{fig:motiv}A shows pairwise gene expression levels of flavin-monooxygenase (FMO) genes to illustrate this phenomenon. For example, genes FMO GS-OX2 and FMO GS-OX5 show a positive correlation in shoots (black dots) but a negative correlation in roots (gray circles). In an idealistic and extreme scenario (Fig.~\ref{fig:motiv}B), suppose that two real-valued random variables $X$ and $Y$ represent the expression levels of two genes. If $X$ and $Y$ have a positive correlation $\rho$ in the shoot tissue but a negative correlation $-\rho$ in the root tissue, and that the two tissues are expected to have equal representation in a study, then $X$ and $Y$ would have a population-level Pearson correlation equal to zero. Real scenarios are usually not so extreme, but many of them exhibit a mixture relationship composed of two linear dependences with mixed signs \citep{li2002genome}, or show the ``Simpson's Paradox," \citep{simpson1951interpretation, pearson1899mathematical, yule1903notes, blyth1972simpson}, where the overall correlation and the conditional correlations have opposite signs. Under such scenarios, Pearson correlation is a misleading measure, as it specifically looks for a single linear dependence. Moreover, these scenarios often lack an index variable (e.g., the shoot/root tissue type) that segregates observations into distinct linear relationships, and numerous variables pairs (e.g., $10^8$ gene pairs) need to be examined to discover unknown, but interesting and interpretable associations. Therefore, an association measure is in much demand to capture such relationships that are decomposable into a (possibly unknown) number of linear dependences, in a powerful and efficient way.

In the literature of scalar-valued association measures, also known as dependence measures, many measures have been developed to capture dependent relationships more general than linear dependence. The first type of measures aims to capture more general functional (i.e., one-to-one) relationships. For monotone relationships, the Spearman's rank correlation \citep{spearman1904proof} and the Kendall's $\tau$ \citep{kendall1938new} are commonly used. For functional relationships more general than monotonicity, there are measures including the maximal correlation efficient \citep{hirschfeld1935connection, gebelein1941statistische, renyi1959measures}, measures based on nonparametric estimation of correlation curves \citep{bjerve1993correlation} or principal curves \citep{delicado2009measuring}, generalized measures of correlation that deals with asymmetrically explained variances and nonlinear relationships \citep{zheng2012generalized}, the $G^2$ statistic derived from a regularized likelihood ratio test for piecewise-linear relationships \citep{wang2017generalized}, and measures for detecting local monotone patterns using count statistics \citep{wang2014gene}. The second type of measures aims to capture general dependence so that they only give zero values to independent random variable pairs. Examples include the maximal correlation coefficient, the Hoeffding's $D$ \citep{hoeffding1948non}, the mutual information \citep{shannon1951mathematical, kraskov2004estimating, cover2012elements}, kernel-based measures such as the Hilbert-Schmidt Independence Criterion (HSIC) \citep{gretton2005measuring}, the distance correlation \citep{szekely2007measuring, szekely2009brownian}, the maximal information coefficient \citep{reshef2011detecting}, and the Heller-Heller-Gorfine (HHG) association test statistic based on ranks of distances \citep{heller2012consistent}. Specifically, the following measures are not restricted to comparing real-valued random variables: the Hoeffding's $D$, the mutual information, the HSIC, the distance correlation, and the HHG test statistic, among which the first four measures have the range $[0,\infty)$ as opposed to having absolute values no greater than one.

The two types of measures both have relative advantages and limitations. Measures of the first type are generally interpretable but cannot capture non-functional (i.e., not one-to-one) relationships that are widespread in the real world. In contrast, measures of the second type, though being versatile and having desirable theoretic properties, do not convey a straightforward interpretation of their captured relationships to practitioners. As our motivating example has shown (Fig. \ref{fig:motiv}), there are widespread relationships that are decomposable into a small number of linear dependences. Since linear dependence is the simplest and most interpretable relationship, a mixture of a small number of linear dependences is also interpretable and often of great interest in scientific research. For example, if researchers observe that one gene positively regulates a vital cancer gene in one cancer subtype but exhibits adverse regulatory effects in another subtype, different treatment strategies may be designed for the two cancer subtypes. However, capturing mixtures of linear dependences remains challenging: the first type of measures often miss them, and the second type of measures cannot distinguish them from other relationships that are less interpretable. Although mixtures of linear models/regressions have been of a broad interest in fields including statistics, economics, social sciences, and machine learning for over 40 years \citep{quandt1978estimating, Murtaph.Raftery.1984, de1989mixtures, jacobs1991adaptive, jones1992fitting, wedel1994review, turner2000estimating, hawkins2001determining, hurn2003estimating, leisch2008modelling, benaglia2009mixtools, scharl2009mixtures}, they did not propose an association measure to capture this particular type of non-functional relationships, and neither do they trivially lead to a reasonable association measure, as we will explain below.

In this work, we propose generalized Pearson correlation squares, for which the squared Pearson correlation is a special case, to capture a mixture of linear dependences. We consider two scenarios: the \textit{specified scenario} where an index variable indicates the line membership of each observation, and the \textit{unspecified scenario} that is more widespread in applications where no index variable is available. For a reasonable generalization, we make our new measures satisfy an essential property---the exchangeability between $X$ and $Y$---embraced by most existing association measures. We also desire our measures to have both population-level parameters and sample-level statistics, just as Pearson correlation, maximal correlation, and distance correlation do, so that statistical inference becomes feasible. To achieve these goals, our development has successfully addressed four critical questions. 
\begin{itemize}
\item First, under the unspecified scenario, can we directly use the existing work on mixtures of linear models/regressions? The answer is no because these models require a specification of which variable is the response and which is the predictor; in other words, they do not consider $X$ and $Y$ symmetric. Except for the degenerate case where only one component exists, i.e., the linear model, these models do not lead to a measure exchangeable between $X$ and $Y$.
\item Second, still under the unspecified scenario, how to assign observations to lines to ensure the exchangeability of $X$ and $Y$? A good assignment should be capable of handling general cases where observations from each line do not follow a specific distribution (required by model-based clustering) or have a spherical shape (required by $K$-means clustering). To handle such cases, we propose a new $K$-lines clustering algorithm in Section \ref{subsec:choose_K}. 
\item Third, how should we define population-level measures to enable proper inference? A critical point is that the specified and unspecified scenarios need different population-level measures; otherwise, it would be impossible to construct unbiased estimators for both scenarios without distributional assumptions. We will elaborate on this point in Section \ref{sec:asym_theory}.  
\item Fourth, when an index variable is available, should we always use it to specify line memberships? Surprisingly, the answer is no because the index variable may be uninformative or irrelevant to the segregation of lines. In that case, it is more informative to directly learn line memberships from data by clustering. We will demonstrate this point in a real data study in Section \ref{sec:arab}. 
\end{itemize}

This paper is organized as follows. In Section \ref{sec:measure}, we define generalized Pearson correlation squares on the sample level, under the line-membership specified and unspecified scenarios. For the unspecified scenario, we develop a $K$-lines clustering algorithm. In Section \ref{sec:asym_theory}, we define the population-level generalized Pearson correlation squares and derive the asymptotic distributions of the corresponding sample-level measures to enable efficient statistical inference.  In Section \ref{sec:simulation}, we conduct simulation studies under various settings to verify the asymptotic distributions and evaluate the finite-sample statistical power of the proposed measures. In Section \ref{sec:real}, we demonstrate the use of the generalized Pearson correlation squares in two real data studies, followed by discussions in Section \ref{sec:discussion}. Supplementary Material includes all the proofs of lemmas and theorems, convergence properties of the $K$-lines algorithm, more simulation results, real data description, and more figures.

\section{Sample-level generalized Pearson correlation squares}\label{sec:measure}
The Pearson correlation coefficient is the most widely used similarity measure to describe the relationship between two random variables $X, Y \in \R$. The sample-level Pearson correlation coefficient $R = \{\sum_{i=1}^n (X_i - \bar X) (Y_i - \bar Y)\} / \{\sum_{i=1}^n (X_i - \bar X)^2 \sum_{i=1}^n (Y_i - \bar Y)^2\}^{1/2}$ is defined based on a sample  $(X_1,Y_1),\ldots,(X_n,Y_n)$ from the joint distribution of $(X, Y)$, where $\bar X = n^{-1}\sum_{i=1}^n X_i$ and $\bar Y = n^{-1}\sum_{i=1}^n Y_i$.
Motivated by the fact that $R^2$, the Pearson correlation square, is commonly used to describe the observed linear dependence in a bivariate sample, we develop generalized Pearson correlation squares to capture a mixture of linear dependences. We will construct the sample-level measures in this section and their population-level counterparts in the next section.

We define the line-membership specified scenario as the case where we also observe an index random variable $Z \in \{1,\ldots,K\}$ that specifies the linear dependence between $X$ and $Y$, and $K$ is the number of linear dependences. In parallel, we define the line-membership unspecified scenario as the case where no index variable is available. In the special case of $K=1$, we have $Z\equiv 1$. There may exist more than one index variable, and correspondingly there could be multiple specified scenarios. For example, in the \textit{Arabidopsis thaliana} gene expression dataset (Table \ref{tabS1:at_data} in the Supplementary Material), there are four index variables (condition, treatment, replicate and tissue), leading to four different specified scenarios. As we will show in Section \ref{sec:arab}, only the specification by the tissue variable leads to a set of linear relationships that fit well to the data (Fig. \ref{fig:motiv} and Figs. \ref{fig:motiv2}--\ref{fig:motiv4} in the Supplementary Material). Hence, a specified scenario should not be always preferred to the unspecified scenario when the goal is to capture an informative mixture of linear dependences.

\subsection{Line-membership specified scenario} \label{subsec:sample_R2gs}
Under the line-membership specified scenario, we consider a sample $(X_i,Y_i,Z_i)$, $i = 1, \ldots, n$, from the joint distribution of $(X, Y, Z) \in \R^2 \times \{1,\ldots,K\}$.

\begin{definition}\label{def:sample_gen_R2}
At the sample level, when observations $\{(X_i,Y_i)\}_{i=1}^n$ have line memberships $\{Z_i\}_{i=1}^n$ specified, the generalized Pearson correlation square is defined as
\begin{align}
	R_{\cG\cS}^2 &= \sum_{k=1}^K \widehat p_{k\cS} \; \widehat\rho_{k\cS}^2\,,
\label{scor2_g}
\end{align}
where the subindex $G$ stands for generalized, $S$ stands for specified, $\widehat p_{k\cS} = \frac{1}{n} \sum_{i=1}^n\1(Z_i=k)$, and
\begin{align*}
\widehat \rho_{k\cS}^2 &= \frac{\left\{\sum_{i=1}^n (X_i - \bar X_{k\cS}) (Y_i - \bar Y_{k\cS}) \1(Z_i=k)\right\}^2}{\left\{\sum_{i=1}^n (X_i - \bar X_{k\cS})^2\1(Z_i=k)\right\}\left\{\sum_{i=1}^n (Y_i - \bar Y_{k\cS})^2\1(Z_i=k)\right\}}\,,
\end{align*}
with $\bar X_{k\cS} = n_{k\cS}^{-1}\sum_{i=1}^n X_i \1(Z_i=k)$, $\bar Y_{k\cS} = n_{k\cS}^{-1}\sum_{i=1}^n Y_i \1(Z_i=k)$, and $n_{k\cS} = \sum_{i=1}^n\1(Z_i=k)$.
\end{definition}

The $R_{\cG\cS}^2$ is a weighted sum of the $R^2$'s of all line components, i.e., $\widehat \rho_{1\cS}^2, \ldots, \widehat \rho_{K\cS}^2$. Variables $X$ and $Y$ are exchangeable in the measure  $R_{\cG\cS}^2$.  Motivated by the above definition, we next define its counterpart under the more common scenario in which no index variable $Z$ is observable.

\subsection{Line-membership unspecified scenario} \label{subsec:sample_R2gu}

We consider a sample $(X_i,Y_i)$, $i = 1, \ldots, n$, from the joint distribution of $(X, Y) \in \R^2$, which we refer to as the line-membership unspecified scenario. As  line-membership information is unavailable, we will first assign each $(X_i,Y_i)$ pair to a line.   Towards that, we define the sample-level unspecified line centers as the $K$ lines that minimize the average squared perpendicular distance of data points to their closest line. We use a shorthand notation $\beta = (a,b,c)\tran$ to denote the line $\{ (x,y)\tran: ax + by + c = 0, \text{ where } a, b, c \in \R \text{ with $a \neq 0$ or $b \neq 0$} \} \subset \R^2$.  Because we wish $X$ and $Y$ to be exchangeable in the new measure, a reasonable  distance from $(x,y)\tran$ to $\bbeta$ is the perpendicular distance $d_\perp: \R^2 \times \R^3 \mapsto \R$:
\begin{equation} \label{def:perp_dist}
d_\perp\left( (x,y)\tran, \bbeta \right) = |a x + b y + c|(a^2+b^2)^{-1/2}\,.
\end{equation}

\begin{definition}\label{def:line_center_u_sample}
Let $B_K= \{\bbeta_1,\ldots,\bbeta_K\}$ be a multiset of $K$ lines with possible repeats. We define the average within-cluster squared perpendicular distance as
\begin{align}\label{eq:W_Pn}
W(B_K, P_n) &= \frac{1}{n}\sum_{i=1}^n \min_{\bbeta \in B_K} d_\perp^2\left( (X_i,Y_i)\tran, \bbeta \right)\,, 
\end{align}
where $P_n$ is the empirical measure by placing mass $n^{-1}$ at each of $(X_1,Y_1), \ldots, (X_n,Y_n)$. Then we define the multiset of sample-level unspecified line centers as
\begin{equation}\label{def:beta_ku_s}
	\widehat B_{K\cU} \in \argmin_{B_K} W(B_K, P_n)\,,
\end{equation}
where the subindex $\cU$ stands for unspecified.   We write each solution to (\ref{def:beta_ku_s}) as $\widehat B_{K\cU} = \{\widehat\bbeta_{1\cU}, \ldots, \widehat\bbeta_{K\cU}\}$, where $\widehat\bbeta_{k\cU} =(\widehat a_{k\cU}, \widehat b_{k\cU}, \widehat c_{k\cU})\tran$ is the $k$-th line center. 
\end{definition}

To find $\widehat B_{K\cU}$, we propose the  $K$-lines clustering algorithm, which is related to and inspired by the well-known $K$-means algorithm \citep{lloyd1982least}. The $K$-means algorithm cannot account for within-cluster correlation structures but only identifies spherical clusters under a distance metric, e.g., the Euclidean distance. In contrast, the $K$-lines algorithm finds clusters that exhibit strong within-cluster correlations; it is specifically designed for applications where two real-valued variables may have distinct correlations in different hidden clusters.

As an iterative procedure, the $K$-lines clustering algorithm includes two alternating steps in each iteration. The recentering step uses the current cluster assignment (i.e., line membership) to update each cluster line center, which minimizes the within-cluster sum of squared perpendicular distances of data points to the line center. The assignment step updates the cluster assignment based on the current cluster line centers: assign every data point to its closest cluster line center in the perpendicular distance. The two steps alternate until the algorithm converges.

\begin{algorithm}\label{alg:K-lines}
\textbf{$K$-lines clustering algorithm}
\begin{tabbing}
   \enspace Assign random initial clusters $\cC_1^{(0)},\ldots,\cC_K^{(0)}$, such that $\cup_{k=1}^K \cC_k^{(0)} = \{1,\ldots,n\}$\\
   \enspace The algorithm proceeds by alternating between two steps. In the $t$-th iteration, $t=1,2,\ldots$ \\
   \qquad Recentering step: Calculate the cluster line centers $\widehat\bbeta_{1\cU}^{(t)},\ldots,\widehat\bbeta_{K\cU}^{(t)}$ based on the \\
   \qquad\qquad cluster assignment $\cC_1^{(t-1)},\ldots,\cC_K^{(t-1)}$ by (\ref{eq:hat_beta_ku})\\
   \qquad Assignment step: Update the cluster assignment for $k=1,\ldots,K$\\
   \qquad\qquad  $\cC_k^{(t)} = \left\{i: d_\perp\left( (X_i,Y_i)\tran, \widehat\bbeta_{k\cU}^{(t)}\right) \le d_\perp\left( (X_i,Y_i)\tran, \widehat\bbeta_{s\cU}^{(t)}\right), \text{ for all } s = 1, \ldots, K \right\}$ \\
   \qquad Stop the iteration when the cluster assignment no longer changes\\
   \enspace Output: Cluster assignment $\cC_1,\ldots,\cC_K$; Sample-level unspecified line centers $\widehat\bbeta_{1\cU},\ldots,\widehat\bbeta_{K\cU}$
\end{tabbing}
\end{algorithm}

The recentering step updates each cluster center using the major axis regression, which minimizes the sum of squares of the perpendicular distance between each point and the regression line. It was shown that the major axis regression line is the first principal component of the sample covariance matrix of $(X, Y)$ \citep{jolliffe1982note, smith2009use}. Given the cluster assignment in the $(t-1)$-th iteration: $\cC_1^{(t-1)},\ldots,\cC_K^{(t-1)}$, the updated $k$-th cluster center is 
\begin{align}\label{eq:hat_beta_ku}
\widehat\bbeta_{k\cU}^{(t)} &= \argmin_{\bbeta} \sum_{i \in \cC_k^{(t-1)}} d_\perp^2\left( (X_i,Y_i)\tran, \bbeta\right) = \left(\widehat u_{12,k},\: -\widehat u_{11,k},\: -\widehat u_{12,k} \bar X_{k\cU} + \widehat u_{11,k} \bar Y_{k\cU}\right)\tran\,,
\end{align}
where $(\widehat u_{11,k}, \widehat u_{12,k})\tran$ is the first principal component of the sample covariance matrix 
\begin{align*}
\left|\cC_k^{(t-1)}\right|^{-1}
	\begin{bmatrix}
	\sum_{i \in \cC_k^{(t-1)}}(X_i-\bar X_{k\cU})^2 & \sum_{i \in \cC_k^{(t-1)}}(X_i-\bar X_{k\cU})(Y_i-\bar Y_{k\cU})\\
	\sum_{i \in \cC_k^{(t-1)}}(X_i-\bar X_{k\cU})(Y_i-\bar Y_{k\cU}) & \sum_{i \in \cC_k^{(t-1)}}(Y_i-\bar Y_{k\cU})^2
	\end{bmatrix}\,,
\end{align*}
with $\bar X_{k\cU} = |\cC_k^{(t-1)}|^{-1}\sum_{i \in \cC_k^{(t-1)}} X_i$ and $\bar Y_{k\cU} = |\cC_k^{(t-1)}|^{-1}\sum_{i \in \cC_k^{(t-1)}} Y_i\,$.

  Similar to the $K$-means clustering algorithm, the $K$-lines clustering algorithm is not guaranteed to find the global minimizer, $\argmin_{B_K} W(B_K, P_n)$. Empirically, we run the $K$-lines clustering algorithm for $M$ times with random initializations and obtain $M$ multisets of unspecified line centers $B_K^{(1)}, \ldots, B_K^{(M)}$. Then we set $\widehat B_{K\cU} \in \argmin_{B_K \in \{B_K^{(1)}, \ldots, B_K^{(M)}\}} W(B_K, P_n)$.

Powered by the $K$-lines algorithm, we introduce the sample surrogate indices $\widehat Z_i \in \{1,\ldots,K\}$, $i=1,\ldots,n$, based on which we then define the sample-level generalized Pearson correlation square for this line-membership unspecified scenario.

\begin{definition}\label{def:tilde_Z_i}
	Suppose that Algorithm \ref{alg:K-lines} outputs $K$ unspecified line centers $\widehat\bbeta_{1\cU}, \ldots, \widehat\bbeta_{K\cU}$. Also suppose that the probability that $(X_i, Y_i)$ is equally close to more than one line centers is zero. For each $(X_i,Y_i)$, we define its sample surrogate index
\begin{equation}\label{eq:tilde_Zi}
	\widehat Z_i = \argmin_{k \in \{1,\ldots,K\}} d_\perp \left( (X_i,Y_i)\tran, \widehat\bbeta_{k\cU} \right) \,,\: i=1,\ldots,n\,.
\end{equation} 
\end{definition}

\begin{definition}\label{def:u_sample_gen_R2}
At the sample level, when observations $\{(X_i,Y_i)\}_{i=1}^n$ have line memberships unspecified, the generalized Pearson correlation square is defined as
\begin{align}
	R_{\cG\cU}^2 &= \sum_{k=1}^K \widehat p_{k\cU} \cdot \widehat\rho_{k\cU}^2\,,
\label{u_scor2_g}
\end{align}
where the subindex $\cG$ stands for generalized, $\cU$ stands for unspecified, $\widehat p_{k\cU} = n^{-1} \sum_{i=1}^n\1(\widehat{Z}_i = k)$, and
\begin{align*}
\widehat \rho_{k\cU}^2 &= \frac{\left\{\sum_{i=1}^n \left(X_i - \bar X_{k\cU}\right) \left(Y_i - \bar Y_{k\cU}\right) \1(\widehat{Z}_i = k)\right\}^2}{\left\{\sum_{i=1}^n \left(X_i - \bar X_{k\cU}\right)^2\1(\widehat{Z}_i = k)\right\}\left\{\sum_{i=1}^n \left(Y_i - \bar Y_{k\cU}\right)^2\1(\widehat{Z}_i = k)\right\}}\,, 
\end{align*} 
with $\bar X_{k\cU} = n_{k\cU}^{-1}\sum_{i=1}^n X_i \1(\widehat{Z}_i = k)$, $\bar Y_{k\cU} = n_{k\cU}^{-1}\sum_{i=1}^n Y_i \1(\widehat{Z}_i = k)$, and $n_{k\cU} = \sum_{i=1}^n\1(\widehat{Z}_i = k)$.
\end{definition}

\subsection{Data-driven choice of $K$ in the unspecified scenario} \label{subsec:choose_K}
When users do not have prior knowledge about the value of $K$, how to choose $K$ becomes an important question in practice. Some methods for choosing $K$ in $K$-means clustering can be adapted. For example, the elbow method, though not being theoretically principled, is visually appealing to practitioners and widely used. It employs a scree plot whose horizontal axis displays a range of $K$ values, and whose vertical axis shows the average within-cluster sum of squared distances corresponding to each $K$. For our $K$-lines algorithm, it is reasonable to use a scree plot to show how $W(B_K, P_n)$, the average within-cluster squared perpendicular distance defined in (\ref{eq:W_Pn}), decreases as $K$ increases. 

Alternatively, when it is reasonable to assume that $(X, Y) \mid (Z=k)$ follows a bivariate Gaussian distribution for all $k=1,\ldots,K$, one may use the Akaike information criterion (AIC) \citep{akaike1998information} to choose $K$. Specifically, AIC is defined as
\begin{align}\label{eq:AIC}
&\text{AIC}(K) = 2(6K-1) - 2 \sum_{i=1}^n \log  p\left(X_i, Y_i \mid \{ \widehat p_{k\cU}, \widehat \bmu_{k\cU}, \widehat \bSigma_{k\cU} \}_{k=1}^K \right)\\
=& 2(6K-1) - \pi^{-1} \sum_{i=1}^n \log \left[ \sum_{k=1}^K \widehat p_{k\cU} \left|\widehat\bSigma_{k\cU}\right|^{-1/2} \exp \left\{ - \frac{1}{2} \left((X_i,Y_i)\tran - \widehat\bmu_{k\cU}\right)\tran \widehat\bSigma_{k\cU}^{-1} \left((X_i,Y_i)\tran - \widehat\bmu_{k\cU}\right) \right\} \right] \,, \notag
\end{align}
where the first term is $2(6K-1)$ because there are $6$ parameters for each component and the component proportions sum to $1$; in the second term, $\widehat p_{k\cU} = |\cC_k|/n$, $\widehat\bmu_{k\cU} = \left( \bar X_{k\cU}, \bar Y_{k\cU} \right)\tran$, and \[\widehat\bSigma_{k\cU} = |\cC_k|^{-1} \left[ \begin{array}{cc}
 		\sum_{i \in \cC_k}(X_i - \bar X_{k\cU})^2
 		&
 		\sum_{i \in \cC_k}(X_i - \bar X_{k\cU})(Y_i - \bar Y_{k\cU})
 		\\
 		\sum_{i \in \cC_k}(X_i - \bar X_{k\cU})(Y_i - \bar Y_{k\cU})
 		&
 		\sum_{i \in \cC_k}(Y_i - \bar Y_{k\cU})^2
 	\end{array}
	\right]\,.\]
We will demonstrate the elbow method and the AIC method in Section \ref{subsec:choose_K_res}. 

\section{Population-level generalized Pearson correlation squares and asymptotics}\label{sec:asym_theory}
In this section, we will define the population-level generalized Pearson correlation squares and derive the asymptotic theory for the corresponding sample-level statistics.  The foundation of our work is the population-level Pearson correlation coefficient: $\rho = \cov(X, Y)/\{\var(X)\var(Y)\}^{-1/2} \in [-1, 1]$, where $\cov(X,Y) = \E[\{X-\E (X)\}\{Y-\E (Y)\}]$, $\var(X) = \E\{[X-\E (X)]^2\}$, and $\var(Y) = \E\{[Y-\E (Y)]^2\}$ denote the covariance between $X$ and $Y$, the variance of $X$, and the variance of $Y$, respectively. We say that $X$ and $Y$ are linearly dependent if $\rho \neq 0$.


\subsection{Line-membership specified scenario} \label{subsec:pop_R2gs}
Under the line-membership specified scenario, we denote $p_{k\cS} = \pr(Z = k)$, $k=1,\ldots, K$. Conditional on $Z=k$, the population-level Pearson correlation between $X$ and $Y$ is $\rho_{k\cS} = \cov(X, Y \mid Z=k)/\{\var(X \mid Z=k) \var(Y \mid Z=k)\}^{1/2}$, if $\var(X \mid Z=k) > 0$ and $\var(Y \mid Z=k) > 0$; otherwise, $\rho_{k\cS} = 0$.
In the special case of $K=1$, $\rho_{1\cS}^2 = \rho^2= \cov^2(X, Y)/\{\var(X)\var(Y)\}$ is the population-level Pearson correlation square that indicates the population-level strength of a linear dependence. Motivated by this, we combine $\rho_{1\cS}^2, \ldots, \rho_{K\cS}^2$ into one measure to indicate the overall strength of $K$ linear dependences.

\begin{definition}\label{def:pop_gen_R2}
At the population level, when the line membership variable $Z$ is specified, the generalized Pearson correlation square between $X$ and $Y$ is defined as
\begin{align}
	\rho_{\cG\cS}^2 &= \E_Z\left(\rho_{Z\cS}^2\right) = \E_Z \left\{\frac{\cov^2(X, Y \mid Z)}{\var(X \mid Z) \var(Y \mid Z)}\right\} = \sum_{k=1}^K p_{k\cS} \; \rho_{k\cS}^2\,.
\label{pcor2_g}
\end{align}
which is a weighted sum of $\rho_{1\cS}^2,\ldots,\rho_{K\cS}^2$, i.e., the strengths of the $K$ linear dependences, with weights as $p_{1\cS},\ldots,p_{K\cS}$. Note that the subindex $\cG$ stands for generalized, and $\cS$ stands for specified.
\end{definition} 

We use $K$ lines to represent the joint distribution of $(X, Y, Z)$, based on the perpendicular distance $d_\perp(\cdot,*)$ defined in (\ref{def:perp_dist}), the conditional means $\mu_{X,k\cS} = \E(X \mid Z=k)$ and $\mu_{Y,k\cS} = \E(Y \mid Z=k)$, and the conditional covariance matrix 
\[ \bSigma_{k\cS} = \begin{bmatrix}
 	\var(X \mid Z=k) & \cov(X, Y \mid Z=k)\\
 	\cov(X, Y \mid Z=k) & \var(Y \mid Z=k)
 	\end{bmatrix}\,.
\]

\begin{definition}\label{def:line_center}
The population-level specified line center of the $k$-th component is 
\begin{equation}\label{def:beta_ks}
	\bbeta_{k\cS} \in \argmin_{\bbeta} \E\left\{ d_\perp^2\left( (X,Y)\tran, \bbeta \right) \mid Z=k\right\}\,.
\end{equation}
\end{definition}
\noindent Let $\bu_{1,k\cS}=(u_{11,k\cS}, u_{12,k\cS})\tran$ be the eigenvector associated with  the largest eigenvalue of $\bSigma_{k\cS}$. Then $\bbeta_{k\cS} = (a_{k\cS}, b_{k\cS}, c_{k\cS})\tran$ is the line 
	 \[\left\{ (x,y)\tran: u_{12,k\cS}\left(x - \mu_{X,k\cS}\right) - u_{11,k\cS}\left(y - \mu_{Y,k\cS}\right) = 0 \right\}\,,\]
where $a_{k\cS} = u_{12,k\cS}$, $b_{k\cS} = -u_{11,k\cS}$, and $c_{k\cS} = -u_{12,k\cS}\mu_{X,k\cS} + u_{11,k\cS}\mu_{Y,k\cS}$.


\subsection{Line-membership unspecified scenario} \label{subsec:pop_R2gu}
Under the line-membership unspecified scenario, we investigate a mixture of $K$ linear dependences between $X$ and $Y$ without observing any index variable $Z$. Motivated by Definition \ref{def:line_center}, we define the population-level line centers for the unspecified scenario as the $K$ lines that minimize the expected squared perpendicular distance of $(X, Y)$ to its closest line.
\begin{definition}\label{def:line_center_u}
We define the expected within-cluster squared perpendicular distance as
\begin{align}\label{eq:W_P}
W(B_K, P) &= \E\left\{ \min_{\bbeta \in B_K} d_\perp^2\left( (X,Y)\tran, \bbeta \right) \right\}\,, 
\end{align}
where $P$ is the joint probability measure of $(X,Y)$. Then we define a multiset of population-level unspecified line centers, $B_{K\cU} = \{\bbeta_{1\cU}, \ldots, \bbeta_{K\cU}\}$, where $\bbeta_{k\cU} =\left(a_{k\cU}, b_{k\cU}, c_{k\cU}\right)\tran$ is the $k$-th line center, as
\begin{equation}\label{def:beta_ku}
	B_{K\cU} \in \argmin_{B_K} W(B_K, P)\,.
\end{equation} 
\end{definition}


Provided that $B_{K\cU}$ is uniquely determined, we define a random surrogate index $\widetilde Z \in \{1,\ldots,K\}$ as the index of the line center to which $(X, Y)$ is closest.

\begin{definition}\label{def:tilde_Z}
	Suppose that the unspecified line centers $\bbeta_{1\cU}, \ldots, \bbeta_{K\cU}$ at the population level are unique. Also, suppose that the probability that $(X, Y)$ is equally close to multiple line centers is zero. We define a random surrogate index as
\begin{equation}\label{eq:tilde_Z}
	\widetilde Z = \argmin_{k \in \{1,\ldots,K\}} d_\perp \left( (X,Y)\tran, \bbeta_{k\cU} \right) \,.
\end{equation} 
\end{definition}

Motivated by $\rho_{\cG\cS}^2$, we define the population-level generalized Pearson correlation square for the line-membership unspecified scenario, based on $(X, Y, \widetilde Z)$.

\begin{definition}\label{def:u_pop_gen_R2}
At the population level, when no line membership variable is specified (the ``unspecified scenario"), the generalized Pearson correlation square between $X$ and $Y$ is defined as
\begin{equation}\label{u_pcor2_g}
\rho_{\cG\cU}^2 = \sum_{k=1}^K p_{k\cU} \; \rho_{k\cU}^2\,,
\end{equation}	
where the subindex $\cG$ stands for generalized, $\cU$ stands for unspecified, $p_{k\cU} = \pr ( \widetilde Z = k )$, and $\rho_{k\cU}^2 = \cov^2(X,Y \mid \widetilde Z = k)/\{\var(X \mid \widetilde Z = k)\, \var(Y \mid \widetilde Z = k)\}$.
\end{definition}

\begin{remark}
Relations and distinctions between the specified and unspecified scenarios. 
\begin{enumerate}
\item $B_{K\cU} \neq B_{K\cS}$, which is reasonable as $B_{K\cU}$ does not have information about $Z$.
\item $\rho_{\cG\cU}^2 \ge \rho_{\cG\cS}^2$. The proof is in the Supplementary Material.	
\item $R_{\cG\cU}^2$ is not an estimator of $\rho_{\cG\cS}^2$; rather, it is an estimator of $\rho_{\cG\cU}^2$. Our construction does not rely on a specific distributional assumption, e.g., bivariate Gaussian mixture model. This property makes the generalized Pearson correlation squares more flexible, just as the Pearson correlation that does not rely on any distributional assumptions either. If the goal were to use $R_{\cG\cU}^2$ as an estimator of $\rho_{\cG\cS}^2$, a specific mixture model must be assumed. Then the $K$-lines algorithm should be replaced by the Expectation-Maximization (EM) algorithm to decide the sample surrogate indices $\widehat{Z}_1, \ldots, \widehat{Z}_n$. When the EM algorithm converges to the global optimum and returns the maximum-likelihood estimates of mixture model parameters, the corresponding $R_{\cG\cU}^2$ will be an asymptotically unbiased estimator of $\rho_{\cG\cS}^2$.
\end{enumerate}	
\end{remark}

To enable statistical inference of the population-level measures $\rho_{\cG\cS}^2$ (\ref{pcor2_g}) and $\rho_{\cG\cU}^2$ (\ref{u_pcor2_g}), we derive the first-order asymptotics of the sample-level measures $R_{\cG\cS}^2$ (\ref{scor2_g}) and $R_{\cG\cU}^2$ (\ref{u_scor2_g}). 

\subsection{Asymptotic distributions of $R_{\cG\cS}^2$ and $R_{\cG\cU}^2$}

We first derive the asymptotics of $R_{\cG\cS}^2$, which is also a foundation for the asymptotics of $R_{\cG\cU}^2$.
\begin{theorem}\label{thm_asym_dist_scor2_gc}
Under the line-membership specified scenario, we define 
\[ \mu_{X^c Y^d,k\cS} = \E\left\{\left(\frac{X - \E(X \mid Z=k)}{\var(X \mid Z=k)^{1/2}}\right)^c \left.\left(\frac{Y - \E(Y \mid Z=k)}{\var(Y \mid Z=k)^{1/2}}\right)^d \right| Z=k\right\}\,, \;  c, d \in \N\,,
\]
Assume $\mu_{X^4,k\cS} < \infty$ and $\mu_{Y^4,k\cS}< \infty$ for all $k=1,\ldots, K$. Then

\begin{equation}
\sqrt{n} \left(R_{\cG\cS}^2 - \rho_{\cG\cS}^2\right) \longrightarrow \cN\left(0, \sum_{k=1}^K \left( A_{k\cS} + B_{k\cS}\right) + 2 \mathop{\sum \sum}_{1 \le k < r \le K} C_{kr\cS}\right) \text{ in distribution, where} \label{asym_dist_scor2_gc}	
\end{equation}
\begin{align*}
	A_{k\cS} &= p_{k\cS} \left[ \rho_{k\cS}^4 \left(\mu_{X^4,k\cS} + 2 \mu_{X^2Y^2,k\cS} + \mu_{Y^4,k\cS}\right)  - 4\rho_{k\cS}^3 \left( \mu_{X^3Y,k\cS} + \mu_{XY^3,k\cS} \right) + 4\rho_{k\cS}^2 \mu_{X^2Y^2,k\cS}\right],\\
	B_{k\cS} &= p_{k\cS} \left(1-p_{k\cS}\right)\rho_{k\cS}^4 \,, \quad \text{and} \quad C_{kr\cS} = - \, p_{k\cS} \, p_{r\cS} \, \rho_{k\cS}^2 \, \rho_{r\cS}^2\,. 
\end{align*}
\end{theorem}

Note that Theorem \ref{thm_asym_dist_scor2_gc} does not rely on any distributional assumptions. When it is applied to the special case where $(X, Y)|Z$ follows a bivariate Gaussian distribution, we obtain a much simpler form of the first-order asymptotic distribution of $R_{\cG\cS}^2$.
\begin{corollary}\label{cor:asym_dist_scor2_gc}
	Under the special case where $(X, Y) \mid (Z=k)$ follows a bivariate Gaussian distribution for all $k=1,\ldots,K$, the asymptotic variance of $\sqrt{n} (R_{\cG\cS}^2 - \rho_{\cG\cS}^2)$ in Theorem \ref{thm_asym_dist_scor2_gc} is simplified and becomes 
\begin{align}\label{asym_dist_gauss_scor2_gc}
\sum_{k=1}^K \left[ 4 \, p_{k\cS} \, \rho_{k\cS}^2 \left( 1 - \rho_{k\cS}^2 \right)^2 + p_{k\cS} \left(1-p_{k\cS}\right)\rho_{k\cS}^4\right] - 2 \mathop{\sum \sum}_{1 \le k < r \le K} p_{k\cS} \, p_{r\cS} \, \rho_{k\cS}^2 \, \rho_{r\cS}^2\,,  
\end{align}
which only depends on $p_{k\cS}$ and $\rho_{k\cS}^2$. 
\end{corollary}


To derive an analog of Theorem \ref{thm_asym_dist_scor2_gc} and Corollary \ref{cor:asym_dist_scor2_gc} for the unspecified scenario, we need to show that each sample surrogate index $\widehat{Z}_i$, $i=1,\ldots,n$, converges in distribution to the random surrogate index $\widetilde Z$. A sufficient condition is the strong consistency of the $K$ sample-level unspecified line centers $\widehat B_{K\cU} = \{ \widehat\bbeta_{1\cU}, \ldots, \widehat\bbeta_{K\cU} \}$ to the $K$ population-level unspecified line centers $B_{K\cU} = \{ \bbeta_{1\cU}, \ldots, \bbeta_{K\cU} \}$. 

\begin{theorem} \label{thm_strong_consistency}
Suppose that $\int \norm{(x,y)\tran}^2 \pr\left( (dx, dy)\tran \right) < \infty$ and that for each $k=1,\ldots,K$ there is a unique multiset $B_{k\cU} = \argmin_{B_k} W(B_k, P)$. Also assume that the globally optimal sample-level unspecified line centers $\widehat{B}_{K\cU} = \argmin_{B_K} W(B_K, P_n)$ is attained and unique. Then as $n \rightarrow \infty$, $\widehat{B}_{K\cU} \rightarrow B_{K\cU}$ almost surely, and $W(\widehat{B}_{K\cU}, P_n) \rightarrow W(B_{K\cU}, P)$ almost surely.
\end{theorem}
The first statement of Theorem \ref{thm_strong_consistency} means that there exists an ordering of the elements in $\widehat B_{K\cU} = \{ \widehat\bbeta_{1\cU}, \ldots, \widehat\bbeta_{K\cU} \}$ and $B_{K\cU} = \left\{ \bbeta_{1\cU}, \ldots, \bbeta_{K\cU} \right\}$ such that as the sample size $n \rightarrow \infty$,
\begin{equation}\label{beta_k_strong_consistency}
\widehat\bbeta_{k\cU} \rightarrow \bbeta_{k\cU} \; \text{ almost surely}\,,\; k=1,\ldots,K\,.	
\end{equation}

Based on Theorems \ref{thm_asym_dist_scor2_gc} and \ref{thm_strong_consistency}, we derive the asymptotic distribution of $R^2_{\cG\cU}$. 

\begin{theorem}\label{thm_asym_dist_scor2_gu}
Under the line-membership unspecified scenario, we define
\[ \mu_{X^c Y^d, k\cU} = \E\left\{\left(\frac{X - E[X|\widetilde Z=k]}{\var(X|\widetilde Z=k)^{1/2}}\right)^c \left.\left(\frac{Y - E[Y|\widetilde Z=k]}{\var(Y|\widetilde Z=k)^{1/2}}\right)^d \right| \widetilde Z=k\right\}\,, \; c, d \in \N\,,
\]
where $\widetilde Z$ is the random surrogate index defined in (\ref{eq:tilde_Z}). Assume $\mu_{X^4, k\cU} < \infty$ and $\mu_{Y^4, k\cU} < \infty$ for all $k=1,\ldots,K$. Then
\begin{equation}\label{asym_dist_scor2_gu}
\sqrt{n} \left( R_{\cG\cU}^2 - \rho_{\cG\cU}^2 \right) \longrightarrow \cN\left(0, \sum_{k=1}^K \left( A_{k\cU} + B_{k\cU}\right) + 2 \mathop{\sum \sum}_{1 \le k < r \le K} C_{kr\cU}\right) \text{ in distribution, where} 
\end{equation}
\begin{align*}
	A_{k\cU} &= p_{k\cU} \left[ \rho_{k\cU}^4 \left(\mu_{X^4,k\cU} + 2 \mu_{X^2Y^2,k\cU} + \mu_{Y^4,k\cU} \right)  - 4\rho_{k\cU}^3 \left( \mu_{X^3Y,k\cU} + \mu_{XY^3,k\cU} \right) + 4\rho_{k\cU}^2 \mu_{X^2Y^2,k\cU} \right],\\
	B_{k\cU} &= p_{k\cU} \left(1-p_{k\cU}\right)\rho_{k\cU}^4\,, \quad \text{and} \quad C_{kr\cU} = - \, p_{k\cU} \, p_{r\cU} \, \rho_{k\cU}^2 \, \rho_{r\cU}^2\,.
\end{align*}
\end{theorem}

Applying Theorem \ref{thm_asym_dist_scor2_gu} to the special case where $(X, Y)|\widetilde Z$ follows a bivariate Gaussian distribution, we obtain a much simpler form of the first-order asymptotic distribution of $R_{\cG\cU}^2$.

\begin{corollary}\label{cor:asym_dist_scor2_gu}
	When $(X, Y) \mid (\widetilde Z=k)$ follows a bivariate Gaussian distribution for all $k=1,\ldots,K$, the asymptotic variance of $\sqrt{n} \left(R_{\cG\cU}^2 - \rho_{\cG\cU}^2\right)$ is simplified and becomes 
\begin{align}\label{asym_dist_gauss_scor2_gu}
\sum_{k=1}^K \left[ 4 \, p_{k\cU} \, \rho_{k\cU}^2 \left( 1 - \rho_{k\cU}^2 \right)^2 + p_{k\cU} \left(1-p_{k\cU}\right)\rho_{k\cU}^4\right] - 2 \mathop{\sum \sum}_{1 \le k < r \le K} p_{k\cU} \, p_{r\cU} \, \rho_{k\cU}^2 \, \rho_{r\cU}^2\,,  	
\end{align}
which only depends on $p_{k\cU}$ and $\rho_{k\cU}^2$. 
\end{corollary}

\begin{remark}
When $K=1$, the asymptotic distributions of $R^2_{\cG\cS}$ and $R^2_{\cG\cU}$ in Theorems \ref{thm_asym_dist_scor2_gc} and \ref{thm_asym_dist_scor2_gu} both reduce to the asymptotic distribution of $R^2$ \citep{ferguson2017course}. In the special case that $K=1$ and $(X,Y)$ follows bivariate Gaussian distribution with correlation $\rho$, the asymptotic distributions of $R^2_{\cG\cS}$ and $R^2_{\cG\cU}$ in Corollaries \ref{cor:asym_dist_scor2_gc} and \ref{cor:asym_dist_scor2_gu} both reduce to $\sqrt{n}(R^2 - \rho^2) \tod \cN\left(0, 4\rho^2\left(1-\rho^2\right)^2\right)$.   
\end{remark}

All the proofs are in the Supplementary Material. In Section \ref{sec:num_verify_simulation}, we will numerically show that the asymptotic distribution in Theorem \ref{thm_asym_dist_scor2_gu} works reasonably well when $K$ is chosen by AIC.

\section{Numerical simulations}\label{sec:simulation}

In this section, we perform simulation studies 
to numerically verify the theoretical results in Section \ref{sec:asym_theory} and to compare our generalized Pearson correlation squares with multiple existing association measures in terms of statistical power. We also demonstrate the effectiveness of our proposed approaches for choosing $K$, the number of line components in the line-membership unspecified scenario.  

\subsection{Numerical verification of theoretical results}\label{sec:num_verify_simulation}

\begin{figure}[htbp]
	\centering
	\includegraphics[width=.8\textwidth]{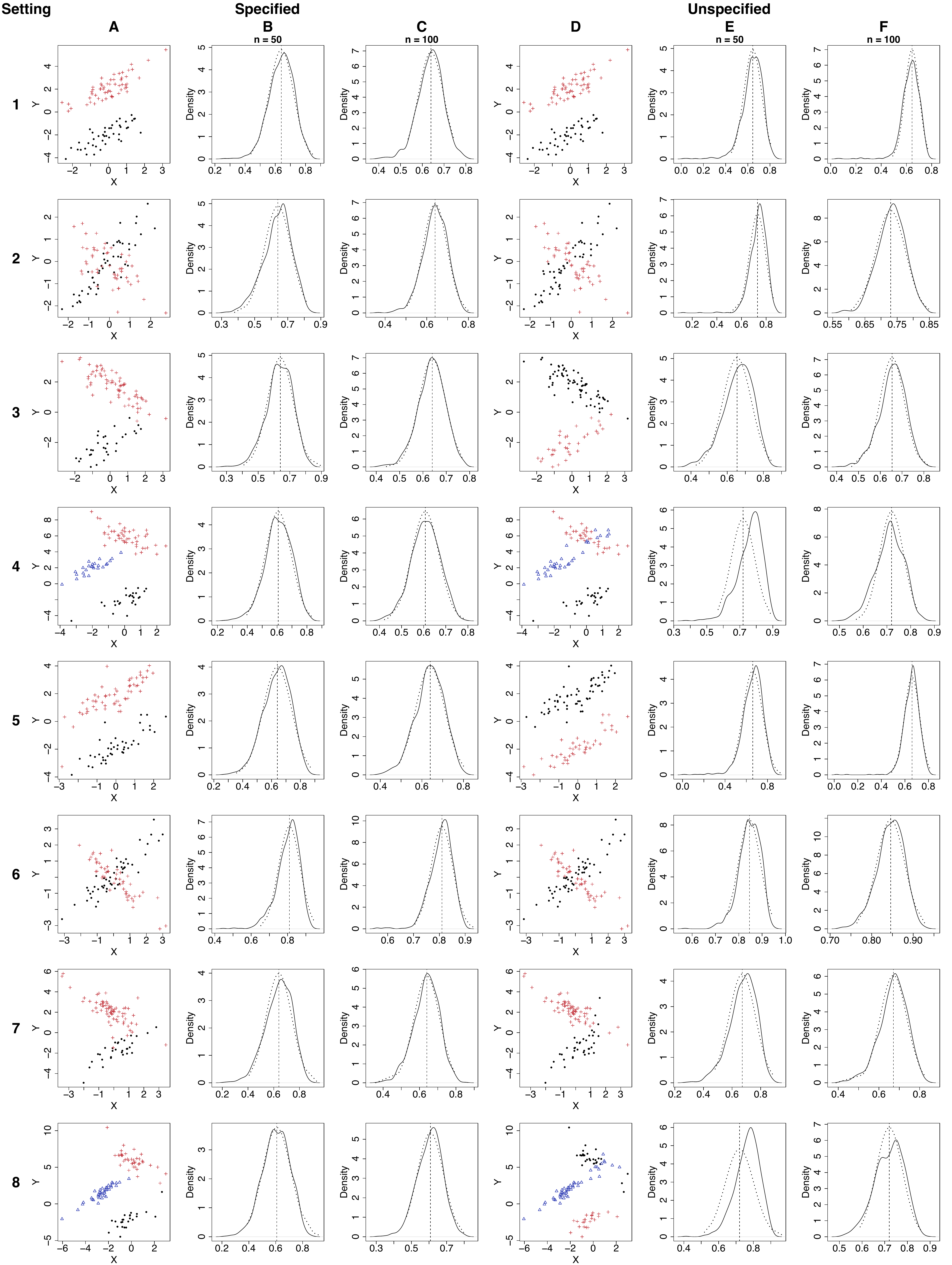}
	\caption{Comparison of the asymptotic distributions and the finite-sample distributions of $R^2_{\cG\cS}$ and $R^2_{\cG\cU}$. A: Example samples with $n=100$; colors and symbols represent values of $Z$. B-C: Finite-sample distributions $n=50$ or $100$ (black solid curves) vs. the asymptotic distribution (black dotted curves) of $R^2_{\cG\cS}$; the vertical dashed lines mark the values of $\rho^2_{\cG\cS}$. D: Example samples with $n=100$; colors and symbols represent values of $\widetilde Z$ inferred by the $K$-lines algorithm. E-F: Finite-sample distributions of $n=50$ or $100$ (black solid curves) vs. the asymptotic distribution (black dotted curves) of $R^2_{\cG\cU}$; the vertical dashed lines mark the values of $\rho^2_{\cG\cU}$.}\label{fig:theory_vs_sim}
\end{figure}

We first compare the asymptotic distributions in Section \ref{sec:asym_theory} with numerically simulated finite-sample distributions under eight settings (Table \ref{tab1:simu_settings}), where $(X, Y) \mid Z$ follows a bivariate Gaussian distribution under the first four settings and a bivariate $t$ distribution under the latter four settings. Under each setting, we generate $B=1000$ samples with sizes $n=50$ or $100$, calculate $R^2_{\cG\cS}$ and $R^2_{\cG\cU}$ on each sample, and compare the simulated finite-sample distributions of $R^2_{\cG\cS}$ and $R^2_{\cG\cU}$ to the corresponding asymptotic distributions. In the first four settings, the asymptotic distributions are from Corollaries \ref{cor:asym_dist_scor2_gc} and \ref{cor:asym_dist_scor2_gu} (the bivariate Gaussian results); in the latter four settings, the asymptotic distributions are from Theorems \ref{thm_asym_dist_scor2_gc} and \ref{thm_asym_dist_scor2_gu} (the general results). The comparison results (Fig. \ref{fig:theory_vs_sim}) show that the finite-sample distributions and the asymptotic results have good agreement, justifying the use of the asymptotic distributions for statistical inference of $\rho^2_{\cG\cS}$ or $\rho^2_{\cG\cU}$ on a finite sample.

\begin{table}[htbp]
\renewcommand{\arraystretch}{0.6}
\begin{tabular}{cccc}
\hline
\bf Setting & $K$ & \bf Population & \bf Parameters \\
\hline
1 & $K=2$ & & $p_1=p_2=0.5$\\
 & & & $\bmu_1 = (0,-2)\tran$, $\bmu_2 = (0,2)\tran$\\
 & & & $\bSigma_1 = \bSigma_2 = \begin{bmatrix}
 1 & 0.8\\
 0.8 & 1	
 \end{bmatrix}
$\\
 & & & \\
2 & $K=2$& \bf Specified: & $p_1=p_2=0.5$\\
 & & $\Prob(Z=k)=p_k$ & $\bmu_1 = \bmu_2 = (0,0)\tran$\\
 & & $(X,Y)|(Z=k) \sim \cN\left(\bmu_k, \bSigma_k\right)$ & $\bSigma_1 = \begin{bmatrix}
 1 & 0.8\\
 0.8 & 1	
 \end{bmatrix}
$, $\bSigma_2 = \begin{bmatrix}
 1 & -0.8\\
 -0.8 & 1	
 \end{bmatrix}
$\\
 & & $k=1,\ldots,K$ & \\
3 & $K=2$& & $p_1=0.3$, $p_2=0.7$\\
 & & \bf Unspecified: & $\bmu_1 = (0,-2)\tran$, $\bmu_2 = (0,2)\tran$\\
 & & $\sum_{k=1}^K p_k \,\cN\left(\bmu_k, \bSigma_k\right)$ & $\bSigma_1 = \begin{bmatrix}
 1 & 0.8\\
 0.8 & 1	
 \end{bmatrix}
$, $\bSigma_2 = \begin{bmatrix}
 1 & -0.8\\
 -0.8 & 1	
 \end{bmatrix}
$\\
 & & & \\
4 & $K=3$& & $p_1=0.25$, $p_2=0.5$, $p_3=0.25$\\
 & & & $\bmu_1 = (0,-2)\tran$, $\bmu_2 = (0,6)\tran$, $\bmu_3 = (-2,2)\tran$\\
 & & & $\bSigma_1 = \begin{bmatrix}
 1 & 0.8\\
 0.8 & 1	
 \end{bmatrix}
$, $\bSigma_2 = \begin{bmatrix}
 1 & -0.7\\
 -0.7 & 1	
 \end{bmatrix}
$, $\bSigma_3 = \begin{bmatrix}
 1 & 0.9\\
 0.9 & 1	
 \end{bmatrix}
$\\
 & & & \\
\hline
5 & $K=2$ & & $p_1=p_2=0.5$, $\nu_1=\nu_2=8$\\
 & & & $\bmu_1 = (0,-2)\tran$, $\bmu_2 = (0,2)\tran$\\
 & & & $\bSigma_1 = \bSigma_2 = \begin{bmatrix}
 1 & 0.8\\
 0.8 & 1	
 \end{bmatrix}
$\\
 & & & \\
6 & $K=2$ & \bf Specified: & $p_1=p_2=0.5$, $\nu_1=\nu_2=8$\\
 & & $\Prob(Z=k)=p_k$ & $\bmu_1 = \bmu_2 = (0,0)\tran$\\
 & & $(X,Y)|(Z=k) \sim t_{\nu_k}\left(\bmu_k, \bSigma_k\right)$ & $\bSigma_1 = \begin{bmatrix}
 1 & 0.8\\
 0.8 & 1	
 \end{bmatrix}
$, $\bSigma_2 = \begin{bmatrix}
 1 & -0.8\\
 -0.8 & 1	
 \end{bmatrix}
$\\
 & & $k=1,\ldots,K$ & \\
7 & $K=2$ & & $p_1=0.3$, $p_2=0.7$, $\nu_1=\nu_2=8$\\
 & & \bf Unspecified: & $\bmu_1 = (0,-2)\tran$, $\bmu_2 = (0,2)\tran$\\
 & & $\sum_{k=1}^K p_k \,t_{\nu_k}\left(\bmu_k, \bSigma_k\right)$ & $\bSigma_1 = \begin{bmatrix}
 1 & 0.8\\
 0.8 & 1	
 \end{bmatrix}
$, $\bSigma_2 = \begin{bmatrix}
 1 & -0.8\\
 -0.8 & 1	
 \end{bmatrix}
$\\
 & & & \\
8 & $K=3$ & & $p_1=0.25$, $p_2=0.5$, $p_3=0.25$\\
 & & & $\nu_1=\nu_2=\nu_3=8$\\
 & & & $\bmu_1 = (0,-2)\tran$, $\bmu_2 = (0,6)\tran$, $\bmu_3 = (-2,2)\tran$\\
 & & & $\bSigma_1 = \begin{bmatrix}
 1 & 0.8\\
 0.8 & 1	
 \end{bmatrix}
$, $\bSigma_2 = \begin{bmatrix}
 1 & -0.7\\
 -0.7 & 1	
 \end{bmatrix}
$, $\bSigma_3 = \begin{bmatrix}
 1 & 0.9\\
 0.9 & 1	
 \end{bmatrix}
$\\
 & & & \\
\hline
\end{tabular}
\caption{Eight settings in simulation studies (Section \ref{sec:simulation}). In the settings $1$--$4$, $\cN(\bmu_k,\bSigma_k)$ represents a bivariate Gaussian distribution with the mean vector $\bmu_k$ and the covariance matrix $\bSigma_k$. In the settings $5$--$8$, $t_{\nu_k}(\bmu_k,\bSigma_k)$ represents a bivariate $t$ distribution with the degrees of freedom $\nu_k$, the location vector $\bmu_k$ and the shape matrix $\bSigma_k$.\label{tab1:simu_settings}}
\end{table}

In practice, $K$ often needs to be found in a data-driven way under the line-membership unspecified scenario. To verify the behavior of $R^2_{\cG\cU}$ when $K$ is chosen by the AIC in \eqref{eq:AIC}, we conduct another simulation study to compare $R^2_{\cG\cU}$'s finite-sample distributions with the asymptotic distributions. The results (Fig. \ref{fig:theory_vs_sim_aic} in the Supplementary Material) show that when $n=100$, the agreement between finite-sample distributions and asymptotic distributions is still reasonably good. This observation justifies the use of our asymptotic results in practice.

However, the asymptotic distributions in Section \ref{sec:asym_theory} involve unobservable parameters in the asymptotic variance terms. 
A classical solution is to plug-in estimates of these parameters. Another common inferential approach is to use the bootstrap, which is computationally more intensive, instead of the closed-form asymptotic distributions. Here we numerically verify whether the plug-in approach works reasonably well for statistical inference of $\rho^2_{\cG\cS}$ and $\rho^2_{\cG\cU}$. Under each of the eight settings, we simulate two samples with sizes $n=50$ and $100$, respectively. We then use each sample to construct a $95\%$ confidence interval (CI) of $\rho^2_{\cG\cS}$ and $\rho^2_{\cG\cU}$ as $R^2_{\cG\cS} \pm 1.96 \text{se}(R^2_{\cG\cS}) $ and $R^2_{\cG\cU} \pm 1.96 \text{se}(R^2_{\cG\cU})$, respectively. We construct the standard errors $\text{se}(R^2_{\cG\cS})$ and $\text{se}(R^2_{\cG\cU})$ in two ways: square roots of (a) the plug-in estimates of the asymptotic variances of $R^2_{\cG\cS}$ and $R^2_{\cG\cU}$, or (b) the bootstrap estimates of $\var(R^2_{\cG\cS})$ and $\var(R^2_{\cG\cU})$. We also calculate the true asymptotic variances
 of  $R^2_{\cG\cS}$ and $R^2_{\cG\cU}$ based on true parameter values and use them to construct the theoretical CIs. The results (Fig. \ref{fig:theory_vs_plugin_vs_boot} in the Supplementary Material) show that the plug-in and bootstrap approaches construct similar CIs on the same sample. When $n$ increases from $50$ to $100$, the CIs constructed by both approaches agree better with the theoretical CIs.

We also evaluate the coverage probabilities of the $95\%$ CIs constructed by the plug-in approach and compare them with those of the theoretical CIs. Table \ref{tab:cover_prob} summarizes the results. The theoretical CIs have coverage probabilities close to $95\%$ under all the eight settings, providing additional verification of the asymptotic distributions. Overall the plug-in confidence intervals have good coverage probabilities, which are increasingly closer to $95\%$ as $n$ increases. Their coverage probabilities are in general closer to $95\%$ under the first four bivariate Gaussians settings than under the last four bivariate $t$ settings. The reason is that mixtures of bivariate Gaussians are more concentrated on $K$ lines and better allow the $K$-lines algorithm to find the sample-level unspecified line centers, thus reducing the unwanted variance due to failed algorithm convergence and making the plug-in variance estimate of $R_{\cG\cU}^2$ more accurate. Comparing the line-membership specified and unspecified scenarios, the plug-in confidence intervals, as expected, have better coverage probabilities under the specified scenario that has less uncertainty. 
Table \ref{tab:cover_prob} also shows that the two plug-in options do not have obvious differences, suggesting that the first plug-in option (``P1''), which uses the asymptotic variances in the special bivariate Gaussian forms (Corollaries \ref{cor:asym_dist_scor2_gc} and \ref{cor:asym_dist_scor2_gu}), is robust and can be used in practice for its simplicity. 

\begin{table}[htbp]
\centering
\begin{tabular}{ccccc|ccc}
\hline
\bf Setting & $n$ & \multicolumn{6}{c}{\bf 95\% CI Coverage Probability}\\
\cline{3-8}
& & \multicolumn{3}{c}{\bf Specified} & \multicolumn{3}{c}{\bf Unspecified}\\
& & Asymp. & P1 & P2 & Asymp. & P1 & P2 \\
\hline
1 & $50$ & $.947$ & $.933$ & & $.954$ & $.916$ & \\
 & $100$ & $.959$ & $.947$ & & $.934$ & $.926$ & \\
\hline
2 & $50$ & $.959$ & $.930$ & & $.959$ & $.924$ & \\
 & $100$ & $.939$ & $.932$ & & $.958$ & $.927$ & \\
\hline
3 & $50$ & $.941$ & $.924$ & & $.929$ & $.881$ & \\
 & $100$ & $.952$ & $.951$ & & $.945$ & $.916$ & \\
\hline
4 & $50$ & $.956$ & $.916$ & & $.921$ & $.775$ & \\
 & $100$ & $.944$ & $.937$ & & $.888$ & $.878$ & \\
\hline
5 & $50$ & $.960$ & $.868$ & $.863$ & $.953$ & $.884$ & $.852$ \\
 & $100$ & $.957$ & $.896$ & $.903$ & $.961$ & $.912$ & $.915$\\
\hline
6 & $50$ & $.942$ & $.906$ & $.897$ & $.948$ & $.888$ & $.869$ \\
 & $100$ & $.949$ & $.900$ & $.917$ & $.965$ & $.900$ & $.898$\\
\hline
7 & $50$ & $.958$ & $.876$ & $.869$ & $.957$ & $.855$ & $.857$ \\
 & $100$ & $.969$ & $.884$ & $.900$ & $.944$ & $.870$ & $.905$\\
\hline
8 & $50$ & $.955$ & $.882$ & $.861$ & $.981$ & $.753$ & $.692$ \\
 & $100$ & $.961$ & $.906$ & $.917$ & $.945$ & $.871$ & $.866$\\
\hline
\end{tabular}
\caption{Coverage probabilities of the $95\%$ CIs of $\rho_{\cG\cS}^2$ or $\rho_{\cG\cU}^2$. Under the unspecified scenario, $\rho_{\cG\cU}^2$ is approximated by $R_{\cG\cU}^2$ calculated from a large sample with size $10,000$. ``Asymp," ``P1," and ``P2" represent the CIs constructed based on the true asymptotic variances, the plug-in estimates of the asymptotic variances in the special bivariate Gaussian forms (Corollaries \ref{cor:asym_dist_scor2_gc} and \ref{cor:asym_dist_scor2_gu}), and the plug-in estimates of the asymptotic variances in the general forms (Theorems \ref{thm_asym_dist_scor2_gc} and \ref{thm_asym_dist_scor2_gu}), respectively. The coverage probabilities are estimated as the percentages of CIs that cover $\rho_{\cG\cS}^2$ or $\rho_{\cG\cU}^2$ in $1000$ simulations. For Settings $1$--$4$, we only consider ``P1". \label{tab:cover_prob}}
\end{table}

\subsection{Use of scree plot and AIC to choose $K$} \label{subsec:choose_K_res}

\begin{figure}[htbp]
\includegraphics[width=\textwidth]{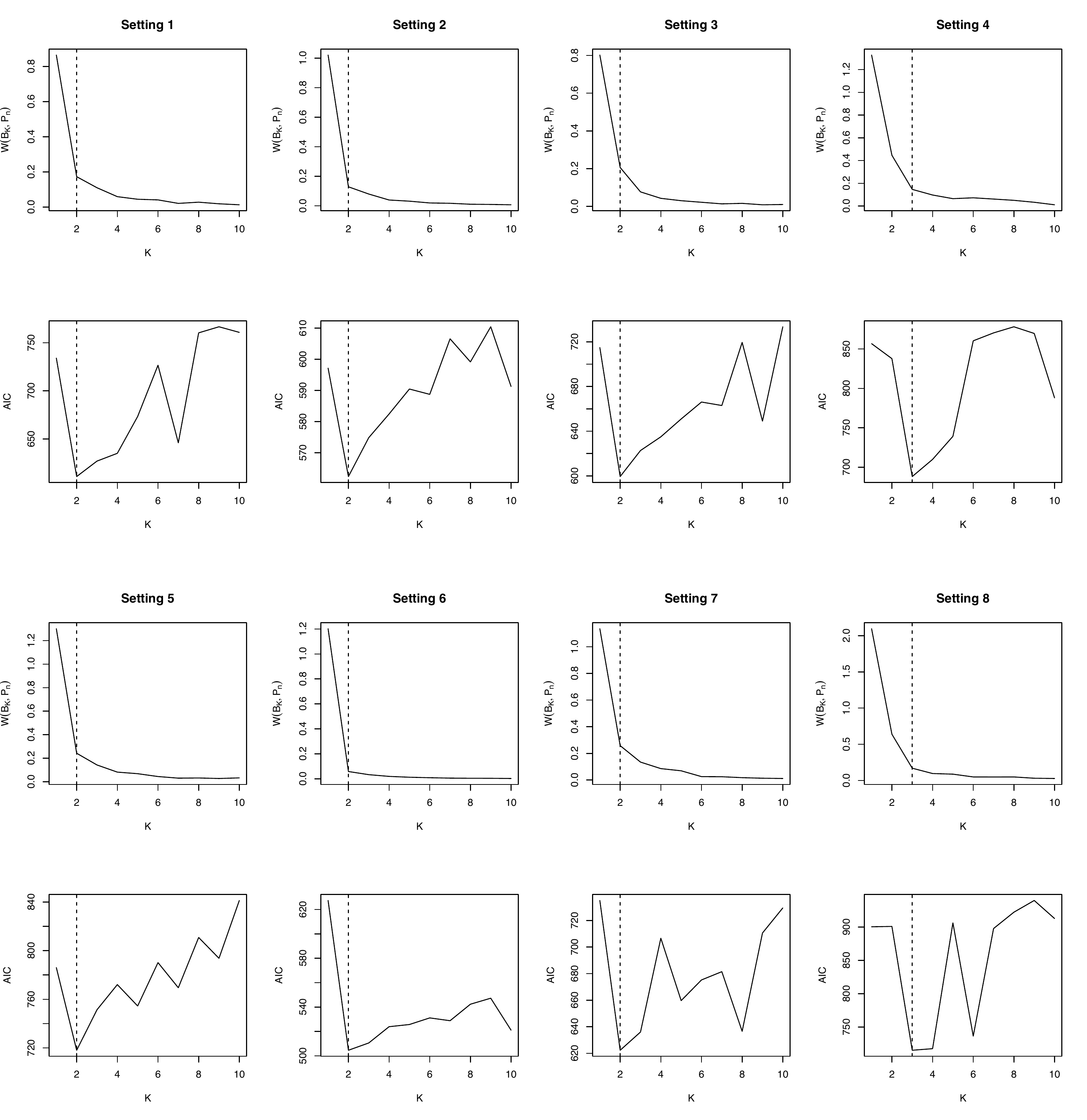}
\caption{The choice of $K$. Under each of the eight simulation settings (Table \ref{tab1:simu_settings}), we simulate a sample of size $n=100$ and calculate $W(B_K, P_n)$ and $AIC(K)$. The dashed vertical lines indicate the true $K$ values used in the simulations.}\label{fig:choose_K}
\end{figure}

Following Section \ref{subsec:sample_R2gu}, here we demonstrate the performance of the scree plot and the AIC in choosing $K$ under the eight simulation settings. For each setting, we simulate a sample of size $n=100$ and evaluate $W(B_K, P_n)$ in (\ref{eq:W_Pn}) and $\text{AIC}(K)$ in (\ref{eq:AIC}) on this sample for $K$ ranging from $1$ to $10$. Figure \ref{fig:choose_K} shows the results. For all the eight settings, the scree plots and the AIC both suggest the correct $K$ values. Even though Settings $5$--$8$ violate the bivariate Gaussian assumption required by the AIC, the AIC results are still reasonable. In practice, users may use the scree plot together with the AIC to decide a reasonable choice of $K$.

\subsection{Power analysis} 
To confirm that $R^2_{\cG\cU}$ is a powerful measure for capturing mixed linear dependences, we conduct a simulation study to compare $R^2_{\cG\cU} (K=2)$ with four existing association measures: the squared Pearson correlation ($R^2$), the maximal correlation (maxCor) \citep{hirschfeld1935connection, gebelein1941statistische, renyi1959measures} estimated by the alternating conditional expectation algorithm \citep{breiman1985estimating}, the distance correlation (dCor) \citep{szekely2007measuring, szekely2009brownian}, and the maximal information coefficient (MIC) \citep{reshef2011detecting} implemented in R package minerva (version 1.4.7). All five measures have values in $[0,1]$. Our simulation procedure follows the study of \cite{simon2014comment}, where each relationship between two real-valued random variables $X$ and $Y$ is composed of a marginal distribution of $X \sim \cN(0, 5^2)$, a noiseless pattern (i.e., relationship) between $X$ and $Y$, and a random error from $\cN(0, \sigma^2)$ added to $Y$. The null hypothesis is that $X$ and $Y$ are independent, while the alternative hypothesis is specified by the noiseless pattern and $\sigma$. Given a sample size $n = 30$, $50$ or $200$, we simulate $B=1000$ samples from the alternative hypothesis. On each of these alternative samples, we randomly permute the $Y$ observations to create a null sample. Then for each $n$ we calculate the five association measures on the $B$ null samples and decide a rejection threshold for each measure as the $(1-\alpha)$ quantile of its $B$ values, where $\alpha=0.05$ is the significance level. Next, we calculate the five association measures on the $B$ alternative samples, compare each measure's $B$ values to its rejection threshold, and estimate the measure's power as the proportion of values above the threshold. Figure \ref{fig:power_values} in the Supplementary Material illustrates each measure's empirical distribution across alternative samples at each $n$ and $\sigma$. All the measures exhibit decreasing variances as $n$ increases.

\begin{figure}[htbp]
\includegraphics[width=\textwidth]{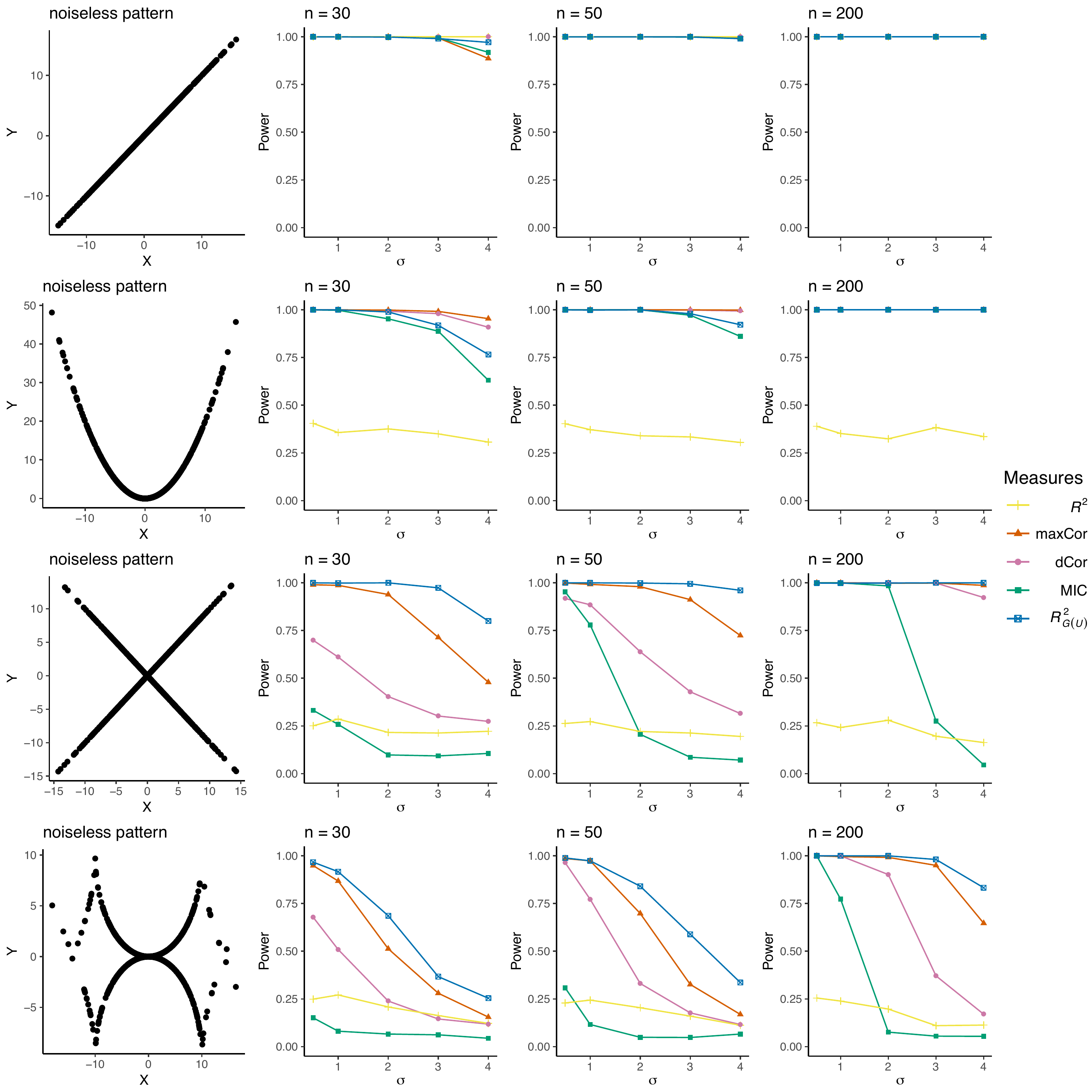}
\caption{Power analysis. Simulation studies that compare the statistical power of five measures: $R^2_{\cG\cU}$ with $K=2$, the squared Pearson correlation ($R^2$), the maximal correlation (maxCor), the distance correlation (dCor), and the maximal information coefficient (MIC). Each noiseless pattern illustrates a relationship between two real-valued random variables $X$ and $Y$ when no noise is added. Under the null hypothesis, the two variables $X$ and $Y$ are independent. Varying alternative hypotheses are formed by the noiseless pattern with noise $\sim \cN(0, \sigma^2)$ added to $Y$ for varying $\sigma$. Under each alternative corresponding to one $\sigma$, we use a permutation test procedure to estimate the power of the five measures given each sample size $n$.}\label{fig:power}
\end{figure}


Figure \ref{fig:power} shows that $R^2_{\cG\cU}$ is the most powerful measure when the pattern is a mixture of positive and negative linear dependences. When the pattern is a mixture of nonlinear relationships that can be approximated by a mixture of linear dependences, $R^2_{\cG\cU}$ is still the most powerful. When the pattern is linear, $R^2$ is expectedly the most powerful, and the other four measures, including $R^2_{\cG\cU}$, also have the perfect power up to $\sigma = 3$ at $n=30$. Under a parabola pattern, maxCor is most powerful, while $R^2$ has low power as expected. Since two intersecting lines can well approximate the parabola pattern, $R^2_{\cG\cU}$ also has good power and is comparable to maxCor, dCor, and MIC, which all aim to capture general dependence.  These results confirm the application potential of $R^2_{\cG\cU}$ in capturing complex but interpretable relationships that can be approximated by mixtures of linear dependences.

\section{Real data applications}\label{sec:real}
\subsection{Gene expression analysis in \textit{Arabidopsis thaliana}}\label{sec:arab}

Back to our motivating example in \textit{Arabidopsis thaliana}, we would like to use this gene expression data \citep{li2008subclade} to demonstrate the use of our generalized Pearson correlation squares in capturing biologically meaningful gene-gene relationships. The glucosinolate (GSL) biosynthesis pathway has been well studied in \textit{Arabidopsis thaliana}, and $31$ genes in this pathway have been experimentally identified \citep{kim2012using}. Since genes in the same pathway are functionally related, their pairwise gene expression relationships should be distinct from their relationships with other genes outside of the pathway. Hence, a powerful association measure should be able to distinguish the pairwise gene-gene relationships within the GSL pathway from the relationships of randomly paired GSL and non-GSL genes.

The data (Table \ref{tabS1:at_data} in the Supplementary Material) contain four index variables: ``condition" (oxidation, wounding, UV-B light, and drought), ``treatment" (yes and no), ``replicate" (1 and 2), and ``tissue" (root and shoot). 
We observe that only the ``tissue" variable is a good indicator of linear dependences, as illustrated in Fig. \ref{fig:motiv} and Figs. \ref{fig:motiv2}--\ref{fig:motiv4} in the Supplementary Material. 

Figure \ref{fig:arab}A shows the values of $R^2$, maxCor, dCor, MIC, and $R^2_{\cG\cU} (K=2)$, all of which do not use index variables, as well as $R^2_{\cG\cS}$, which uses the index variable as ``condition," ``treatment," ``replicate," or ``tissue." All these measures were computed for GSL gene pairs and randomly paired GSL and non-GSL genes.  Among these measures, only $R^2_{\cG\cU} (K=2)$ and $R^2_{\cG\cS} (\text{tissue})$ show stronger relationships within the GSL pathway than in random gene pairs.
These results demonstrate that $R^2_{\cG\cS}$ is a useful and powerful measure when a good index variable is available; otherwise, $R^2_{\cG\cU}$ is advantageous in capturing complex but interpretable gene-gene relationships without knowledge of index variables. 

Besides, we verify the performance of our $K$-lines clustering algorithm on this real dataset. For every GSL gene pair, we compare the two clusters identified by the algorithm with the two tissue types (root and shoot) using Fisher's exact test. The results show that $72\%$, $63\%$, or $59\%$ of the gene pairs have $p$-values under $0.05$, $0.005$, or $0.0005$, suggesting that the algorithm finds a reasonable separation of tissue types for most gene pairs.

\subsection{Cell-cycle gene expression analysis on single-cell RNA sequencing data}
In the second application, we study a single-cell RNA-sequencing dataset that includes gene expression levels in $182$ mouse cells, which span three cell-cycle stages: G1, G2M, and S \citep{buettner2015computational}. Information on the data accession and preprocessing is available in the Supplementary Material. Our exploratory data analysis suggests that the cell stage is not a good index variable for linear dependences, so we do not include  $R^2_{\cG\cS}$ but apply $R^2$, maxCor, dCor, MIC, and $R_{\cG\cU}^2 (K=2)$ to study pairwise relationships among $625$ mouse cell-cycle genes. 

We randomly select $614$ cell-cycle gene pairs
, compute the five measures on each pair, and summarize the empirical distribution of each measure. In Fig. \ref{fig:arab}B, $R_{\cG\cU}^2 (K=2)$ is the only measure that shows a two-mode distribution, revealing that certain cell-cycle gene pairs have strong relationships that are distinct from the rest. This result is biologically reasonable because some cell cycle genes are known to function together, while others do not. In contrast, $R^2$ has dominantly low values, indicating its low power in detecting complex gene-gene relationships from a mix of cells in multiple cell-cycle stages. The other three measures, maxCor, dCor, and MIC, though having overall larger values than $R^2$, still do not show a clear division of cell-cycle gene pairs into the strongly related ones and others. Hence, from a scientific discovery perspective, $R_{\cG\cU}^2 (K=2)$ reveals the most information among the five measures.

Although experimental validation is required to systematically verify the strongly related genes found by $R_{\cG\cU}^2 (K=2)$, as a preliminary study, here we search the literature to investigate the top gene pairs ranked by $R_{\cG\cU}^2$. For example, the gene pair \textit{Cdc25b}--\textit{Lats2} has the highest $R_{\cG\cU}^2$ value, and their physical interaction was previously reported \citep{mukai2015lats1}. Moreover, \textit{Lats2} appears in $17$ gene pairs, among which $11$ pairs are among the top $25\%$ pairs in $R_{\cG\cU}^2$ values. This result is consistent with the fact that \textit{Lats2} is an essential regulator that interacts with many cell-cycle genes \citep{yabuta2007lats2}.

\begin{figure}[htp]
\includegraphics[width=\textwidth]{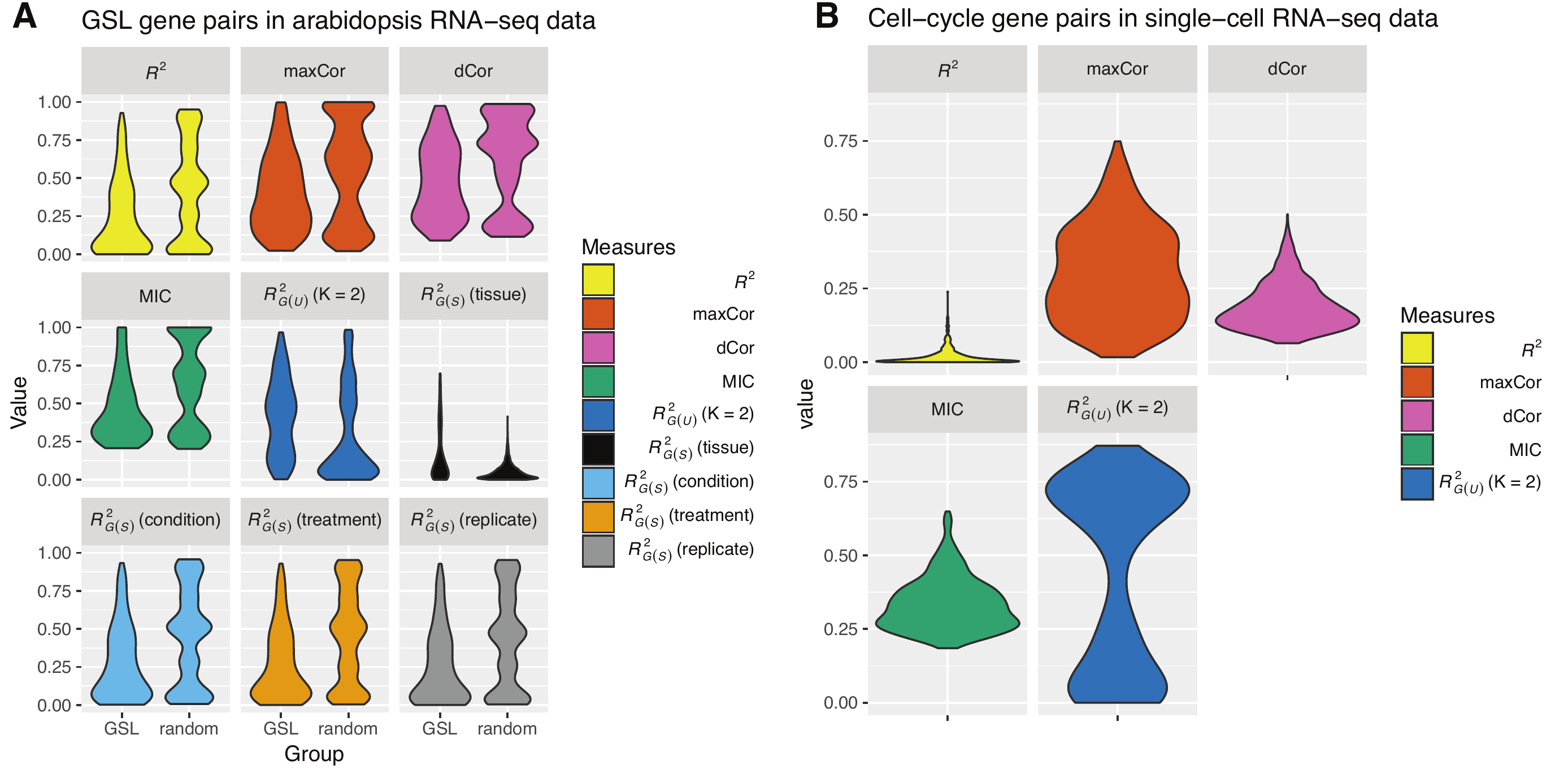}
\caption{Real data applications of $R_{\cG\cS}^2$ and $R_{\cG\cU}^2$ to capturing gene-gene relationships. A: Gene expression analysis in \textit{Arabidopsis}. We compare nine measures, including five unspecified measures ($R^2$, maxCor, dCor, MIC, and $R_{\cG\cU}^2$ with $K=2$) and four $R^2_{\cG\cS}$ measures with different index variables, in terms of measuring pairwise gene relationships within the GSL pathway (``GSL) vs. relationships between a GSL gene and a randomly paired non-GSL gene (``random"). $R^2_{\cG\cU} (K=2)$ and $R^2_{\cG\cS}  (\text{tissue})$ are the only two measures indicating that the gene pairs within the GSL pathway have stronger relationships than the random GSL-nonGSL gene pairs do. B: Cell-cycle gene expression analysis based on single-cell RNA sequencing data. The five unspecified measures used in A are applied to capture the relationships of $1000$ known cell-cycle gene pairs. $R^2_{\cG\cU} (K=2)$ shows that some pairs of cell-cycle genes have strong relationships, which would likely be missed by the other four measures.}\label{fig:arab}
\end{figure}


\section{Discussions}\label{sec:discussion}

The generalized Pearson correlation squares, for the first time to our knowledge, extend the classic and popular Pearson correlation to capturing heterogeneous linear relationships. 
This new suite of measures 
has broad potential use in scientific applications. In addition to gene expression analysis, statistical genetics, for example, is another important application domain. 
It was reported that across human subpopulations, a genotype might exhibit heterogeneous associations with a phenotype \citep{haiman2007multiple,wheeler2017impact}. However, known subpopulations, e.g., race, gender, and geography, cannot explain some observed mixtures of associations. Hence, $R^2_{\cG\cU}$ and the $K$-lines algorithm will be useful. 

A future direction is to extend the generalized Pearson correlation squares to be rank-based. This extension will make the measures robust to outliers and capable of capturing a mixture of monotone relationships.

%
%
%
%

\section*{Software}
We have implemented the inference of the specified and unspecified generalized Pearson correlation squares, the $K$-lines algorithm, as well as the choice of $K$, in an R package \verb+gR2+ available at \verb+GitHub+: https://github.com/lijy03/gR2.

\clearpage

\begin{center}
\Large{Supplementary Material for ``Generalized Pearson correlation squares for capturing mixtures of bivariate linear dependences"}	
\end{center}

\setcounter{section}{0}
\renewcommand{\thesection}{S\arabic{section}}
\renewcommand{\thesubsection}{S\arabic{section}.\arabic{subsection}}
\setcounter{table}{0}
\renewcommand{\thetable}{S\arabic{table}}
\setcounter{figure}{0}
\renewcommand{\thefigure}{S\arabic{figure}}

\section{Proofs}\label{sec:proofs}

\subsection{Proof of Theorem \ref{thm_asym_dist_scor2_gc}}
\begin{proof}
We derive the asymptotic distribution of the sample-level specified generalized Pearson correlation square, $R_{\cG\cS}^2$, in this proof. For notation simplicity, we drop the subscript ``$\cS$" in $p_{k\cS}$ and $\rho_{k\cS}$ in in Section \ref{subsec:pop_R2gs}, and $\widehat p_{k\cS}$, $\widehat\rho^2_{k\cS}$, $\bar X_{k\cS}$, and $\bar Y_{k\cS}$ in Definition \ref{def:sample_gen_R2}, following which 
we have
\begin{align*}
R_{\cG\cS}^2 = \sum_{k=1}^K \widehat{p}_{k} \, \widehat\rho^2_{k}	= \sum_{k=1}^K \widehat{p}_{k} \,\frac{\widehat\sigma_{XY,k}^2}{\widehat\sigma_{X,k}^2 \,\widehat\sigma_{Y,k}^2} \,,
\end{align*}
where 
\begin{align*}
\widehat p_k &= \frac{1}{n} \sum_{i=1}^n \1(Z_i = k)\,,\\
\widehat\sigma_{XY,k} &= \frac{1}{n \widehat p_k} \sum_{i=1}^n (X_i - \bar X_k) (Y_i - \bar Y_k) \,\1(Z_i = k)\,,\\
\widehat\sigma_{X,k}^2 &= \frac{1}{n \widehat p_k} \sum_{i=1}^n (X_i - \bar X_k)^2 \,\1(Z_i = k)\,,\\
\widehat\sigma_{Y,k}^2 &= \frac{1}{n \widehat p_k} \sum_{i=1}^n (Y_i - \bar Y_k)^2 \,\1(Z_i = k)\,,
\end{align*}
with $\bar{X}_k = \frac{1}{n \widehat p_k}\sum_{i=1}^n X_i \, \1(Z_i = k)$ and $\bar{Y}_k = \frac{1}{n \widehat p_k}\sum_{i=1}^n Y_i \, \1(Z_i = k)$.
 

To derive the asymptotic distribution of $R_{\cG\cS}^2$, we need to first derive the joint asymptotic distribution of 
\[ \widehat{\theta} = \left(\widehat p_1, \, \widehat\sigma_{XY,1}, \, \widehat\sigma_{X,1}^2, \, \widehat\sigma_{Y,1}^2, \, \cdots, \, \widehat p_K, \, \widehat\sigma_{XY,K}, \, \widehat\sigma_{X,K}^2, \, \widehat\sigma_{Y,K}^2 \right)\tran \in \R^{4K}\,.
\]

Without loss of generality, we assume
\begin{equation*}
\begin{aligned}[c]
\mu_{X,k} &= \E[X|Z=k] = 0\,,\\	
\sigma_{X,k}^2 &= \var[X|Z=k] = 1\,,\\	
\end{aligned}
\qquad
\begin{aligned}[c]
\mu_{Y,k} &= \E[Y|Z=k] = 0\,,\\
\sigma_{Y,k}^2 &= \var[Y|Z=k] = 1\,.
\end{aligned}	
\end{equation*}
Hence,
\begin{equation*}
\begin{aligned}[c]
\E[X] &= 0\,,\\
\mu_{X^2,k} &= \E[X^2|Z=k] = 1\,,\\	
\sigma_{XY,k} &= \cov[X,Y|Z=k] = \rho_k\,,\\
\end{aligned}
\qquad
\begin{aligned}[c]
\E[Y] &= 0\,,\\
\mu_{Y^2,k} &= \E[Y^2|Z=k] = 1\,,\\
\mu_{XY,k} &= \E[XY|Z=k] = \rho_k\,,
\end{aligned}
\end{equation*}
where $\rho_k = \rho_{k\cS}$ defined in Section \ref{subsec:pop_R2gs}. 

To facilitate the rest of the proof, we define the following sample moments.
\begin{equation*}
\begin{aligned}[c]
M_{k} &:= \frac{1}{n} \sum_{i=1}^n \1(Z_i=k)\,,\\
M_{X,k} &:= \frac{1}{n} \sum_{i=1}^n X_i \, \1(Z_i=k) \,,\\	
M_{Y,k} &:= \frac{1}{n} \sum_{i=1}^n Y_i \, \1(Z_i=k) \,,\\
\end{aligned}
\qquad
\begin{aligned}[c]
M_{XY,k} &:= \frac{1}{n} \sum_{i=1}^n X_i Y_i \, \1(Z_i=k) \,,\\
M_{X^2,k} &:= \frac{1}{n} \sum_{i=1}^n X_i^2 \, \1(Z_i=k) \,,\\
M_{Y^2,k} &:= \frac{1}{n} \sum_{i=1}^n Y_i^2 \, \1(Z_i=k) \,,
\end{aligned}
\end{equation*}
for $k = 1,\ldots,K$. Then
\begin{align*}
\widehat p_k &= M_{k} \,, \quad \bar X_k = M_{X,k} / M_k \,, \quad \bar Y_k = M_{Y,k} / M_k\\
\widehat\sigma_{XY,k} &= \frac{1}{n \widehat p_k} \sum_{i=1}^n (X_i - \bar X_k) (Y_i - \bar Y_k) \,\1(Z_i = k) \\
	&= \frac{1}{n \widehat p_k} \sum_{i=1}^n (X_i Y_i - X_i \bar Y_k - \bar X_k Y_i + \bar X_k \bar Y_k) \,\1(Z_i = k) \\
	&= \left( \frac{1}{n \widehat p_k} \sum_{i=1}^n X_i Y_i \, \1(Z_i=k) \right) - \bar Y_k \left( \frac{1}{n \widehat p_k} \sum_{i=1}^n X_i \, \1(Z_i=k) \right) \\
	& \tab - \bar X_k \left( \frac{1}{n \widehat p_k} \sum_{i=1}^n Y_i \, \1(Z_i=k) \right) + \bar X_k \bar Y_k \left( \frac{1}{n \widehat p_k} \sum_{i=1}^n \, \1(Z_i=k) \right) \\
	&= M_{XY,k}/M_k - M_{X,k}\,M_{Y,k}/M_k^2\,,\\
\widehat\sigma_{X,k}^2 &= \frac{1}{n\widehat p_k} \sum_{i=1}^n (X_i - \bar X_k)^2 \,\1(Z_i = k)\\
	&= \frac{1}{n \widehat p_k} \sum_{i=1}^n \left(X_i^2 - 2 X_i \bar X_k + (\bar X_k)^2 \right) \,\1(Z_i = k) \\
	&= \left( \frac{1}{n \widehat p_k} \sum_{i=1}^n X_i^2 \, \1(Z_i=k) \right) - 2\bar X_k \left( \frac{1}{n \widehat p_k} \sum_{i=1}^n X_i \, \1(Z_i=k) \right) + (\bar X_k)^2 \left( \frac{1}{n \widehat p_k} \sum_{i=1}^n \, \1(Z_i=k) \right) \\
	&= M_{X^2,k}/M_k - (M_{X,k})^2/M_k^2\,.\\
\text{Similarly,}\\
\widehat\sigma_{Y,k}^2 &= M_{Y^2,k}/M_k - (M_{Y,k})^2/M_k^2\,.
\end{align*}

Hence, 
\[
\widehat{\theta} = \begin{pmatrix}
 	\widehat p_{1}\\
 	\widehat\sigma_{XY,1}\\
 	\widehat\sigma_{X,1}^2\\
 	\widehat\sigma_{Y,1}^2\\
 	\vdots\\
 	\widehat p_{K}\\
 	\widehat\sigma_{XY,K}\\
 	\widehat\sigma_{X,K}^2\\
 	\widehat\sigma_{Y,K}^2
 \end{pmatrix}
 = \begin{pmatrix}
 	M_{1}\\
 	M_{XY,1}/M_1 - M_{X,1}\,M_{Y,1}/M_{1}^2\\
 	M_{X^2,1}/M_1 - (M_{X,1})^2/M_{1}^2\\
 	M_{Y^2,1}/M_1 - (M_{Y,1})^2/M_{1}^2\\
 	\vdots\\
 	M_{K}\\
 	M_{XY,K}/M_k - M_{X,K}\,M_{Y,K}/M_{K}^2\\
 	M_{X^2,K}/M_k - (M_{X,K})^2/M_{K}^2\\
 	M_{Y^2,K}/M_k - (M_{Y,K})^2/M_{K}^2
 \end{pmatrix}
 = g \begin{pmatrix}
 	M_1\\
 	M_{X,1}\\
 	M_{Y,1}\\
 	M_{X^2,1}\\
 	M_{Y^2,1}\\
 	M_{XY,1}\\
 	\vdots\\
 	M_K\\
 	M_{X,K}\\
 	M_{Y,K}\\
 	M_{X^2,K}\\
 	M_{Y^2,K}\\
 	M_{XY,K}	
 \end{pmatrix}\,.
\]
Given any $\bxi = (x_1, y_1, z_1, u_1, v_1, w_1, \cdots, x_K, y_K, z_K, u_K, v_K, w_K)\tran \in \R^{6K}$, the function $g: \R^{6K} \rightarrow \R^{4K}$ is
\begin{equation} g(\bxi) = g \begin{pmatrix}
 	x_1\\
 	y_1\\
 	z_1\\
 	u_1\\
 	v_1\\
 	w_1\\
 	\vdots\\
 	x_K\\
 	y_K\\
 	z_K\\
 	u_K\\
 	v_K\\
 	w_K\\	
 \end{pmatrix}
 = \begin{pmatrix}
 	x_{1}\\
 	w_1/x_1 - y_1 z_1/x_1^2\\
 	u_1/x_1 - y_1^2/x_1^2\\
 	v_1/x_1 - z_1^2/x_1^2\\
 	\vdots\\
 	x_{K}\\
 	w_K/x_K - y_K z_K/x_K^2\\
 	u_K/x_K - y_K^2/x_K^2\\
 	v_K/x_K - z_K^2/x_K^2\\
 \end{pmatrix}\,,\label{function_g}
\end{equation}
with Jacobian
\[ D g(\bxi) =
\begin{bmatrix}
D g(\bxi)_1 & & \\
& \ddots & \\
& & D g(\bxi)_K	
\end{bmatrix}_{(4K)\times(6K)}\,,
\]
where
\begin{align*}
	D g(\bxi)_k &=  
	\begin{bmatrix}
 	1 & 0 & 0 & 0 & 0 & 0 \\	
 	-w_k/x_k^2 + 2y_kz_k/x_k^3 & -z_k/x_k^2 & -y_k/x_k^2 & 0 & 0 & 1/x_k \\
 	-u_k/x_k^2 + 2y_k^2/x_k^3 & -2y_k/x_k^2 & 0 & 1/x_k & 0 & 0 \\
 	-v_k/x_k^2 + 2z_k^2/x_k^3 & 0 & -2z_k/x_k^2 & 0 & 1/x_k & 0 
 \end{bmatrix}\,,\; k =1,\ldots,K\,.
\end{align*}

To derive the joint asymptotic distribution of $\widehat\btheta$, we need the joint asymptotic distribution of 
\begin{align*}
	\widehat{\mu} &= (M_1, \, M_{X,1}, \, M_{Y,1}, \, M_{X^2,1}, \, M_{Y^2,1}, \, M_{XY,1}, \, \cdots, M_K, \, M_{X,K}, \, M_{Y,K}, \, M_{X^2,K}, \, M_{Y^2,K}, \, M_{XY,K})\tran \\
	&= (\widehat\bmu_1\tran, \cdots, \widehat\bmu_K\tran)\tran \,,
\end{align*} 
where
\[ \widehat\bmu_k = \left(M_k, \, M_{X,k}, \, M_{Y,k}, \, M_{X^2,k}, \, M_{Y^2,k}, \, M_{XY,k} \right)\tran\,,
\]
because $\widehat\btheta = g(\widehat\bmu)$.

For notation simplicity, we denote $Z_{[k]} = \1(Z=k)$ and $Z_{[k]i} = \1(Z_i=k)$, and thus $\widehat\bmu_k$ can be expressed as
\[ \widehat\bmu_k = \left(\frac{1}{n}\sum_{i=1}^n Z_{[k]i}\,,\, \frac{1}{n}\sum_{i=1}^n X_iZ_{[k]i}\,,\, \frac{1}{n}\sum_{i=1}^n Y_iZ_{[k]i}\,,\, \frac{1}{n}\sum_{i=1}^n X_i^2Z_{[k]i}\,,\, \frac{1}{n}\sum_{i=1}^n Y_i^2Z_{[k]i}\,,\, \frac{1}{n}\sum_{i=1}^n X_iY_iZ_{[k]i} \right)\tran.
\]
Hence, the joint asymptotic distribution of $\widehat{\mu}$ can be derived based on the central limit theorem.

The joint moments satisfy \begin{align*}
\E\left[X^a Y^b Z_{[k]}\right] &= \pr\left(Z_{[k]}=1\right)\E\left[\left.X^a Y^b Z_{[k]}\right|Z_{[k]}=1\right] + \pr\left(Z_{[k]}=0\right)\E\left[\left.X^a Y^b Z_{[k]}\right|Z_{[k]}=0\right]	\\
	&= p_k \, \mu_{X^aY^b,k} + (1-p_k) \cdot 0\\ 
	&= p_k \, \mu_{X^aY^b,k}\,,\; a,b \in \R\,,
\end{align*}
and thus specifically,
\begin{equation*}
\begin{aligned}[c]
\E\left[XZ_{[k]}\right] &= p_k \, \mu_{X,k} = 0\,,\\
\E\left[X^2Z_{[k]}\right] &= p_k \, \mu_{X^2,k} = p_k\,,\\
\E\left[XYZ_{[k]}\right] &= p_k \, \mu_{XY,k} = p_k\,\rho_k\,,\\
\E\left[X^3Z_{[k]}\right] &= p_k \, \mu_{X^3,k}\,,\\
\E\left[X^2YZ_{[k]}\right] &= p_k \, \mu_{X^2Y,k}\,,\\
\E\left[X^4Z_{[k]}\right] &= p_k \, \mu_{X^4,k}\,,\\
\E\left[X^3YZ_{[k]}\right] &= p_k \, \mu_{X^3Y,k}\,,\\
\E\left[X^2Y^2Z_{[k]}\right] &= p_k \, \mu_{X^2Y^2,k}\,.\\
\end{aligned}
\qquad
\begin{aligned}[c]
\E\left[YZ_{[k]}\right] &= p_k \, \mu_{Y,k} = 0\,,\\
\E\left[Y^2Z_{[k]}\right] &= p_k \, \mu_{Y^2,k} = p_k\,,\\
\\
\E\left[Y^3Z_{[k]}\right] &= p_k \, \mu_{Y^3,k}\,,\\
\E\left[XY^2Z_{[k]}\right] &= p_k \, \mu_{XY^2,k}\,,\\
\E\left[Y^4Z_{[k]}\right] &= p_k \, \mu_{Y^4,k}\,,\\
\E\left[XY^3Z_{[k]}\right] &= p_k \, \mu_{XY^3,k}\,,\\
\\
\end{aligned}	
\end{equation*}

By the central limit theorem, the joint asymptotic distribution of $\widehat\bmu$ is
\begin{equation}
	\sqrt{n} ( \widehat{\mu} - \mu) \tod \cN(0, \Sigma)\,,\label{asym_clt}
\end{equation}
where
\begin{align*}
\mu &= (\bmu_1,\cdots,\bmu_K)\tran\,,
\end{align*}
with
\begin{align*}
\bmu_k &= \left(\E\left[Z_{[k]}\right],\, \E\left[XZ_{[k]}\right] ,\, \E\left[YZ_{[k]}\right] ,\, \E\left[X^2Z_{[k]}\right] ,\, \E\left[Y^2Z_{[k]}\right] ,\, \E\left[XYZ_{[k]}\right]	\right)\tran\\
	&= \left(p_k, \, 0, \, 0, \, p_k, \, p_k, \, p_k\rho_k\right)\tran\,,
\end{align*}
and
\begin{align*}
\Sigma &= \begin{bmatrix}
	\Sigma_{11} &  \cdots & \Sigma_{1K}\\
	\vdots & \ddots & \vdots\\
	\Sigma_{K1} &  \cdots & \Sigma_{KK}\\	
 \end{bmatrix}_{(6K)\times(6K)}\,.
\end{align*}
For $k=1,\ldots,K$, $\Sigma_{kk}$ is the variance-covariance matrix of $\left(Z_{[k]}, XZ_{[k]}, YZ_{[k]}, X^2Z_{[k]}, Y^2Z_{[k]}, XYZ_{[k]}\right)\tran$:
\setlength{\arraycolsep}{1.5pt}
\begin{align*}
& \Sigma_{kk}= \\	
& \footnotesize\begin{bmatrix}
 	\var(Z_{[k]}) & \cov(Z_{[k]}, XZ_{[k]}) & \cov(Z_{[k]}, YZ_{[k]}) & \cov(Z_{[k]}, X^2Z_{[k]}) & \cov(Z_{[k]}, Y^2Z_{[k]}) & \cov(Z_{[k]}, XYZ_{[k]})\\
 	\cov(XZ_{[k]}, Z_{[k]}) & \var(XZ_{[k]}) & \cov(XZ_{[k]}, YZ_{[k]}) & \cov(XZ_{[k]}, X^2Z_{[k]}) & \cov(XZ_{[k]}, Y^2Z_{[k]}) & \cov(XZ_{[k]}, XYZ_{[k]})\\
 	\cov(YZ_{[k]},Z_{[k]}) & \cov(YZ_{[k]}, XZ_{[k]}) & \var(YZ_{[k]}) & \cov(YZ_{[k]}, X^2Z_{[k]}) & \cov(YZ_{[k]}, Y^2Z_{[k]}) & \cov(YZ_{[k]}, XYZ_{[k]})\\
 	\cov(X^2Z_{[k]},Z_{[k]}) & \cov(X^2Z_{[k]}, XZ_{[k]}) & \cov(X^2Z_{[k]}, YZ_{[k]}) & \var(X^2Z_{[k]}) & \cov(X^2Z_{[k]}, Y^2Z_{[k]}) & \cov(X^2Z_{[k]}, XYZ_{[k]})\\
 	\cov(Y^2Z_{[k]},Z_{[k]}) & \cov(Y^2Z_{[k]}, XZ_{[k]}) & \cov(Y^2Z_{[k]}, YZ_{[k]}) & \cov(Y^2Z_{[k]},X^2Z_{[k]}) & \var(Y^2Z_{[k]}) & \cov(Y^2Z_{[k]}, XYZ_{[k]})\\
 	\cov(XYZ_{[k]},Z_{[k]}) & \cov(XYZ_{[k]}, XZ_{[k]}) & \cov(XYZ_{[k]}, YZ_{[k]}) & \cov(XYZ_{[k]}, X^2Z_{[k]}) & \cov(XYZ_{[k]}, Y^2Z_{[k]}) & \var(XYZ_{[k]})\\
 \end{bmatrix}\,,
 \normalsize
\end{align*}
\setlength{\arraycolsep}{6pt}
where
\begin{align*}
	&\var\left(Z_{[k]}\right) = p_k\left(1-p_k\right)\,,\\
	&\cov\left(Z_{[k]}, XZ_{[k]}\right) = \E\left[XZ_{[k]}^2\right] - \E\left[Z_{[k]}\right]\E\left[XZ_{[k]}\right] = p_k\,,\\ 
	&\cov\left(Z_{[k]}, YZ_{[k]}\right) = p_k\,,\\
	&\cov\left(Z_{[k]}, X^2Z_{[k]}\right) = \E\left[X^2Z_{[k]}^2\right] - \E\left[Z_{[k]}\right]\E\left[X^2Z_{[k]}\right] = \E\left[X^2Z_{[k]}\right] - p_k^2 = p_k\left(1-p_k\right)\,,\\ 
	&\cov\left(Z_{[k]}, Y^2Z_{[k]}\right) = p_k\left(1-p_k\right)\,,\\ 
	&\cov\left(Z_{[k]}, XYZ_{[k]}\right) = \E\left[XYZ_{[k]}^2\right] - \E\left[Z_{[k]}\right]\E\left[XYZ_{[k]}\right] = \E\left[XYZ_{[k]}\right] - p_k^2\,\rho_k = p_k\left(1-p_k\right)\rho_k\,,\\
	&\var\left(XZ_{[k]}\right) = \E\left[X^2Z_{[k]}^2\right] - \left(\E\left[XZ_{[k]}\right]\right)^2 = \E\left[X^2Z_{[k]}\right] - 0 = p_k\,,\\ 
	&\cov\left(XZ_{[k]}, YZ_{[k]}\right) = \E\left[XYZ_{[k]}^2\right] - \E\left[XZ_{[k]}\right]\E\left[YZ_{[k]}\right] = \E\left[XYZ_{[k]}\right] - 0 = p_k\,\rho_k\,,\\
	&\cov\left(XZ_{[k]}, X^2Z_{[k]}\right) = \E\left[X^3Z_{[k]}^2\right] - \E\left[XZ_{[k]}\right]\E\left[X^2Z_{[k]}\right] = \E\left[X^3Z_{[k]}\right] - 0 = p_k\,\mu_{X^3,k}\,,\\
	&\cov\left(XZ_{[k]}, Y^2Z_{[k]}\right) = \E\left[XY^2Z_{[k]}^2\right] - \E\left[XZ_{[k]}\right]\E\left[Y^2Z_{[k]}\right] = \E\left[XY^2Z_{[k]}\right] - 0 = p_k\,\mu_{XY^2,k}\,,\\
	&\cov\left(XZ_{[k]}, XYZ_{[k]}\right) = \E\left[X^2YZ_{[k]}^2\right] - \E\left[XZ_{[k]}\right]\E\left[XYZ_{[k]}\right] = \E\left[X^2YZ_{[k]}\right] - 0 = p_k\,\mu_{X^2Y,k}\,,\\
	&\var\left(YZ_{[k]}\right) = \E\left[Y^2Z_{[k]}^2\right] - \left(\E\left[YZ_{[k]}\right]\right)^2 = \E\left[Y^2Z_{[k]}\right] - 0 = p_k\,,\\
	&\cov\left(YZ_{[k]}, X^2Z_{[k]}\right) = \E\left[X^2YZ_{[k]}^2\right] - \E\left[YZ_{[k]}\right]\E\left[X^2Z_{[k]}\right] = \E\left[X^2YZ_{[k]}\right] - 0 = p_k\,\mu_{X^2Y,k}\,,\\
	&\cov\left(YZ_{[k]}, Y^2Z_{[k]}\right) = \E\left[Y^3Z_{[k]}^2\right] - \E\left[YZ_{[k]}\right]\E\left[Y^2Z_{[k]}\right] = \E\left[Y^3Z_{[k]}\right] - 0 = p_k\,\mu_{Y^3,k}\,,\\
	&\cov\left(YZ_{[k]}, XYZ_{[k]}\right) = \E\left[XY^2Z_{[k]}^2\right] - \E\left[YZ_{[k]}\right]\E\left[XYZ_{[k]}\right] = \E\left[XY^2Z_{[k]}\right] - 0 = p_k\,\mu_{XY^2,k}\,,\\
	&\var\left(X^2Z_{[k]}\right) = \E\left[X^4Z_{[k]}^2\right] - \left(\E\left[X^2Z_{[k]}\right]\right)^2 = \E\left[X^4Z_{[k]}\right] - p_k^2 = p_k\, \mu_{X^4,k} - p_k^2\,,\\
	&\cov\left(X^2Z_{[k]}, Y^2Z_{[k]}\right) = \E\left[X^2Y^2Z_{[k]}^2\right] - \E\left[X^2Z_{[k]}\right]\E\left[Y^2Z_{[k]}\right] = \E\left[X^2Y^2Z_{[k]}\right] - p_k^2\\
	& \tab = p_k\,\mu_{X^2Y^2,k} - p_k^2\,,\\
	&\cov\left(X^2Z_{[k]}, XYZ_{[k]}\right) = \E\left[X^3YZ_{[k]}^2\right] - \E\left[X^2Z_{[k]}\right]\E\left[XYZ_{[k]}\right] = \E\left[X^3YZ_{[k]}\right] - p_k^2\,\rho_k \\
	& \tab = p_k\,\mu_{X^3Y,k} - p_k^2\,\rho_k\,,\\
	&\var\left(Y^2Z_{[k]}\right) = \E\left[Y^4Z_{[k]}^2\right] - \left(\E\left[Y^2Z_{[k]}\right]\right)^2 = \E\left[Y^4Z_{[k]}\right] - p_k^2 = p_k\,\mu_{Y^4,k} - p_k^2\,,\\
\end{align*}
\begin{align*}
	&\cov\left(Y^2Z_{[k]}, XYZ_{[k]}\right) = \E\left[XY^3Z_{[k]}^2\right] - \E\left[Y^2Z_{[k]}\right]\E\left[XYZ_{[k]}\right] = \E\left[XY^3Z_{[k]}\right] - p_k^2\,\rho_k \\
	& \tab = p_k\,\mu_{XY^3,k} - p_k^2\,\rho_k\,,\\
	&\var\left(XYZ_{[k]}\right) = \E\left[X^2Y^2Z_{[k]}^2\right] - \left(\E\left[XYZ_{[k]}\right]\right)^2 = \E\left[X^2Y^2Z_{[k]}\right] - p_k^2\,\rho_k^2 = p_k\,\mu_{X^2Y^2,k} - p_k^2\,\rho_k^2\,.
\end{align*}
That is,
\begin{align*}
& \Sigma_{kk} \\	
=& \begin{bmatrix}
 	p_k(1-p_k) & p_k & p_k & p_k(1-p_k) & p_k(1-p_k) & p_k(1-p_k)\rho_k\\
 	p_k & p_k & p_k\rho_k & p_k\mu_{X^3,k} & p_k\mu_{XY^2,k} & p_k\mu_{X^2Y,k}\\
 	p_k & p_k\rho_k & p_k & p_k\mu_{X^2Y,k} & p_k\mu_{Y^3,k} & p_k\mu_{XY^2,k}\\
 	p_k(1-p_k) & p_k\mu_{X^3,k} & p_k\mu_{X^2Y,k} & p_k \mu_{X^4,k} - p_k^2 & p_k \mu_{X^2Y^2,k} - p_k^2 & p_k \mu_{X^3Y,k} - p_k^2\rho_k\\
 	p_k(1-p_k) & p_k\mu_{XY^2,k} & p_k\mu_{Y^3,k} & p_k \mu_{X^2Y^2,k} - p_k^2 & p_k \mu_{Y^4,k} - p_k^2 & p_k \mu_{XY^3,k} - p_k^2\rho_k\\
 	p_k(1-p_k)\rho_k & p_k\mu_{X^2Y,k} & p_k\mu_{XY^2,k} & p_k \mu_{X^3Y,k} - p_k^2\rho_k & p_k \mu_{XY^3,k} - p_k^2\rho_k & p_k \mu_{X^2Y^2,k} - p_k^2\rho_k^2\\
 \end{bmatrix}\,.
\end{align*}
For $1 \le k \neq r \le K$, $\Sigma_{ks}$ is the covariance matrix of $\left(Z_{[k]}, XZ_{[k]}, YZ_{[k]}, X^2Z_{[k]}, Y^2Z_{[k]}, XYZ_{[k]}\right)\tran$ and \\$\left(Z_{[r]}, XZ_{[r]}, YZ_{[r]}, X^2Z_{[r]}, Y^2Z_{[r]}, XYZ_{[r]}\right)\tran$:
\setlength{\arraycolsep}{1.5pt}
\begin{align*}
& \Sigma_{ks}= \\	
& \footnotesize\begin{bmatrix}
 	\cov(Z_{[k]}, Z_{[r]}) & \cov(Z_{[k]}, XZ_{[r]}) & \cov(Z_{[k]}, YZ_{[r]}) & \cov(Z_{[k]}, X^2Z_{[r]}) & \cov(Z_{[k]}, Y^2Z_{[r]}) & \cov(Z_{[k]}, XYZ_{[r]})\\
 	\cov(XZ_{[k]}, Z_{[r]}) & \cov(XZ_{[k]}, XZ_{[r]}) & \cov(XZ_{[k]}, YZ_{[r]}) & \cov(XZ_{[k]}, X^2Z_{[r]}) & \cov(XZ_{[k]}, Y^2Z_{[r]}) & \cov(XZ_{[k]}, XYZ_{[r]})\\
 	\cov(YZ_{[k]}, Z_{[r]}) & \cov(YZ_{[k]}, XZ_{[r]}) & \cov(YZ_{[k]}, YZ_{[r]}) & \cov(YZ_{[k]}, X^2Z_{[r]}) & \cov(YZ_{[k]}, Y^2Z_{[r]}) & \cov(YZ_{[k]}, XYZ_{[r]})\\
 	\cov(X^2Z_{[k]},Z_{[r]}) & \cov(X^2Z_{[k]}, XZ_{[r]}) & \cov(X^2Z_{[k]}, YZ_{[r]}) & \cov(X^2Z_{[k]}, X^2Z_{[r]}) & \cov(X^2Z_{[k]}, Y^2Z_{[r]}) & \cov(X^2Z_{[k]}, XYZ_{[r]})\\
 	\cov(Y^2Z_{[k]}, Z_{[r]}) & \cov(Y^2Z_{[k]}, XZ_{[r]}) & \cov(Y^2Z_{[k]}, YZ_{[r]}) & \cov(Y^2Z_{[k]},X^2Z_{[r]}) & \cov(Y^2Z_{[k]}, Y^2Z_{[r]}) & \cov(Y^2Z_{[k]}, XYZ_{[r]})\\
 	\cov(XYZ_{[k]},Z_{[r]}) & \cov(XYZ_{[k]}, XZ_{[r]}) & \cov(XYZ_{[k]}, YZ_{[r]}) & \cov(XYZ_{[k]}, X^2Z_{[r]}) & \cov(XYZ_{[k]}, Y^2Z_{[r]}) & \var(XYZ_{[k]}, XYZ_{[r]})\\
 \end{bmatrix}\,,
 \normalsize
\end{align*}
\setlength{\arraycolsep}{6pt}where
\begin{align*}
	& \cov\left(Z_{[k]}, Z_{[r]}\right) = \E\left[Z_{[k]}Z_{[r]}\right] - \E\left[Z_{[k]}\right]\E\left[Z_{[r]}\right] = 0 - p_k\,p_r = - p_k\,p_r\,,\\
	& \cov\left(Z_{[k]}, XZ_{[r]}\right) = \E\left[XZ_{[k]}Z_{[r]}\right] - \E\left[Z_{[k]}\right]\E\left[XZ_{[r]}\right] = 0 - p_k \cdot 0 = 0\,,\\ 
	& \cov\left(Z_{[k]}, YZ_{[r]}\right) = 0\,,\\ 
	& \cov\left(Z_{[k]}, X^2Z_{[r]}\right) = \E\left[X^2Z_{[k]}Z_{[r]}\right] - \E\left[Z_{[k]}\right]\E\left[X^2Z_{[r]}\right] = 0 - p_k\,p_r = - p_k\,p_r\,,\\ 
	& \cov\left(Z_{[k]}, Y^2Z_{[r]}\right) = - p_k\,p_r\,,\\ 
	& \cov\left(Z_{[k]}, XYZ_{[r]}\right) = \E\left[XYZ_{[k]}Z_{[r]}\right] - \E\left[Z_{[k]}\right]\E\left[XYZ_{[r]}\right] = 0 - p_k\,p_r\,\rho_r = - p_k\,p_r\,\rho_r\,,\\
	& \cov\left(XZ_{[k]}, Z_{[r]}\right) = \E\left[XZ_{[k]}Z_{[r]}\right] - \E\left[XZ_{[k]}\right]\E\left[Z_{[r]}\right] = 0 - 0 = 0\,,\\
	& \cov\left(XZ_{[k]}, XZ_{[r]}\right) = \E\left[X^2Z_{[k]}Z_{[r]}\right] - \E\left[XZ_{[k]}\right]\E\left[XZ_{[r]}\right] = 0 - 0 = 0\,,\\ 
	& \cov\left(XZ_{[k]}, YZ_{[r]}\right) = \E\left[XYZ_{[k]}Z_{[r]}\right] - \E\left[XZ_{[k]}\right]\E\left[YZ_{[r]}\right] = 0 - 0 = 0\,,\\
	& \cov\left(XZ_{[k]}, X^2Z_{[r]}\right) = \E\left[X^3Z_{[k]}Z_{[r]}\right] - \E\left[XZ_{[k]}\right]\E\left[X^2Z_{[r]}\right] = 0 - 0 = 0\,,\\
\end{align*}
\begin{align*}
	& \cov\left(XZ_{[k]}, Y^2Z_{[r]}\right) = \E\left[XY^2Z_{[k]}Z_{[r]}\right] - \E\left[XZ_{[k]}\right]\E\left[Y^2Z_{[r]}\right] = 0 - 0 = 0\,,\\
	& \cov\left(XZ_{[k]}, XYZ_{[r]}\right) = \E\left[X^2YZ_{[k]}Z_{[r]}\right] - \E\left[XZ_{[k]}\right]\E\left[XYZ_{[r]}\right] = 0 - 0 = 0\,,\\
	& \cov\left(YZ_{[k]}, Z_{[r]}\right) = \E\left[YZ_{[k]}Z_{[r]}\right] - \E\left[YZ_{[k]}\right]\E\left[Z_{[r]}\right] = 0 - 0 = 0\,,\\
	& \cov\left(YZ_{[k]}, XZ_{[r]}\right) = \E\left[XYZ_{[k]}Z_{[r]}\right] - \E\left[YZ_{[k]}\right]\E\left[XZ_{[r]}\right] = 0 - 0 = 0\,,\\
	& \cov\left(YZ_{[k]}, YZ_{[r]}\right) = \E\left[Y^2Z_{[k]}Z_{[r]}\right] - \E\left[YZ_{[k]}\right]\E\left[YZ_{[r]}\right] = 0 - 0 = 0\,,\\
	& \cov\left(YZ_{[k]}, X^2Z_{[r]}\right) = \E\left[X^2YZ_{[k]}Z_{[r]}\right] - \E\left[YZ_{[k]}\right]\E\left[X^2Z_{[r]}\right] = 0 - 0 = 0\,,\\
	& \cov\left(YZ_{[k]}, Y^2Z_{[r]}\right) = \E\left[Y^3Z_{[k]}Z_{[r]}\right] - \E\left[YZ_{[k]}\right]\E\left[Y^2Z_{[r]}\right] = 0 - 0 = 0\,,\\
	& \cov\left(YZ_{[k]}, XYZ_{[r]}\right) = \E\left[XY^2Z_{[k]}Z_{[r]}\right] - \E\left[YZ_{[k]}\right]\E\left[XYZ_{[r]}\right] = 0 - 0 = 0\,,\\
	& \cov\left(X^2Z_{[k]}, Z_{[r]}\right) = \E\left[X^2Z_{[k]}Z_{[r]}\right] - \E\left[X^2Z_{[k]}\right]\E\left[Z_{[r]}\right] = 0 - p_k\,p_r = - p_k\,p_r\,,\\
	& \cov\left(X^2Z_{[k]}, XZ_{[r]}\right) = \E\left[X^3Z_{[k]}Z_{[r]}\right] - \E\left[X^2Z_{[k]}\right]\E\left[XZ_{[r]}\right] = 0 - 0 = 0\,,\\
	& \cov\left(X^2Z_{[k]}, YZ_{[r]}\right) = \E\left[X^2YZ_{[k]}Z_{[r]}\right] - \E\left[X^2Z_{[k]}\right]\E\left[YZ_{[r]}\right] = 0 - 0 = 0\,,\\
	& \cov\left(X^2Z_{[k]}, X^2Z_{[r]}\right) = \E\left[X^4Z_{[k]}Z_{[r]}\right] - \E\left[X^2Z_{[k]}\right]\E\left[X^2Z_{[r]}\right] = 0 - p_k\,p_r = - p_k\,p_r\,,\\
	& \cov\left(X^2Z_{[k]}, Y^2Z_{[r]}\right) = \E\left[X^2Y^2Z_{[k]}Z_{[r]}\right] - \E\left[X^2Z_{[k]}\right]\E\left[Y^2Z_{[r]}\right] = 0 - p_k\,p_r = - p_k\,p_r\,,\\
	& \cov\left(X^2Z_{[k]}, XYZ_{[r]}\right) = \E\left[X^3YZ_{[k]}Z_{[r]}\right] - \E\left[X^2Z_{[k]}\right]\E\left[XYZ_{[r]}\right] = 0 - p_k\,p_r\,\rho_r = - p_k\,p_r\,\rho_r\,,\\
	& \cov\left(Y^2Z_{[k]}, Z_{[r]}\right) = \E\left[Y^2Z_{[k]}Z_{[r]}\right] - \E\left[Y^2Z_{[k]}\right]\E\left[Z_{[r]}\right] = 0 - p_k\,p_r = - p_k\,p_r\,,\\
	& \cov\left(Y^2Z_{[k]}, XZ_{[r]}\right) = \E\left[XY^2Z_{[k]}Z_{[r]}\right] - \E\left[Y^2Z_{[k]}\right]\E\left[XZ_{[r]}\right] = 0 - 0 = 0\,,\\
	& \cov\left(Y^2Z_{[k]}, YZ_{[r]}\right) = \E\left[Y^3Z_{[k]}Z_{[r]}\right] - \E\left[Y^2Z_{[k]}\right]\E\left[YZ_{[r]}\right] = 0 - 0 = 0\,,\\
& \cov\left(Y^2Z_{[k]}, X^2Z_{[r]}\right) = \E\left[X^2Y^2Z_{[k]}Z_{[r]}\right] - \E\left[Y^2Z_{[k]}\right]\E\left[X^2Z_{[r]}\right] = 0 - p_k\,p_r = - p_k\,p_r\,,\\
	& \cov\left(Y^2Z_{[k]}, Y^2Z_{[r]}\right) = \E\left[Y^4Z_{[k]}Z_{[r]}\right] - \E\left[Y^2Z_{[k]}\right]\E\left[Y^2Z_{[r]}\right] = 0 - p_k\,p_r = - p_k\,p_r\,,\\
	& \cov\left(Y^2Z_{[k]}, XYZ_{[r]}\right) = \E\left[XY^3Z_{[k]}Z_{[r]}\right] - \E\left[Y^2Z_{[k]}\right]\E\left[XYZ_{[r]}\right] = 0 - p_k\,p_r\,\rho_r = - p_k\,p_r\,\rho_r\,,\\
& \cov\left(XYZ_{[k]}, Z_{[r]}\right) = \E\left[XYZ_{[k]}Z_{[r]}\right] - \E\left[XYZ_{[k]}\right]\E\left[Z_{[r]}\right] = 0 - p_k\,\rho_k\,p_r = - p_k\,p_r\,\rho_k\,,\\
	& \cov\left(XYZ_{[k]}, XZ_{[r]}\right) = \E\left[X^2YZ_{[k]}Z_{[r]}\right] - \E\left[XYZ_{[k]}\right]\E\left[XZ_{[r]}\right] = 0 - 0 = 0\,,\\
	& \cov\left(XYZ_{[k]}, YZ_{[r]}\right) = \E\left[XY^2Z_{[k]}Z_{[r]}\right] - \E\left[XYZ_{[k]}\right]\E\left[YZ_{[r]}\right] = 0 - 0 = 0\,,\\
	& \cov\left(XYZ_{[k]}, X^2Z_{[r]}\right) = \E\left[X^3YZ_{[k]}Z_{[r]}\right] - \E\left[XYZ_{[k]}\right]\E\left[X^2Z_{[r]}\right] = 0 - p_k\,\rho_k\,p_r = - p_k\,p_r\,\rho_k\,,\\
	& \cov\left(XYZ_{[k]}, Y^2Z_{[r]}\right) = \E\left[XY^3Z_{[k]}Z_{[r]}\right] - \E\left[XYZ_{[k]}\right]\E\left[Y^2Z_{[r]}\right] = 0 - p_k\,\rho_k\,p_r = - p_k\,p_r\,\rho_k\,,\\
	& \cov\left(XYZ_{[k]}, XYZ_{[r]}\right) = \E\left[X^2Y^2Z_{[k]}Z_{[r]}\right] - \E\left[XYZ_{[k]}\right]\E\left[XYZ_{[r]}\right] = 0 - p_k\,\rho_k\,p_r\,\rho_r = - p_k\,p_r\,\rho_k\,\rho_r\,.
\end{align*}
That is,
\begin{align*}
\Sigma_{kr} &= \begin{bmatrix}
 	-p_k\,p_r & 0 & 0 & -p_k\,p_r & -p_k\,p_r & -p_k\,p_r\,\rho_r\\
 	0 & 0 & 0 & 0 & 0 & 0\\
 	0 & 0 & 0 & 0 & 0 & 0\\
 	-p_k\,p_r & 0 & 0 & -p_k\,p_r & -p_k\,p_r & -p_k\,p_r\,\rho_r\\
 	-p_k\,p_r & 0 & 0 & -p_k\,p_r & -p_k\,p_r & -p_k\,p_r\,\rho_r\\
 	-p_k\,p_r\,\rho_k & 0 & 0 & -p_k\,p_r\,\rho_k & -p_k\,p_r\,\rho_k & - p_k\,p_r\,\rho_k\,\rho_r\\
 \end{bmatrix}\,.
\end{align*}

Applying Cram\'er's Theorem \citep{ferguson2017course} to function $g$ (Equation (\ref{function_g})) and the joint asymptotic distribution of $\widehat{\mu}$ (Equation (\ref{asym_clt})), we have the joint asymptotic distribution of $\widehat{\theta} = g(\widehat{\mu})$ as
\begin{equation}
\sqrt{n}\left( \widehat{\btheta} - \btheta \right) \tod \cN\left(0, \bOmega\right)\,,	\label{asym_beta}
\end{equation}
where
\[
\btheta = g(\bmu) = g \begin{pmatrix}
 	p_1\\
 	0\\
 	0\\
 	p_1\\
 	p_1\\
 	p_1 \rho_1\\
 	\vdots\\
 	p_K\\
 	0\\
 	0\\
 	p_K\\
 	p_K\\
 	p_K \rho_K\\	
 \end{pmatrix}
 = \begin{pmatrix}
 	p_{1}\\
 	\rho_1\\
 	1\\
 	1\\
 	\vdots\\
 	p_{K}\\
 	\rho_K\\
 	1\\
 	1\\
 \end{pmatrix}\,,
\]
and
\[ \bOmega = Dg(\bmu)\,\bSigma\, \left(Dg(\bmu)\right)\tran\,.
\]

Evaluating the Jacobian of $g$ at $\bmu$, we have
\[
Dg(\bmu) = \begin{bmatrix}
D g(\bmu)_1 & & \\
& \ddots & \\
& & D g(\bmu)_K	
\end{bmatrix}\,,  
\]
with
\[
Dg(\bmu)_k =
\begin{bmatrix}
 	1 & 0 & 0 & 0 & 0 & 0 \\	
 	-\rho_k/p_k & 0 & 0 & 0 & 0 & 1/p_k \\
 	-1/p_k & 0 & 0 & 1/p_k & 0 & 0 \\
 	-1/p_k & 0 & 0 & 0 & 1/p_k & 0 
\end{bmatrix}\,.
\]

Hence,
\begin{align*}
\bOmega &= 
\begin{bmatrix}	\Omega_{11} &  \cdots & \Omega_{1K}\\
	\vdots & \ddots & \vdots\\
	\Omega_{K1} &  \cdots & \Omega_{KK}\\	
 \end{bmatrix}_{(4K)\times(4K)}  \,,	
\end{align*}
where for $k=1,\ldots,K$,
\begin{align*}
\bOmega_{kk} &=	Dg(\bmu)_k\,\bSigma_{kk}\, \left(Dg(\bmu)_k\right)\tran\\
	&= 
	\begin{bmatrix}
 	p_k(1-p_k) & 0 & 0 & 0\\
 	0 & \frac{-\rho_k^2+\mu_{X^2Y^2,k}}{p_k} & \frac{-\rho_k+\mu_{X^3Y,k}}{p_k} & \frac{-\rho_k+\mu_{XY^3,k}}{p_k}\\
 	0 & \frac{-\rho_k+\mu_{X^3Y,k}}{p_k} & \frac{-1+\mu_{X^4,k}}{p_k} & \frac{-1+\mu_{X^2Y^2,k}}{p_k}\\
 	0 & \frac{-\rho_k+\mu_{XY^3,k}}{p_k} & \frac{-1+\mu_{X^2Y^2,k}}{p_k} & \frac{-1+\mu_{Y^4,k}}{p_k}
	\end{bmatrix}\,,
\end{align*}
and for $1\le k \neq r \le K$,
\begin{align*}
\bOmega_{kr} &=	Dg(\bmu)_k\,\bSigma_{kr}\, \left(Dg(\bmu)_r\right)\tran = 
	\begin{bmatrix}
 	-p_k\,p_r & 0 & 0 & 0\\
 	0 & 0 & 0 & 0\\
 	0 & 0 & 0 & 0\\
 	0 & 0 & 0 & 0
	\end{bmatrix}\,.
\end{align*}

With the joint asymptotic distribution of $\widehat\btheta = (\widehat p_1, \, \widehat\sigma_{XY,1}, \, \widehat\sigma_{X,1}^2, \, \widehat\sigma_{Y,1}^2, \, \cdots, \, \widehat p_K, \, \widehat\sigma_{XY,K}, \, \widehat\sigma_{X,K}^2, \, \widehat\sigma_{Y,K}^2 )\tran$, our final step is to derive the asymptotic distribution of $R_{\cG\cS}^2$. 

Given any $\bEta = (x_1, y_1, z_1, u_1, \cdots, x_K, y_K, z_K, u_K)\tran \in \R^{4K}$, we define a function $h: \R^{4K} \rightarrow \R$ as
\[ h(\bEta) = \sum_{k=1}^K x_k \frac{y_k^2}{z_ku_k}.
\]
Hence, \[ R_{\cG\cS}^2  = h(\widehat\btheta)\,.\]

The gradient of $h$ is
\[ \nabla h(\bEta) = \left(\frac{y_1^2}{z_1u_1},\, 2x_1 \frac{y_1}{z_1u_1},\, -x_1 \frac{y_1^2}{z_1^2u_1},\, -x_1 \frac{y_1^2}{z_1u_1^2},\, \cdots,\,  \frac{y_K^2}{z_Ku_K},\, 2x_K \frac{y_K}{z_Ku_K},\, -x_K \frac{y_K^2}{z_K^2u_K},\, -x_K \frac{y_K^2}{z_Ku_K^2}\right)\tran\,.
\]

Applying Cram\'er's Theorem \citep{ferguson2017course} to the joint asymptotic distribution of $\widehat\btheta$ (Equation (\ref{asym_beta})), we have
\[ \sqrt{n}\left( R_{\cG\cS}^2 - h(\btheta) \right) \tod \cN\left(0, \left(\nabla h(\btheta)\right)\tran \bOmega\, \nabla h(\btheta)\right)\,.
\]

Given $\btheta = (p_1,\, \rho_1,\, 1,\, 1, \cdots, 	p_K,\, \rho_K,\, 1, 1)\tran
$, it is easy to see that
\[ h(\btheta) = \sum_{k=1}^K p_k \rho_k^2 = \rho_{\cG\cS}^2\,,
\]
based on the definition of the population measure $\rho_{\cG\cS}^2$ in Equation (\ref{pcor2_g}).

Evaluating the gradient of $h$ at $\btheta$, we have
\[ \nabla h(\btheta) = \left( (\nabla h(\btheta)_1)\tran ,\, \cdots, (\nabla h(\btheta)_K)\tran \right)\tran\,, \text{ with }  \nabla h(\btheta)_k = \left(\rho_k^2,\, 2p_k\rho_k,\,-p_k\rho_k^2,\,-p_k\rho_k^2\right)\tran\,.
\]
Hence,
\begin{align*}
\left(\nabla h(\btheta)\right)\tran \bOmega\, \nabla h(\btheta) = \sum_{k=1}^K \sum_{r=1}^K (\nabla h(\btheta)_k)\tran \bOmega_{kr} \nabla h(\btheta)_r	\,.
\end{align*}
For $k = 1,\ldots,K$,
\begin{align*}
&(\nabla h(\btheta)_k)\tran \bOmega_{kk} \nabla h(\btheta)_k \\
=& \left(\rho_k^2,\, 2p_k\rho_k,\,-p_k\rho_k^2,\,-p_k\rho_k^2\right)	\footnotesize\begin{bmatrix}
 	p_k(1-p_k) & 0 & 0 & 0\\
 	0 & \frac{-\rho_k^2+\mu_{X^2Y^2,k}}{p_k} & \frac{-\rho_k+\mu_{X^3Y,k}}{p_k} & \frac{-\rho_k+\mu_{XY^3,k}}{p_k}\\
 	0 & \frac{-\rho_k+\mu_{X^3Y,k}}{p_k} & \frac{-1+\mu_{X^4,k}}{p_k} & \frac{-1+\mu_{X^2Y^2,k}}{p_k}\\
 	0 & \frac{-\rho_k+\mu_{XY^3,k}}{p_k} & \frac{-1+\mu_{X^2Y^2,k}}{p_k} & \frac{-1+\mu_{Y^4,k}}{p_k}
	\end{bmatrix}
	\normalsize\begin{pmatrix}
	\rho_k^2\\
	2p_k\rho_k\\
	-p_k\rho_k^2\\
	-p_k\rho_k^2	
	\end{pmatrix}\\
=& \:p_k \left[ \rho_k^4 \left(\mu_{X^4,k} + 2 \mu_{X^2Y^2,k} + \mu_{Y^4,k}\right) - 4\rho_k^3 \left( \mu_{X^3Y,k} + \mu_{XY^3,k} \right) + 4\rho_k^2 \mu_{X^2Y^2,k}\right] + p_k (1-p_k)\rho_k^4\\
=& \: A_{k\cS} + B_{k\cS}\,.
\end{align*}
For $1\le k \neq r \le K$,
\begin{align*}
&(\nabla h(\btheta)_k)\tran \bOmega_{kr} \nabla h(\btheta)_r \\
=& \left(\rho_k^2,\, 2\rho_k,\,-\rho_k^2,\,-\rho_k^2\right)	\begin{bmatrix}
 	-p_k\,p_r & 0 & 0 & 0\\
 	0 & 0 & 0 & 0\\
 	0 & 0 & 0 & 0\\
 	0 & 0 & 0 & 0
	\end{bmatrix}
	\begin{pmatrix}
	\rho_r^2\\
	2\rho_r\\
	-\rho_r^2\\
	-\rho_r^2	
	\end{pmatrix}\\
=& \:-p_k\,p_r\,\rho_k^2\,\rho_r^2\\
=& \:C_{kr\cS}\,.
\end{align*}
Hence,
\begin{align*}
\left(\nabla h(\btheta)\right)\tran \bOmega\, \nabla h(\btheta) &= \sum_{k=1}^K (\nabla h(\btheta)_k)\tran \bOmega_{kk} \nabla h(\btheta)_k + 2 \mathop{\sum \sum}_{1 \le k < r \le K} (\nabla h(\btheta)_k)\tran \bOmega_{kr} \nabla h(\btheta)_r	\\
	&= \sum_{k=1}^K \left( A_{k\cS} + B_{k\cS}\right) + 2 \mathop{\sum \sum}_{1 \le k < r \le K} C_{kr\cS}\,.
\end{align*}
Therefore, the asymptotic distribution of $R_{\cG\cS}^2$ is 
\[
\sqrt{n} \left(R_{\cG\cS}^2 - \rho_{\cG\cS}^2\right) \tod \cN\left(0, \sum_{k=1}^K \left( A_{k\cS} + B_{k\cS}\right) + 2 \mathop{\sum \sum}_{1 \le k < r \le K} C_{kr\cS}\right)\,,
\]
which completes the proof.
\end{proof}

\subsection{Proof of Corollary \ref{cor:asym_dist_scor2_gc}}\label{A.3}
\begin{proof}
Following the proof of Theorem \ref{thm_asym_dist_scor2_gc}, where we assume without loss of generality that
\begin{align*}
\mu_{X,k} &= \E[X|Z=k] = 0\,,\\	
\mu_{Y,k} &= \E[Y|Z=k] = 0\,,\\
\sigma_{X,k}^2 &= \var[X|Z=k] = 1\,,\\	
\sigma_{Y,k}^2 &= \var[Y|Z=k] = 1\,.
\end{align*}
Then in this special case, with the same notations used in the proof of Theorem \ref{thm_asym_dist_scor2_gc}, we have
\[ (X, Y)|(Z=k) \sim \cN\left(0, \begin{bmatrix}
 	1 & \rho_k\\
 	\rho_k & 1
 	\end{bmatrix}
\right)\,,
\]
with the following moments
\begin{equation*}
\begin{aligned}[c]
	\mu_{X^4,k} &= \E[X^4|Z=k] = 3\,,\\
	\mu_{X^3Y,k} &= \E[X^3Y|Z=k] = 3\rho_k\,,\\
	\mu_{X^2Y^2,k} &= \E[X^2Y^2|Z=k] = (1+2\rho_k^2)\,.
\end{aligned}
\qquad	
\begin{aligned}[c]
	\mu_{Y^4,k} &= \E[Y^4|Z=k] = 3\,,\\
	\mu_{XY^3,k} &= \E[XY^3|Z=k] = 3\rho_k\,,
\end{aligned}
\end{equation*}
Hence,
\begin{align*}
A_k &= p_k \left[ \rho_k^4 \left(\mu_{X^4,k} + 2 \mu_{X^2Y^2,k} + \mu_{Y^4,k}\right) - 4\rho_k^3 \left( \mu_{X^3Y,k} + \mu_{XY^3,k} \right) + 4\rho_k^2 \mu_{X^2Y^2,k}\right]\\
	&= p_k\left( 8\rho_k^4  + 4\rho_k^6 - 24 \rho_k^4  + 8 \rho_k^4 + 4 \rho_k^2 \right)\\
	&= 4 \, p_k \, \rho_k^2 \left( 1 - \rho_k^2 \right)^2\,.
\end{align*}
Provided that
\begin{align*}
B_k &= p_k (1-p_k)\rho_k^4\,,\\
C_{kr} &= -p_k\,p_r\,\rho_k^2\,\rho_r^2\,, \qquad k\neq r\,,	
\end{align*}
the joint asymptotic distribution of $R_{\cG\cS}^2$ in (\ref{asym_dist_scor2_gc}) becomes (\ref{asym_dist_gauss_scor2_gc}), which completes the proof.
\end{proof}


\subsection{Proof of Theorem \ref{thm_strong_consistency}}

The first statement of Theorem \ref{thm_strong_consistency} means that there exists an ordering of the elements in $\widehat B_{K\cU} = \left\{ \widehat\bbeta_{1\cU}, \ldots, \widehat\bbeta_{K\cU} \right\}$ and $B_{K\cU} = \left\{ \bbeta_{1\cU}, \ldots, \bbeta_{K\cU} \right\}$ such that as the sample size $n \rightarrow \infty$,
\begin{equation}\label{beta_k_strong_consistency}
\widehat\bbeta_{k\cU} \rightarrow \bbeta_{k\cU} \; \text{ almost surely}\,,\; k=1,\ldots,K\,.	
\end{equation}

Theorem \ref{thm_strong_consistency} only hinges on three conditions. The first condition \[\int \norm{(x,y)\tran}^2 P\left( (dx, dy)\tran \right) < \infty \,,\]
where $||\cdot||$ is the $\ell_2$ norm in $\R^2$, holds for many common probability measures $P$ on $\R^2$. This is to ensure that $W(B_k, P)$ is finite for every $B_k$, $k=1,\ldots,K$, because for each $\bbeta = (a,b,c)\tran \in \R^3$, where $a \neq 0$ or $b \neq 0$ (without loss of generality, we assume $a \neq 0$ and $|c| < \infty$ below),

\begin{align}\label{eq:dist_finite}
\int d_\perp^2 \left((x,y)\tran, \bbeta\right) P\left( (dx, dy)\tran \right) 
&\le \int \norm{(x,y)\tran - \left( -c/a, 0 \right)\tran }^2 P\left( (dx, dy)\tran \right) \notag\\
&\le \int \left( \norm{(x,y)\tran} + \norm{\left( -c/a, 0 \right)\tran}\right)^2 P\left( (dx, dy)\tran \right) \notag\\
&\le 4c^2/a^2 + 2 \int_{\norm{(x,y)\tran} \ge |c/a|} \norm{(x,y)\tran}^2 P\left( (dx, dy)\tran \right) \notag\\
&< \infty \,,
\end{align}
where the first inequality holds by the definition of $d_\perp \left((x,y)\tran, \bbeta\right)$ (\ref{def:perp_dist}), because $(-c/a,0)\tran$ is a point on the line $\left\{(x,y)\tran: ax+by+c=0\right\}$ defined by $\bbeta$. 
The second uniqueness condition on $B_{k\cU}$ is needed for the inductive argument in the proof.

We first prove the following lemma.
\begin{lemma}\label{lemma:triangle_inequality}
Given two points $(x, y)\tran, (x', y')\tran \in \R^2$ and a line $\bbeta=(a,b,c)\tran$ where $a,b,c \in \R$ with $a \neq 0$ or $b \neq 0$, the following inequality holds.
\begin{equation}
	d_\perp\left( (x,y)\tran, \bbeta \right) \le \norm{(x,y)\tran - (x',y')\tran} + d_\perp\left( (x',y')\tran, \bbeta \right)\,,
\end{equation}
where $\norm{\cdot}$ represents the $\ell_2$ norm.	
\end{lemma}

\begin{proof}
By the definition of the perpendicular distance $d_\perp(\cdot,*)$ in Equation (\ref{def:perp_dist}), we have for $\bbeta = (a,b,c)\tran$, where $a, b, c \in \R$ with $a \neq 0$ or $b \neq 0$, that
\[ d_\perp\left( (x,y)\tran, \bbeta \right) \le \norm{(x,y)\tran - (x_0,y_0)\tran}\,, \,\forall (x_0,y_0)\tran \text{ s.t. } ax_0 + by_0 + c = 0\,,
\] 	
and
\[ d_\perp\left( (x',y')\tran, \bbeta \right) = \norm{(x',y')\tran - \left(\frac{b(bx' - ay') - ac}{a^2+b^2}, \frac{a(-bx' + ay') - bc}{a^2+b^2}\right)\tran}\,,
\]
where $\left(\frac{b(bx' - ay') - ac}{a^2+b^2}, \frac{a(-bx' + ay') - bc}{a^2+b^2}\right)\tran$ is a point on the line corresponding to $\bbeta$.
Therefore,
\begin{align*}
	d_\perp\left( (x,y)\tran, \bbeta \right) &\le \norm{(x,y)\tran - \left(\frac{b(bx' - ay') - ac}{a^2+b^2}, \frac{a(-bx' + ay') - bc}{a^2+b^2}\right)\tran}\\
	& \le \norm{(x,y)\tran - (x',y')\tran} + \norm{(x',y')\tran - \left(\frac{b(bx' - ay') - ac}{a^2+b^2}, \frac{a(-bx' + ay') - bc}{a^2+b^2}\right)\tran}\\
	& \le \norm{(x,y)\tran - (x',y')\tran} + d_\perp\left( (x',y')\tran, \bbeta \right)\,,
\end{align*}
which completes the proof of Lemma \ref{lemma:triangle_inequality}.
\end{proof}

Leveraging Lemma \ref{lemma:triangle_inequality} and the strong consistency result of the $K$-means clustering \citep{pollard1981strong}, we derive the strong consistency of the $K$ sample-level unspecified line centers found by the $K$-lines clustering algorithm (Algorithm \ref{alg:K-lines}). A key technical challenge we need to tackle in this proof is the handling of two types of points, where the \textit{data} type refers to the $n$ data points $(X_1,Y_1)\tran,\ldots, (X_n,Y_n)\tran$, and the \textit{center} type corresponds to the $K$ lines $\bbeta_1=(a_1,b_1,c_1)\tran, \ldots, \bbeta_K=(a_K,b_K,c_K)\tran$, where $a_k \neq 0$ or $b_k \neq 0$, $k=1,\ldots,K$, as cluster centers. In this proof, without loss of generality, we assume $0 < c_k < \infty$, $k=1,\ldots,K$, because we can always find a vertical shift of the data and the $K$ lines to satisfy this condition. To simplify the following discussion, we define a \textit{center point} $(\eta, \xi)\tran$ as an equivalent representation of a line $\bbeta =(a, b, c)\tran$, such that
\begin{align}\label{def:center_point}
	\left\{
	\begin{array}{l}
		\eta = -ac/(a^2+b^2)\\
		\xi = -bc/(a^2+b^2)
	\end{array}
	\right. \text{ and } \left\{
	\begin{array}{l}
		a = -\eta/\sqrt{\eta^2 + \xi^2} \cdot u\\
		b = -\xi/\sqrt{\eta^2 + \xi^2} \cdot u\\
		c = \sqrt{\eta^2 + \xi^2} \cdot u
	\end{array}
	\right. \text{with } u \neq 0 \,,
\end{align}
so that $\norm{(\eta, \xi)\tran} = d_\perp\left( (0,0)\tran, \bbeta \right)$, where the perpendicular distance $d_\perp(\cdot,*)$ is defined in Equation (\ref{def:perp_dist}).

We define the distance between a data point $(x,y)\tran$ and a center point $(\eta,\xi)\tran$ not as the usual Euclidean distance but as
\begin{align*}
	d'\left( (x,y)\tran, (\eta, \xi)\tran \right) &:= d_\perp\left((x,y)\tran, \bbeta \right) = \frac{|-\eta x - \xi y + \eta^2 + \xi^2|}{\sqrt{\eta^2 + \xi^2}} \,.
\end{align*}
Note that $d'(\cdot,*)$ is not a metric, and $d'\left( (x,y)\tran, (\eta, \xi)\tran \right) = \norm{(x,y)\tran - (\eta, \xi)\tran}$ if and only if $(x,y)\tran = (0,0)\tran$.

The result in Lemma \ref{lemma:triangle_inequality} now translates to
\begin{equation}\label{lemma:triangle_inequality_2}
	d'\left( (x,y)\tran, (\eta, \xi)\tran \right) \le \norm{(x,y)\tran - (x',y')\tran} + d'\left( (x',y')\tran, (\eta, \xi)\tran \right).	
\end{equation}

For the rest of the proof, we denote the set of $K$ cluster centers using two equivalent representations:
\begin{enumerate}
\item $B_K' = \left\{ (\eta_1,\xi_1)\tran, \ldots, (\eta_K,\xi_K)\tran \right\}$, where each cluster center is represented by a center point.
\item $B_K = \left\{ \bbeta_1, \ldots, \bbeta_K \right\}$, where each line $\bbeta_k = (a_k, b_k, c_k)\tran$, where $a_k \neq 0$ or $b_k \neq 0$.
\end{enumerate}
The one-to-one relationship between $(\eta_k, \xi_k)\tran$ and $(a_k, b_k, c_k)\tran$ follows from (\ref{def:center_point}). Hence, there is a one-to-one correspondence between $B_K'$ and $B_K$, and we define
\begin{align}\label{def:W_prime}
W'(B_K', P) &:= \int \min_{(\eta,\xi)\tran \in B_K'} d'^2\left( (x,y)\tran, (\eta,\xi)\tran\right) P\left((dx, dy)\tran\right) \notag\\	
&= \int \min_{\bbeta \in B_K} d_\perp^2 \left( (x,y)\tran, \bbeta \right)P\left((dx, dy)\tran\right) \notag\\
&= W(B_K,P)\,.
\end{align}

We define the set of center points to which the origin data point $(0,0)\tran$ has a distance no greater than $M>0$ as
\begin{align*}
	\cB(M) := \left\{(\eta,\xi)\tran: d'\left( (0,0)\tran, (\eta,\xi)\tran \right) \le M \right\} &= \left\{(\eta,\xi)\tran: \eta^2 + \xi^2 \le M^2 \right\}\,,
\end{align*}
which is a closed ball of radius $M$ and centered at the origin in $\R^2$.

The first step consists of finding an $M > 0$ so large that, when $n$ is large enough, at least one center point in $\widehat B_{K\cU}' = \left\{ (\widehat\eta_1,\widehat\xi_1)\tran, \ldots, (\widehat\eta_K,\widehat\xi_K)\tran \right\}$ (the alternative representation of $\widehat B_{K\cU} = \left\{ \widehat\bbeta_{1\cU}, \ldots, \widehat\bbeta_{K\cU} \right\}$ defined in Equation (\ref{def:beta_ku_s})) is contained in the set $\cB(M)$.  We will prove this in the following  using a counterexample. 

\begin{itemize}
\item We find an $r > 0$ so that the ball $\cB(r) \subset \R^2$ of radius $r$ and centered at the origin has a positive $P$ measure. For the purposes of this first step, it will suffice that $M$ be large enough to make 
\begin{align}\label{eq:M_req1}
	 (M-r)^2 \, P(\cB(r)) > \int d'^2\left( (x,y)\tran, (1,1)\tran\right) P\left( (dx, dy)\tran \right)\,,\\
	 \text{(1st requirement on $M$)}\notag
\end{align}
whose right hand side is equal to $\int d_\perp \left((x,y)\tran, (-1,-1,2)\tran\right)^2 P\left( (dx, dy)\tran \right)$, which is finite by (\ref{eq:dist_finite}). 
Define $B_0' = \left\{ (1,1)\tran \right\}$, a set with one center point. Then (\ref{eq:M_req1}) becomes 
\begin{equation}\label{eq:M_req1a}
(M-r)^2 \, P(\cB(r)) > W'(B_0', P)\,.	
\end{equation}

\item By the strong law of large numbers (SLLN),
\begin{align} \label{eq:origin_conv}
W'(B_0', P_n) &=\int d'^2\left( (x,y)\tran, (1,1)\tran\right) P_n\left((dx, dy)\tran\right) \notag\\
&\rightarrow \int d'^2\left( (x,y)\tran, (1,1)\tran\right) P\left((dx, dy)\tran\right) \text{ a.s. }\notag\\
&= W'(B_0', P)\,.
\end{align}

\item If, for infinitely many values of $n$, no centers in $\widehat B_{K\cU}'$ were contained in $\cB(M)$,\\ i.e., $\min_{(\eta, \xi)\tran \in \widehat B_{K\cU}'} d' \left((0, 0)\tran, (\eta, \xi)\tran \right) > M $, then
\begin{equation}\label{eq:contra1}
	\limsup_n W'(\widehat B_{K\cU}', P_n) \ge \lim_n (M-r)^2 \, P_n(\cB(r)) = (M-r)^2 \, P(\cB(r)) \text{ a.s.}\,,
\end{equation}
where the first inequality holds because for any data point $(X_i,Y_i)\tran \in \cB(r)$, which has a positive $P$ measure, the following holds by (\ref{lemma:triangle_inequality_2}):
\begin{align*}
\min_{(\eta,\xi)\tran \in B_n'}d'\left((X_i,Y_i)\tran, (\eta,\xi)\tran \right) &\ge  \min_{(\eta,\xi)\tran \in B_n'}d'\left((0,0)\tran, (\eta,\xi)\tran\right) - \norm{(X_i,Y_i)\tran - (0,0)\tran}\\ 
&> M - r \,. 
\end{align*} 

\item Hence by (\ref{eq:M_req1a}), (\ref{eq:origin_conv}) and (\ref{eq:contra1}), 
\begin{equation}
\limsup_n W'(\widehat B_{K\cU}', P_n) > \lim_n W'(B_0', P_n)\,,	
\end{equation}
which implies that $W(\widehat B_{K\cU}, P_n) = W'(\widehat B_{K\cU}', P_n) > W'(B_0', P_n) = W(B_0, P_n)$ infinitely often, where $B_0 = \left\{ (-1,2)\tran \right\}$ is the set of one line corresponding to $B_0'$. This contradicts the  definition of $\widehat B_{K\cU}$ in (\ref{def:beta_ku_s}) of Definition \ref{def:line_center_u_sample}: $W(\widehat B_{K\cU}, P_n) \le W(B_0, P_n)$ for any set $B_0$ containing at most $K$ lines. Without loss of generality, we therefore assume that $\widehat B_{K\cU}'$ always contains at least one center point in $\cB(M)$. We denote one of such center points as $(\widehat\eta_1, \widehat\xi_1)\tran$.	
\end{itemize}

If $K=1$, the second step in this proof can be skipped; if $K>1$, then we will show that, for $n$ large enough, the set $\cB(5M)$ contains all the center points in $\widehat B_{K\cU}'$.  
\begin{itemize}
\item For the purposes of an inductive argument, we assume that the conclusions of the theorem are valid when applied to globally optimal clustering with $1,2,\ldots,K-1$ centers. If $\widehat B_{K\cU}'$ were not eventually contained in $\cB(5M)$, we could delete the center points in $\widehat B_{K\cU}'$ that are outside of $\cB(5M)$ to obtain a set of at most $K-1$ center points that would reduce $W'(\cdot, P_n)$ below its minimum over all sets of $K-1$ centers, i.e., $W'(\widehat B_{K-1\cU}', P_n)$, which results in a contradiction.

\item To obtain such a contradiction, we need $M$ large enough to ensure that
\begin{align}\label{eq:M_req2}
4\int_{\norm{(x,y)\tran}\ge 2M} \norm{(x,y)\tran}^2	 P\left( (dx, dy)\tran \right) < \epsilon\,,\\
\text{(2nd requirement on $M$)}\notag
\end{align}
where $\epsilon > 0$ is chosen to satisfy $\epsilon < W'(B_{K-1\cU}', P) - W'(B_{K\cU}', P)$, which is positive by the uniqueness condition on $B_{k\cU}$ and thus $B_{k\cU}'$. 

\item Suppose that $\widehat B_{K\cU}'$ contains at least one center point outside $\cB(5M)$. Denote one of such center points as $(\widehat\eta_K, \widehat\xi_K)\tran$. What would be the effect on $W'(\cdot, P_n)$ of deleting such center points in $\widehat B_{K\cU}'$? At worst, one of the centers that are known to lie in $\cB(M)$, e.g., $(\widehat\eta_1, \widehat\xi_1)\tran$, might have to accept into its own cluster all of the data points presently assigned to cluster centers outside $\cB(5M)$. Denote one of these data points as $(X_i,Y_i)\tran$. We argue that $(X_i,Y_i)\tran$ must have a Euclidean distance at least $2M$ from the origin data point $(0,0)\tran$; otherwise $(X_i,Y_i)\tran$ would have been closer to the cluster center $(\widehat\eta_1, \widehat\xi_1)\tran$ than to $(\widehat\eta_K, \widehat\xi_K)\tran$. The reason is as follows. By (\ref{lemma:triangle_inequality_2}),
\begin{align*}
d'\left((X_i,Y_i)\tran, (\widehat\eta_1, \widehat\xi_1)\tran\right)	&\le \norm{(X_i,Y_i)\tran - (0,0)\tran} + d'\left((0,0)\tran, (\widehat\eta_1, \widehat\xi_1)\tran\right)\\ 
	&\le \norm{(X_i,Y_i)\tran} + M\,,\\
\end{align*}
\begin{align*}
d'\left((X_i,Y_i)\tran, (\widehat\eta_K, \widehat\xi_K)\tran\right)	&\ge d'\left((0,0)\tran, (\widehat\eta_K, \widehat\xi_K)\tran\right) - \norm{(X_i,Y_i)\tran - (0,0)\tran}\\ 
	&\ge 5M - \norm{(X_i,Y_i)\tran}\,.
\end{align*}
If $\norm{(X_i,Y_i)\tran} < 2M$, then 
\[ d'\left((X_i,Y_i)\tran, (\widehat\eta_1, \widehat\xi_1)\tran\right) < 3M < d'\left((X_i,Y_i)\tran, (\widehat\eta_K, \widehat\xi_K)\tran\right)\,.
\] 
That is, if $(X_i,Y_i)\tran$ were assigned to the cluster with center $(\widehat\eta_K, \widehat\xi_K)\tran$, then $\norm{(X_i,Y_i)\tran} \ge 2M$.

\item The extra contribution to $W'(\cdot, P_n)$ due to deleting centers outside $\cB(5M)$ would therefore be at most
\begin{align}\label{eq:extra_contri}
& \int_{\norm{(x,y)\tran} \ge 2M}	d'^2\left((x,y)\tran, (\widehat\eta_1, \widehat\xi_1)\tran\right) P_n\left( (dx, dy)\tran \right) \notag\\
\le& \int_{\norm{(x,y)\tran} \ge 2M} \left( \norm{(x,y)\tran - (0,0)\tran} + d'\left((0,0)\tran, (\widehat\eta_1, \widehat\xi_1)\tran\right) \right)^2 P_n\left( (dx, dy)\tran \right) \notag\\
\le& 4\int_{\norm{(x,y)\tran} \ge 2M} \norm{(x,y)\tran}^2 P_n\left( (dx, dy)\tran \right)\,.
\end{align}	

\item The set $\widehat B_{K\cU}'^-$ obtained by deleting from $\widehat B_{K\cU}'$ all centers outside $\cB(5M)$ is a candidate for minimizing $W'(\cdot, P_n)$ over sets of $K-1$ or fewer centers; it is therefore beaten by the optimal set $\widehat B_{K-1\cU}'$. Thus
\begin{align}\label{eq:extra_contri2}
W'(\widehat B_{K\cU}'^-, P_n) &\ge W'(\widehat B_{K-1\cU}', P_n) \notag\\
&= W(\widehat B_{K-1\cU}, P_n) \notag\\
&\rightarrow W(B_{K-1\cU}, P) = W'(B_{K-1\cU}', P) \, \text{ a.s.} \,	
\end{align}
by the inductive hypothesis. If $\widehat B_{K\cU}' \not\subseteq \cB(5M)$ along some subsequence $\{n_j\}$ of sample sizes, we therefore obtain
\begin{align*}
&W'(B_{K-1\cU}', P) \\
\le & \liminf_j W'(\widehat B_{K\cU}'^-, P_{n_j}) \text{ a.s. \hspace{2.1in} by (\ref{eq:extra_contri2})}\\
\le & \limsup_n \left[ W'(\widehat B_{K\cU}', P_n) + 	4\int_{\norm{(x,y)\tran} \ge 2M} \norm{(x,y)\tran}^2 P_n\left( (dx, dy)\tran \right) \right] \text{\hfill by (\ref{eq:extra_contri})}\\
\le & \limsup_n W'(B_{K\cU}', P_n) + 4\int_{\norm{(x,y)\tran} \ge 2M} \norm{(x,y)\tran}^2 P\left( (dx, dy)\tran \right) \text{ a.s.}\\
< & W'(B_{K\cU}', P) + \epsilon \text{\hspace{2.8in} by (\ref{eq:M_req2})}\\
< & W'(B_{K-1\cU}', P)\,,
\end{align*}
a contradiction.
\end{itemize}

We now know that, for $n$ large enough, it suffices to search for $\widehat B_{K\cU}'$ among the class of sets $\Xi_K = \left\{ B' \subseteq \cB(5M): |B'| \le K \right\}$. For the final requirement on $M$, we assume it is large enough to ensure that $\Xi_K$ contains $B_{K\cU}'$. Therefore, the function $W'(\cdot, P)$ achieves its unique minimum on $\Xi_K$ at $B_{K\cU}'$. Under the topology induced by the Hausdorff metric, $\Xi_K$ is compact (this follows from the compactness of $\cB(5M)$) and, as is proved in \citep{pollard1981strong}, the map $B' \rightarrow W'(B', P)$ is continuous on $\Xi_K$. The function $W'(\cdot, P)$ therefore enjoys an even stronger minimization property on $\Xi_K$: given any neighborhood $\cN$ of $B_{K\cU}'$, there exists an $\delta > 0$, depending on $\cN$, such that
\begin{align*}
W'(B', P) \ge W'(B_{K\cU}', P) + \delta\,,	 \: \forall B' \in \Xi_K \backslash \cN\,.
\end{align*}
The proof can now be completed by an appeal to a uniform  SLLN:
\begin{align*}
\sup_{B' \in \Xi_K} |W'(B', P_n) - W'(B', P)| \rightarrow 0 \text{ a.s.}	
\end{align*}
This result is proved in \citep{pollard1981strong}. We need to show that $\widehat B_{K\cU}'$ is eventually inside the neighborhood $\cN$. It is enough to check that $W'(\widehat B_{K\cU}', P) < W'(B_{K\cU}', P) + \delta$ eventually. This follows from
\begin{align*}
W'(\widehat B_{K\cU}', P_n) \le W'(B_{K\cU}', P_n)	
\end{align*}
and
\begin{align*}
W'(\widehat B_{K\cU}', P_n) - W'(\widehat B_{K\cU}', P) \rightarrow 0 \text{ a.s.}
\end{align*}
and
\begin{align*}
W'(B_{K\cU}', P_n) - W'(B_{K\cU}', P) \rightarrow 0 \text{ a.s.}
\end{align*}
Therefore, $\widehat B_{K\cU}' \rightarrow B_{K\cU}'$ a.s. That is, $\widehat B_{K\cU} \rightarrow B_{K\cU}$ a.s. 

Similarly, for $n$ large enough,
\begin{align*}
W(\widehat B_{K\cU}, P_n) &= W'(\widehat B_{K\cU}', P_n) \\
	&= \inf\{ W'(B', P_n): B' \in \Xi_K \}	\\ 
	&\rightarrow \inf\{ W'(B', P): B' \in \Xi_K \}\\ 
	&= W'(B_{K\cU}', P) = W(B_{K\cU}, P)\text{ a.s.}
\end{align*}


\subsection{Proof of $\rho_{\cG\cU}^2 \ge \rho_{\cG\cS}^2$}\label{pf:supervised_vs_unsupervised}
\begin{proof}
Denote by $Z$ the index variable under the specified scenario and by $\widetilde{Z}$ the surrogate index variable under the unspecified scenario. Both $Z, \widetilde{Z} \in \{1,\ldots,K\}$. Given Definition \ref{def:pop_gen_R2} of $\rho_{\cG\cS}^2$ and Definition \ref{def:u_pop_gen_R2} of $\rho_{\cG\cU}^2$, if we denote
\[ \rho^2(Z) = \frac{\cov^2(X, Y|Z)}{\var(X|Z) \var(Y|Z)}\,,
\]
then
\begin{equation}\label{eq:equiv_exp}
	\rho_{\cG\cS}^2 = \E_Z\left[\rho^2(Z)\right] \text{ and }
\rho_{\cG\cU}^2 = \E_{\widetilde{Z}}\left[\rho^2(\widetilde{Z})\right]\,.
\end{equation}

Without loss of generality, we assume $\rho(Z), \rho(\widetilde{Z}) \ge 0$, $\var(X|Z) = \var(Y|Z) = \var(X|\widetilde{Z}) = \var(Y|\widetilde{Z}) = 1$ for all $Z, \widetilde{Z}$. 

Denote by $\lambda_1(Z)$ and $\lambda_2(Z)$ the first and second eigenvalues of the correlation matrix of $(X, Y)|Z$, denoted by $\bSigma(Z)$. An important property about the two eigenvalues is
\[
	\lambda_1(Z) + \lambda_2(Z) = 2 \text{ and }
	\lambda_1(Z) = 1 + \rho(Z)\,,
\]
where the second equation was proved in \citep{morrison1967multivariate}. Hence
\begin{equation}\label{eq:eigenvalue_properties}
\lambda_2(Z) = 1 - \rho(Z) \in [0,1]\,.	
\end{equation}

$\lambda_2(Z)$ is equal to the variance of the projection of $(X, Y)|Z$ onto its second principal component, which is perpendicular to its first principal component, i.e., the line center defined in Definition \ref{def:line_center} and denoted by $\bbeta(Z)$ here. Hence,
\begin{equation*}
\lambda_2(Z) = \E\left[ d_\perp^2 \left( (X,Y)\tran, \bbeta(Z) \right) \bigg| Z \right]\,.	
\end{equation*}
Denote by $B_K(Z)$ the set of $K$ population line centers corresponding to $(X, Y, Z)$. Then
\begin{align}\label{eq:eigenvalue_comp} 
	\E_Z[\lambda_2(Z)] &\ge \E_{(X,Y,Z)}\left[ \min_{\bbeta \in B_K(Z)} d_\perp^2\left((X,Y)\tran, \bbeta\right) \right] \\
	&\ge \E_{(X,Y,\widetilde{Z})}\left[ \min_{\bbeta \in B_K(\widetilde{Z})} d_\perp^2\left((X,Y)\tran, \bbeta\right) \right] = \E_{\widetilde{Z}}[\lambda_2(\widetilde{Z})] \,, \notag
\end{align}
where the second inequality and the last equality hold by Definition \ref{def:line_center_u} of the unspecified population line centers and Definition \ref{def:tilde_Z} of $\widetilde{Z}$.

By (\ref{eq:equiv_exp}), (\ref{eq:eigenvalue_properties}) and (\ref{eq:eigenvalue_comp}),
\begin{align*}
\rho_{\cG\cS}^2 - \rho_{\cG\cU}^2 &= \E_Z\left[\rho^2(Z)\right] - \E_{\widetilde{Z}}\left[\rho^2(\widetilde{Z})\right]\\
&= \E_Z\left[ (1-\lambda_2(Z))^2\right] - \E_{\widetilde{Z}}\left[ (1-\lambda_2(\widetilde{Z}))^2\right]\\
&= \E_{(Z, \widetilde Z)}\left[\left(\lambda_2(\widetilde{Z}) - \lambda_2(Z)\right) \left(2 - \lambda_2(Z) - \lambda_2(\widetilde{Z})\right) \right]\\
&\le 2 \E_{(Z, \widetilde Z)}\left[\lambda_2(\widetilde{Z}) - \lambda_2(Z) \right]\\
&\le 0\,,
\end{align*}
which completes the proof.
\end{proof}

\subsection{Proof of Theorem \ref{thm_asym_dist_scor2_gu}}

\begin{proof}
We derive the asymptotic distribution of the sample-level unspecified generalized Pearson correlation square, $R_{\cG\cU}^2$, in this proof. For notation simplicity, we drop the subscript ``$\cU$" in $p_{k\cU}$, $\rho^2_{k\cU}$, $\widehat p_{k\cU}$, $\widehat\rho^2_{k\cU}$, $\bar X_{k\cU}$, and $\bar Y_{k\cU}$ in Definition \ref{def:u_sample_gen_R2}, following which 
we have
\begin{align}\label{def:R2_GU}
R_{\cG\cU}^2 = \sum_{k=1}^K \widehat{p}_{k} \, \widehat\rho^2_{k}	= \sum_{k=1}^K \widehat{p}_{k} \,\frac{\widehat\sigma_{XY,k}^2}{\widehat\sigma_{X,k}^2 \,\widehat\sigma_{Y,k}^2} \,,
\end{align}
where 
\begin{align*}
\widehat p_k &= \frac{1}{n} \sum_{i=1}^n \1\left(\widehat{Z}_i = k\right)\,,\\
\widehat\sigma_{XY,k} &= \frac{1}{n \widehat p_k} \sum_{i=1}^n (X_i - \bar X_k) (Y_i - \bar Y_k) \,\1\left(\widehat{Z}_i = k\right)\,,\\
\widehat\sigma_{X,k}^2 &= \frac{1}{n \widehat p_k} \sum_{i=1}^n (X_i - \bar X_k)^2 \,\1\left(\widehat{Z}_i = k\right)\,,\\
\widehat\sigma_{Y,k}^2 &= \frac{1}{n \widehat p_k} \sum_{i=1}^n (Y_i - \bar Y_k)^2 \,\1\left(\widehat{Z}_i = k\right)\,,
\end{align*}
with $\bar{X}_k = \frac{1}{n \widehat p_k}\sum_{i=1}^n X_i \, \1\left(\widehat{Z}_i = k\right)$ and $\bar{Y}_k = \frac{1}{n \widehat p_k}\sum_{i=1}^n Y_i \, \1\left(\widehat{Z}_i = k\right)$.

Similar to the proof of Theorem \ref{thm_asym_dist_scor2_gc}, to derive the asymptotic distribution of $R_{\cG\cU}^2$, we need to first derive the joint asymptotic distribution of 
\[ \widehat{\theta} = \left(\widehat p_1, \, \widehat\sigma_{XY,1}, \, \widehat\sigma_{X,1}^2, \, \widehat\sigma_{Y,1}^2, \, \cdots, \, \widehat p_K, \, \widehat\sigma_{XY,K}, \, \widehat\sigma_{X,K}^2, \, \widehat\sigma_{Y,K}^2 \right)\tran \in \R^{4K}\,,
\]
which depends on the joint asymptotic distribution of
\[ \widehat\bmu = \left(M_1, M_{X,1}, M_{Y,1}, M_{X^2,1}, M_{Y^2,1}, M_{XY,1}, \cdots, M_K, M_{X,K}, M_{Y,K}, M_{X^2,K}, M_{Y^2,K}, M_{XY,K}\right)\tran \in \R^{6K}\,,
\]
where
\begin{equation*}
\begin{aligned}[c]
M_{k} &:= \frac{1}{n} \sum_{i=1}^n \1\left(\widehat{Z}_i = k\right)\,,\\
M_{X,k} &:= \frac{1}{n} \sum_{i=1}^n X_i \, \1\left(\widehat{Z}_i = k\right) \,,\\	
M_{Y,k} &:= \frac{1}{n} \sum_{i=1}^n Y_i \, \1\left(\widehat{Z}_i = k\right) \,,\\
\end{aligned}
\qquad
\begin{aligned}[c]
M_{XY,k} &:= \frac{1}{n} \sum_{i=1}^n X_i Y_i \, \1\left(\widehat{Z}_i = k\right) \,,\\
M_{X^2,k} &:= \frac{1}{n} \sum_{i=1}^n X_i^2 \, \1\left(\widehat{Z}_i = k\right) \,,\\
M_{Y^2,k} &:= \frac{1}{n} \sum_{i=1}^n Y_i^2 \, \1\left(\widehat{Z}_i = k\right) \,,
\end{aligned}
\end{equation*}
$k=1,\ldots,K$.

To derive the joint asymptotic distribution of $\widehat\bmu$, we need to resolve the non i.i.d. nature of $(X_1,Y_1,\widehat{Z}_1), \ldots, (X_n,Y_n,\widehat{Z}_n)$, which is due to fact that $\widehat{Z}_i$ depends on the $K$ unspecified sample line centers $\widehat B_{K\cU}$ and thus the whole sample $(X_1,Y_1),\ldots,(X_n,Y_n)$. Given the $K$ unspecified population line centers $B_{K\cU} = \{\bbeta_{1\cU}, \ldots, \bbeta_{K\cU}\}$ defined in (\ref{def:beta_ku}) in Definition \ref{def:line_center_u}, we define the (unobservable) realization of $\widetilde Z$, defined in (\ref{eq:tilde_Z}), based on the sample $(X_1,Y_1),\ldots,(X_n,Y_n)$ as
\begin{align}\label{def:tilde_Z_i_star}
\widetilde Z_i := \argmin_{k\in \{1,\ldots,K\}} d_\perp \left( (X_i, Y_i)\tran, \bbeta_{k\cU} \right)\,.	
\end{align}

Because $(X_1,Y_1),\ldots,(X_n,Y_n)$ are i.i.d. and $B_{K\cU}$ is fixed, we have $(X_1,Y_1,\widetilde Z_1), \ldots, (X_n,Y_n,\widetilde Z_n)$ as i.i.d. from the joint distribution of $(X, Y, \widetilde Z)$.

Define
%
\[ \widehat\bmu^* = (M_1^*, M_{X,1}^*, M_{Y,1}^*, M_{X^2,1}^*, M_{Y^2,1}^*, M_{XY,1}^*, \cdots, M_K^*, M_{X,K}^*, M_{Y,K}^*, M_{X^2,K}^*, M_{Y^2,K}^*, M_{XY,K}^*)\tran \in \R^{6K}\,,
\]
where
\begin{equation*}
\begin{aligned}[c]
M_{k}^* &:= \frac{1}{n} \sum_{i=1}^n \1\left(\widetilde Z_i = k\right)\,,\\
M_{X,k}^* &:= \frac{1}{n} \sum_{i=1}^n X_i \, \1\left(\widetilde Z_i = k\right) \,,\\	
M_{Y,k}^* &:= \frac{1}{n} \sum_{i=1}^n Y_i \, \1\left(\widetilde Z_i = k\right) \,,\\
\end{aligned}
\qquad
\begin{aligned}[c]
M_{XY,k}^* &:= \frac{1}{n} \sum_{i=1}^n X_i Y_i \, \1\left(\widetilde Z_i = k\right) \,,\\
M_{X^2,k}^* &:= \frac{1}{n} \sum_{i=1}^n X_i^2 \, \1\left(\widetilde Z_i = k\right) \,,\\
M_{Y^2,k}^* &:= \frac{1}{n} \sum_{i=1}^n Y_i^2 \, \1\left(\widetilde Z_i = k\right) \,,
\end{aligned}
\end{equation*}
$k=1,\ldots,K$.

Because of the i.i.d. nature of $(X_1,Y_1,\widetilde Z_1), \ldots, (X_n,Y_n,\widetilde Z_n)$, following the proof of Theorem \ref{thm_asym_dist_scor2_gc}, we can derive the joint asymptotic distribution of $\widehat{\bmu}^*$, which we denote as
\begin{equation}\label{hat_mu_star_dist}
\sqrt{n}\left(\widehat{\bmu}^* - \bmu\right) \tod \cN(0, \bSigma)\,,
\end{equation}
where $\bSigma$ has the same form as in (\ref{asym_clt}), except that all the notations related to $Z$ under the specified scenario are now related to $\widetilde Z$ under the current unspecified scenario. 


Next we will derive the asymptotic distribution of $\widehat\bmu$ based on (\ref{hat_mu_star_dist}). By Theorem \ref{thm_strong_consistency}, we have (\ref{beta_k_strong_consistency}): $\widehat\bbeta_{k\cU} \rightarrow \bbeta_{k\cU}$ a.s., $k=1,\ldots,K$. By the definitions of of $\widehat{Z}_i$ (\ref{def:tilde_Z_i}), which is determined by $(X_i, Y_i)$ and $\widehat B_{k\cU} = \left\{\widehat\bbeta_{1\cU}, \ldots, \widehat\bbeta_{K\cU}\right\}$, and $\widetilde Z_i$ (\ref{def:tilde_Z_i_star}), which is determined by $(X_i, Y_i)$ and $B_{k\cU} = \left\{\bbeta_{1\cU}, \ldots, \bbeta_{K\cU}\right\}$, we have
\begin{align}
\1\left(\widehat{Z}_i = k\right) - \1\left(\widetilde Z_i = k\right) \overset{P}\longrightarrow 0\,, \, \forall i=1,2,\ldots\,.	
\end{align}
Hence,
\begin{align*}
\left(M_k - M_k^*\right) &= \frac{1}{n} \sum_{i=1}^n \left[\1\left(\widehat{Z}_i = k\right) - \1\left(\widetilde Z_i = k\right)\right] \overset{P}\longrightarrow 0\,,\\
\left(M_{X,k} - M_{X,k}^*\right) &= \frac{1}{n} \sum_{i=1}^n X_i \left[\1\left(\widehat{Z}_i = k\right) - \1\left(\widetilde Z_i = k\right)\right] \overset{P}\longrightarrow 0\,,\\
\left(M_{Y,k} - M_{Y,k}^*\right) &= \frac{1}{n} \sum_{i=1}^n Y_i \left[\1\left(\widehat{Z}_i = k\right) - \1\left(\widetilde Z_i = k\right)\right] \overset{P}\longrightarrow 0\,,\\
\left(M_{XY,k} - M_{XY,k}^*\right) &= \frac{1}{n} \sum_{i=1}^n X_i Y_i \left[\1\left(\widehat{Z}_i = k\right) - \1\left(\widetilde Z_i = k\right)\right] \overset{P}\longrightarrow 0\,,\\
\left(M_{X^2,k} - M_{X^2,k}^*\right) &= \frac{1}{n} \sum_{i=1}^n X_i^2 \left[\1\left(\widehat{Z}_i = k\right) - \1\left(\widetilde Z_i = k\right)\right] \overset{P}\longrightarrow 0\,,\\
\left(M_{Y^2,k} - M_{Y^2,k}^*\right) &= \frac{1}{n} \sum_{i=1}^n Y_i^2 \left[\1\left(\widehat{Z}_i = k\right) - \1\left(\widetilde Z_i = k\right)\right] \overset{P}\longrightarrow 0\,,
\end{align*}
$k=1,\ldots,K$. That is,
\begin{align}\label{hat_mu_minus_hat_mu_star}
\left(\widehat\bmu - \widehat\bmu^*\right) \overset{P}\longrightarrow 0	
\end{align}

Therefore, given (\ref{hat_mu_star_dist}) and (\ref{hat_mu_minus_hat_mu_star}) and by the Slutsky Theorem, we have the asymptotic distribution of $\widehat\bmu$ as
\begin{equation}\label{hat_mu_dist}
\sqrt{n}\left(\widehat{\bmu} - \bmu\right) \tod \cN(0, \bSigma)\,.
\end{equation} 

Then given (\ref{hat_mu_dist}) and by applying Cramer's Theorem \citep{ferguson2017course}, we can derive the joint asymptotic distribution of $\widehat{\btheta}$, which we denote as
\begin{equation}\label{hat_theta_star_dist}
\sqrt{n}\left(\widehat{\btheta} - \btheta\right) \tod \cN(0, \bOmega)\,.	
\end{equation}
where $\bOmega$ has the same form as in (\ref{asym_beta}), except that all the notations related to $Z$ under the specified scenario are now related to $\widetilde Z$ under the current unspecified scenario. 

Finally, given (\ref{hat_theta_star_dist}) and by applying Cramer's Theorem \citep{ferguson2017course} again, we can derive the asymptotic distribution of $R_{\cG\cU}^{2}$:
\begin{equation*}
\sqrt{n} \left( R_{\cG\cU}^{2} - \rho_{\cG\cU}^2 \right) \tod \cN\left(0, \sum_{k=1}^K \left( A_{k\cU} + B_{k\cU}\right) + 2 \mathop{\sum \sum}_{1 \le k < r \le K} C_{kr\cU}\right)\,,	
\end{equation*}
where $A_{k\cU}$, $B_{k\cU}$, and $C_{kr\cU}$ are defined in (\ref{asym_dist_scor2_gu}). The detailed derivation steps follow from the proof of Theorem \ref{thm_asym_dist_scor2_gc}.	
\end{proof}

Applying Theorem \ref{thm_asym_dist_scor2_gu} to the special case where $(X, Y)|\widetilde Z$ follows a bivariate Gaussian distribution, we obtain a much simpler form of the first-order asymptotic distribution of $R_{\cG\cU}^2$.

\subsection{Proof of Corollary \ref{cor:asym_dist_scor2_gu}}
\begin{proof}
Given Theorem \ref{thm_asym_dist_scor2_gu}, the proof of Corollary \ref{cor:asym_dist_scor2_gu} follows from the proof of Corollary \ref{cor:asym_dist_scor2_gu} in \ref{A.3}.	
\end{proof}

\section{Convergence properties of the $K$-lines algorithm}\label{sec:conv_K_lines}

Motivated by \cite{bottou1995convergence}, we study the convergence properties of our proposed $K$-lines clustering algorithm (Algorithm \ref{alg:K-lines}). Below we show that the $K$-lines algorithm can be related to a gradient descent algorithm and an expectation-maximization (EM) style algorithm.

\subsection{Relating $K$-lines to a gradient descent algorithm}

Given a sample $\{(X_i, Y_i)\}_{i=1}^n$, the $K$-lines algorithm computes $K$ sample line centers $\widehat B_{K\cU} = \left\{ \widehat\bbeta_{1\cU}, \ldots, \widehat\bbeta_{K\cU} \right\}$, which minimize the  loss function $W(B_K, P_n)$ defined in (\ref{eq:W_Pn}), the average squared perpendicular distance between each data point and its closest sample line center. Denoting $B_K = \left\{ \bbeta_1, \ldots, \bbeta_K \right\}$, minimizing $W(B_K, P_n)$ is equivalent to minimizing
\begin{align}
L\left( B_K \right) := \sum_{i=1}^n \frac{1}{2} \min_k d_\perp^2 \left( (X_i, Y_i)\tran, \bbeta_k \right) = \sum_{i=1}^n \frac{1}{2} d_\perp^2 \left( (X_i, Y_i)\tran, \bbeta_{s_i(B_K)} \right)\,,	
\end{align}
where $s_i(B_K)$ is the index of the closest line center to the $i$-th data point. Note that
$W(B_K, P_n) = \frac{2}{n}  L\left( B_K \right)$.

We can then derive a gradient descent algorithm based on $L\left( B_K \right)$ for $\bbeta_k = (a_k, b_k, c_k)\tran$:
\begin{align}
\Delta \bbeta_k = \sum_{i=1}^n \left\{
\begin{array}{ll}
	-\epsilon_t 
 \begin{pmatrix}
 	\frac{(a_k X_i + b_k Y_i + c_k)X_i}{a_k^2+b_k^2} - \frac{(a_k X_i + b_k Y_i + c_k)^2 a_k}{(a_k^2+b_k^2)^2} \\
 	\frac{(a_k X_i + b_k Y_i + c_k)Y_i}{a_k^2+b_k^2} - \frac{(a_k X_i + b_k Y_i + c_k)^2 b_k}{(a_k^2+b_k^2)^2} \\
 	\frac{a_k X_i + b_k Y_i + c_k}{a_k^2+b_k^2}
 \end{pmatrix}
  & \text{if } k = s_i(B_K) \\
	(0, 0, 0)\tran & \text{otherwise}
\end{array}
\right.,
\end{align}
where $\epsilon_t$ is the learning rate is to be specified \citep{kohonen2012self}.
	
\subsection{Relating $K$-lines to an EM algorithm} 

Although $K$-lines does not fit in a probabilistic framework, its derivation is similar to that of the EM algorithm, except that the soft-thresholding of EM is changed to hard-thresholding.

The EM algorithm has the principle of introducing additional hidden variables to simplify the optimization problem. Since these hidden variables are unobservable, the maximization step (i.e., the M step) maximizes an auxiliary function calculated in the expectation step (i.e., the E step), which averages  over the possible values of the hidden variables given the parameter estimates from the previous iteration. In our unspecified scenario, the hidden variables are the assignments $s_1(B_K), \ldots, s_n(B_K)$ of the data points to the sample line centers.  Instead of considering the expectation of $L(B_K)$ over the distribution on these hidden variables as in the EM algorithm, the $K$-lines algorithm calculates the values of the hidden variables that maximize the negative loss given the parameter estimates from the previous iteration. That is, the ``$Q$ function" to be maximized in the M step after $(t-1)$ ($t=1, 2, \ldots$) iterations becomes:
\begin{align}
Q\left(B_K, B_K^{(t-1)}\right) := -\sum_{i=1}^n \frac{1}{2} d_\perp^2 \left( (X_i, Y_i)\tran, \bbeta_{s_i\left(B_K^{(t-1)}\right)} \right)\,.
\end{align}
The next step is to find
\begin{align}\label{eq:Q_update}
B_K^{(t)}	:= \argmax_{B_K} Q\left(B_K, B_K^{(t-1)}\right)\,.
\end{align}
The solution $B_K^{(t)} = \left\{ \bbeta_1^{(t)}, \ldots, \bbeta_K^{(t)} \right\}$ consists of $\bbeta_k^{(t)}$ calculated by (\ref{eq:hat_beta_ku}), where $\cC_k^{(k-1)}$ represents the index set of data points that are closer to $\bbeta_k^{(t-1)}$ than other sample line centers in perpendicular distance. The $K$-lines algorithm iterates by replacing $B_K^{(t-1)}$ by $B_K^{(t)}$ using the update equation (\ref{eq:Q_update}) until convergence. Since $s_i\left(B_K^{(t-1)}\right)$ is by definition the best assignment of the $i$-th data point to the closet line center in $B_K^{(t-1)}$, we have the following inequality:
\begin{align*}
-L\left(B_K^{(t)}\right) - Q\left(B_K^{(t)}, B_K^{(t-1)}\right)	= \frac{1}{2} \sum_{i=1}^n d_\perp^2 \left( (X_i, Y_i)\tran, \bbeta_{s_i\left(B_K^{(t-1)}\right)}^{(t)} \right) - d_\perp^2 \left( (X_i, Y_i)\tran, \bbeta_{s_i\left(B_K^{(t)}\right)}^{(t)} \right) \ge 0\,.
\end{align*}
Using this result, the identity $-L\left(B_K^{(t-1)}\right) = Q\left(B_K^{(t-1)}, B_K^{(t-1)}\right)$, and the definition of  $B_K^{(t)}$ in (\ref{eq:Q_update}), we have the following inequality:
\begin{align}
L\left(B_K^{(t)}\right) - L\left(B_K^{(t-1)}\right) &= L\left(B_K^{(t)}\right) + Q\left(B_K^{(t)}, B_K^{(t-1)}\right) - Q\left(B_K^{(t)}, B_K^{(t-1)}\right) - L\left(B_K^{(t-1)}\right)\\
&\le - Q\left(B_K^{(t)}, B_K^{(t-1)}\right) + Q\left(B_K^{(t-1)}, B_K^{(t-1)}\right) \overset{(\ref{eq:Q_update})}\le 0 \notag \,.
\end{align}
Hence, each iteration decreases the loss function until $B_K^{(t)} = B_K^{(t-1)}$, denoted by $B_K^*$. Since the assignments $s_1(B_K), \ldots, s_n(B_K)$ are discrete, there is an open neighborhood of $B_K^*$ in which the assignments are constant. Based on their definitions, the functions $-L(\cdot)$ and $Q(\cdot, B_K^*)$ are equal in this neighborhood. Therefore, $B_K^*$, the maximum of the function $Q(\cdot, B_K^*)$, is also a local minimum  of the loss function $L(\cdot)$.

We can further show that the M step of an EM algorithm under a bivariate Gaussian mixture model is the same as the M step of the $K$-lines algorithm in (\ref{eq:Q_update}). We assume that $(X_i, Y_i) \overset{\text{i.i.d.}}\sim \sum_{k=1}^K p_k \,\cN\left( \bmu_k, \bSigma_k \right)$ with mean vectors $\bmu_k$ and covariance matrices $\bSigma_k$, $k=1,\ldots,K$, and we denote $\Theta = \left\{ p_k, \bmu_k, \bSigma_k \right\}_{k=1}^K$. The hidden variables are $Z_i \overset{\text{i.i.d.}}\sim \text{Multinomial}\left(K, (p_1,\ldots,p_K)\tran\right)$. Suppose that in the $t$-th iteration of the EM algorithm, the E step computes $\delta_{ik}^{(t)} := \Prob\left(Z_i = k ~|~ (X_i, Y_i), \Theta^{(t-1)}\right)$. The next M step would then update the value of $\bSigma_k$ as
\begin{align}\label{eq:bSigma_k_t}
\bSigma_k^{(t)} = \left[
	\begin{array}{cc}
		\frac{\sum_{i=1}^n \delta_{ik}^{(t)} \left(X_i - \bar X_k^{(t)}\right)^2}{\sum_{i=1}^n \delta_{ik}^{(t)}} & \frac{\sum_{i=1}^n \delta_{ik}^{(t)} \left(X_i - \bar X_k^{(t)}\right) \left(Y_i - \bar Y_k^{(t)}\right)}{\sum_{i=1}^n \delta_{ik}^{(t)}} \\
		\frac{\sum_{i=1}^n \delta_{ik}^{(t)} \left(X_i - \bar X_k^{(t)}\right) \left(Y_i - \bar Y_k^{(t)}\right)}{\sum_{i=1}^n \delta_{ik}^{(t)}} & \frac{\sum_{i=1}^n \delta_{ik}^{(t)} \left(Y_i - \bar Y_k^{(t)}\right)^2}{\sum_{i=1}^n \delta_{ik}^{(t)}}
	\end{array}
\right]\,,	
\end{align}
where $\bar X_k^{(t)} = \frac{\sum_{i=1}^n \delta_{ik}^{(t)} X_i}{\sum_{i=1}^n \delta_{ik}^{(t)}}$ and $\bar Y_k^{(t)} = \frac{\sum_{i=1}^n \delta_{ik}^{(t)} Y_i}{\sum_{i=1}^n \delta_{ik}^{(t)}}$. 

If we modify the EM algorithm by replacing the E step in every iteration by that of the $K$-lines algorithm, i.e., define 
\begin{align}
\delta_{ik}^{(t)} := \1\left\{ d_\perp \left( (X_i, Y_i)\tran, \bbeta_k^{(t-1)} \right) < d_\perp \left( (X_i, Y_i)\tran, \bbeta_r^{(t-1)} \right), \forall r \neq k \right\} 
\end{align}
in a hard-thresholding way (assuming that every data point is closest to only one sample line center in perpendicular distance), the M step still updates $\bSigma_k^{(t)}$ as in (\ref{eq:bSigma_k_t}).
Then by (\ref{eq:hat_beta_ku}), $\bbeta_k^{(t)} = \left(\widehat u_{12,k}^{(t)},\: -\widehat u_{11,k}^{(t)},\: -\widehat u_{12,k}^{(t)}\bar X_{k}^{(t)} + \widehat u_{11,k}^{(t)} \bar Y_{k}^{(t)} \right)\tran$, where $\left(\widehat u_{11,k}^{(t)}, \widehat u_{12,k}^{(t)}\right)\tran$ is the eigenvector corresponding to the largest eigenvalue of $\bSigma_k^{(t)}$,  minimizes the function $\sum_{i=1}^n \delta_{ik}^{(t)} d_\perp^2 \left( (X_i, Y_i)\tran, * \right)$. Hence, $B_K^{(t)} = \left\{ \bbeta_1^{(t)}, \ldots, \bbeta_K^{(t)} \right\}$ is the same solution as in (\ref{eq:Q_update}).

Xu and Jordan showed in their 1994 classic paper that the EM algorithm approximate the Newton optimization algorithm in the case of a Gaussian mixture model \citep{xu1996convergence}. In other words, the EM algorithm has fast convergence in this special case. Given the relevance of our $K$-lines algorithm to the bivariate Gaussian EM algorithm, with the M step stays the same, we can say that our $K$-lines algorithm is also approximately the Newton algorithm.

\section{More simulation results}\label{sec:more simulation}


Fig. \ref{fig:theory_vs_sim_aic} shows the comparison of finite sample distributions of $R^2_{\cG\cU}$, when $K$ is chosen by the AIC in \eqref{eq:AIC}, and the asymptotic distributions. The results show that the agreement between finite-sample distributions and asymptotic distributions deteriorates when $K$ is chosen by the AIC instead of being set to the true value in the data generative model. This is expected, as more uncertainty is introduced into the procedure when $K$ is unknown. However, we observe from Fig. \ref{fig:theory_vs_sim_aic} that, when $n=100$, the agreement between finite-sample distributions and asymptotic distributions is still reasonably good. This observation further justifies the use of our asymptotic results in practice.

In our construction of confidence intervals of $\rho^2_{\cG\cS}$ and $\rho^2_{\cG\cU}$, we construct the standard errors $\text{se}(R^2_{\cG\cS})$ and $\text{se}(R^2_{\cG\cU})$ in two ways: square roots of (1) the plug-in estimates of the asymptotic variances of $R^2_{\cG\cS}$ and $R^2_{\cG\cU}$, or (2) the bootstrap estimates of $\var(R^2_{\cG\cS})$ and $\var(R^2_{\cG\cU})$. In the first four settings as mixtures of bivariate Gaussians, we use the asymptotic variances from Corollaries \ref{cor:asym_dist_scor2_gc} and \ref{cor:asym_dist_scor2_gu}, and we perform the parametric bootstrap to obtain the bootstrap estimates. In the latter four settings as mixtures of bivariate $t$ distributions, we compare the non-parametric bootstrap approach with two plug-in options: the first is to plug in the asymptotic variances of the special bivariate Gaussian forms in Corollaries \ref{cor:asym_dist_scor2_gc} and \ref{cor:asym_dist_scor2_gu}, and the second is to plug in the asymptotic variances of the general forms in Theorems \ref{thm_asym_dist_scor2_gc} and \ref{thm_asym_dist_scor2_gu}. Fig. \ref{fig:theory_vs_plugin_vs_boot} shows that the plug-in and bootstrap approaches construct similar CIs on the same sample. When $n$ increases from $50$ to $100$, the CIs constructed by both approaches agree better with the theoretical CIs based on the true asymptotic variances, as expected. In addition, under Settings $5$--$8$, the two plug-in options result in similar CIs, suggesting that the simpler plug-in approach based on Corollaries \ref{cor:asym_dist_scor2_gc} and \ref{cor:asym_dist_scor2_gu} is robust.

\begin{figure}[htbp]
\centering
\includegraphics[height=1\textheight]{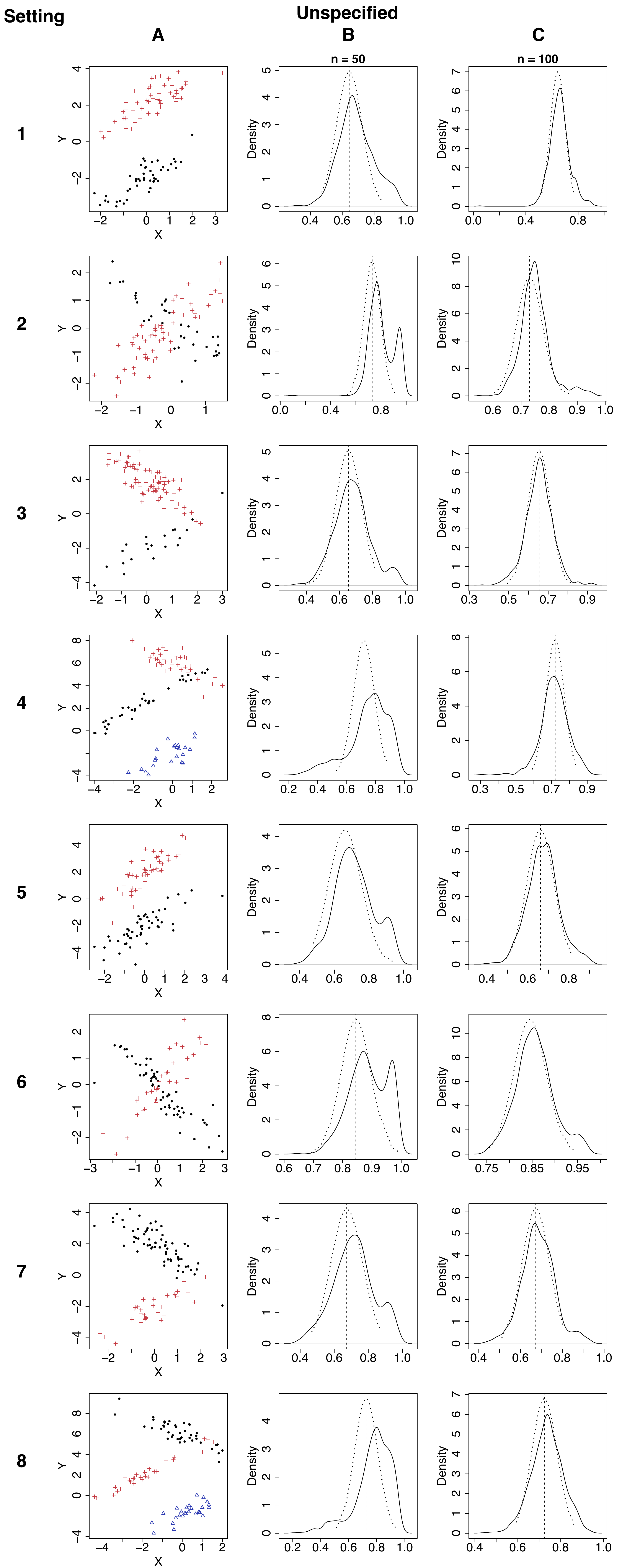}
\caption{(Continued on the following page.) \label{fig:theory_vs_sim_aic}}
\end{figure}
\begin{figure}[htbp]
  \contcaption{Comparison of the asymptotic distributions (theoretical results, same as those in Figure \ref{fig:theory_vs_sim}) and the finite-sample distributions of the sample-level unspecified generalized Pearson correlation square $R^2_{\cG\cU}$, when $K$ is chosen by the AIC (Equation (\ref{eq:AIC})) in a data-driven way, under the eight simulation settings in Table \ref{tab1:simu_settings}. A: For each setting, a scatterplot shows a sample with $n=100$ under the unspecified scenario; different colors and symbols represent different values of $\widetilde Z$ inferred by the $K$-lines algorithm (Algorithm \ref{alg:K-lines}). B-C: Finite-sample distributions of $n=50$ or $100$ (black solid curves) vs. the asymptotic distribution (black dotted curves) of $R^2_{\cG\cU}$ in Corollary \ref{cor:asym_dist_scor2_gu} (Settings $1$--$4$) or Theorem \ref{thm_asym_dist_scor2_gu} (Settings $5$--$8$); the vertical dashed lines mark the values of $\rho^2_{\cG\cU}$.}
\end{figure}

\begin{figure}[htbp]
	\includegraphics[width=1\textwidth]{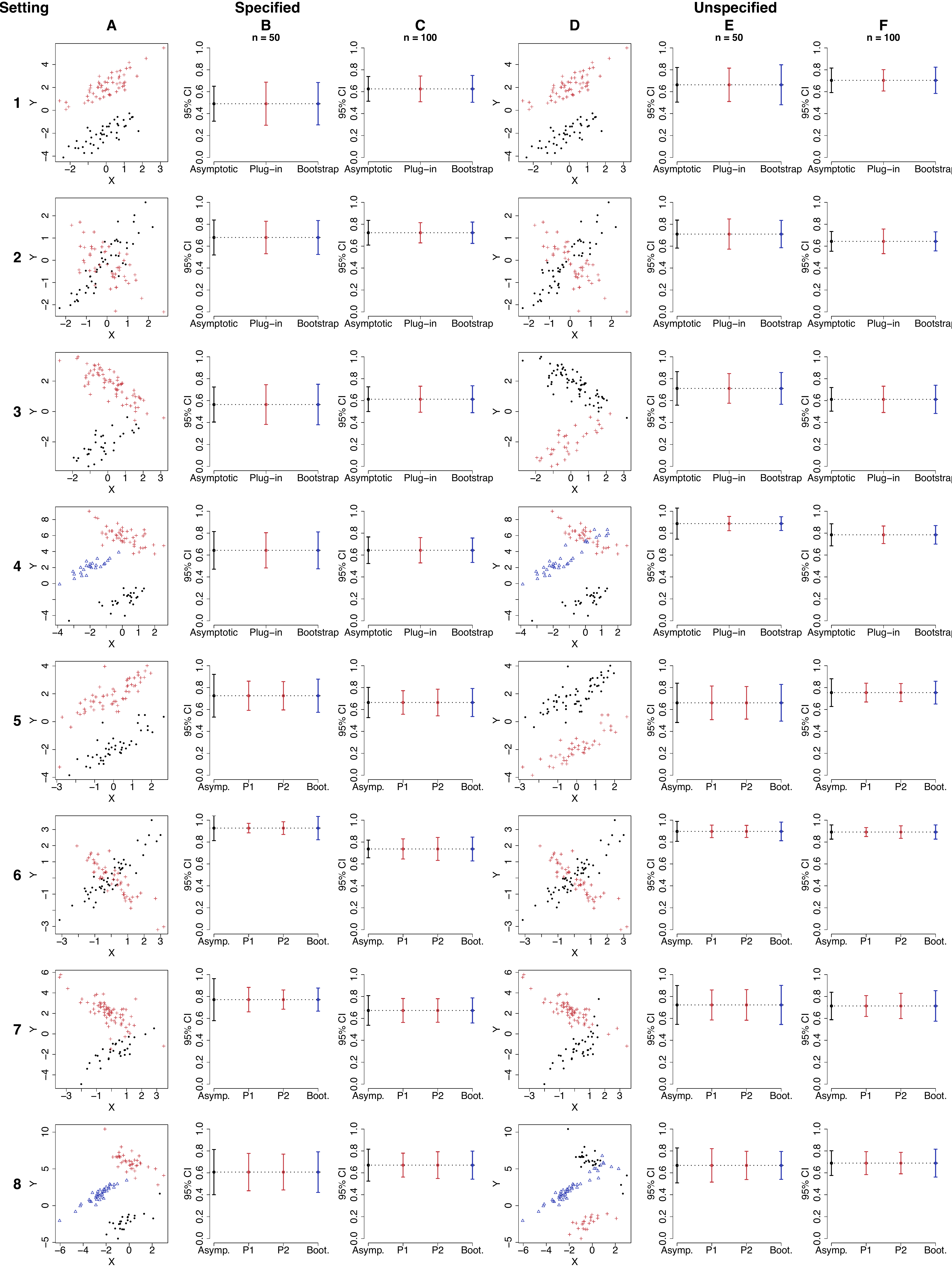}
	\caption{(Continued on the following page.)}\label{fig:theory_vs_plugin_vs_boot}
\end{figure}
\begin{figure}[htbp]
  \contcaption{Comparison of the $95\%$ confidence intervals of $\rho^2_{\cG\cS}$ and $\rho^2_{\cG\cU}$ under the eight simulation settings (Table \ref{tab1:simu_settings}). A: For each setting, a scatterplot shows a sample with $n=100$ under the specified scenario; different colors and symbols represent different values of $Z$. B-C: Confidence intervals of $\rho^2_{\cG\cS}$ based one sample ($n=50$ or $100$) using one of the three distributions: (1) black: the asymptotic distribution using the true asymptotic variance of $R^2_{\cG\cS}$ in Corollary \ref{cor:asym_dist_scor2_gc} (Settings $1$--$4$) or Theorem \ref{thm_asym_dist_scor2_gc} (Settings $5$--$8$), (2) red: the asymptotic distribution using the plug-in asymptotic variance  of $R^2_{\cG\cS}$ in Corollary \ref{cor:asym_dist_scor2_gc} (Settings $1$--$4$ and ``P1" under Settings $5$--$8$) or Theorem \ref{thm_asym_dist_scor2_gc} (``P2" under Settings $5$--$8$), (3) blue: the parametric (Settings $1$--$4$) or non-parametric (Settings $5$--$8$) bootstrap based on $1000$ simulations. D: For each setting, a scatterplot shows a sample with $n=100$ under the unspecified scenario; different colors and symbols represent different values of $\widetilde Z$ inferred by the $K$-lines algorithm (Algorithm \ref{alg:K-lines}). E-F: Confidence intervals of $\rho^2_{\cG\cU}$ based one sample ($n=50$ or $100$) using one of the three distributions: (1) black: the asymptotic distribution using the true asymptotic variance of $R^2_{\cG\cU}$ in Corollary \ref{cor:asym_dist_scor2_gu} (Settings $1$--$4$) or Theorem \ref{thm_asym_dist_scor2_gu} (Settings $5$--$8$), (2) red: the asymptotic distribution using the plug-in asymptotic variance  of $R^2_{\cG\cU}$ in Corollary \ref{cor:asym_dist_scor2_gu} (Settings $1$--$4$ and ``P1" under Settings $5$--$8$) or Theorem \ref{thm_asym_dist_scor2_gu} (``P2" under Settings $5$--$8$), (3) blue: the parametric (Settings $1$--$4$) or non-parametric (Settings $5$--$8$) bootstrap based on $1000$ simulations.}
\end{figure}

\begin{figure}[htbp]
\includegraphics[width=1.1\textwidth]{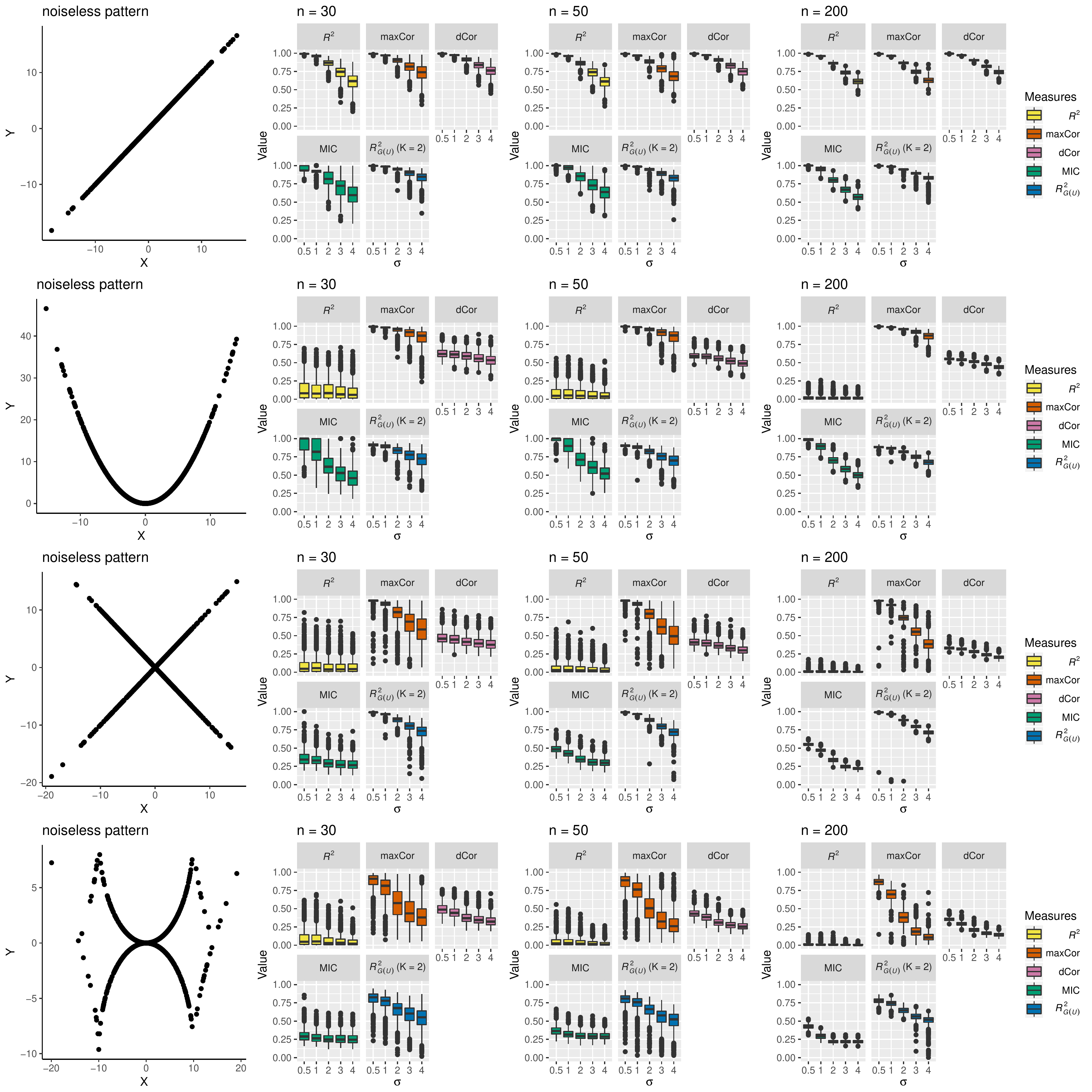}
\caption{Values of the five measures: squared Pearson correlation ($R^2$), maximal correlation (maxCor), distance correlation (dCor), maximal information coefficient (MIC), and $R^2_{\cG\cU}$ with $K=2$, in $1000$ simulations at each sample size $n$ and noise standard deviation $\sigma$. \label{fig:power_values}}
\end{figure}


\section{\textit{Arabidopsis thaliana} gene expression data}\label{sec:arab_data}
In the application of our generalized Pearson correlation squares to \textit{Arabidopsis thaliana} gene expression analysis, we used four public microarray datasets from the AtGenExpress Consortium \[\texttt{www.arabidopsis.org/portals/expression/microarray/ATGenExpress.jsp},\] summarized in Table~\ref{tabS1:at_data}.

\begin{table}[htp]
\centering
\begin{tabular}{cccc}
	\hline
	\bf Dataset & \bf Submission Number & \bf Number of Genes & \bf Sample Size\\
	\hline
	Oxidation & ME00340 & $22,657$ & $52$\\
	Wounding & ME00330 & $22,657$ & $60$ \\
	UV-B light & ME00329 & $22,657$ & $60$ \\
	Drought & ME00338 & $22,657$ & $60$\\
	\hline
\end{tabular}
\caption{Four \textit{Arabidopsis thaliana} microarray gene expression datasets from the AtGenExpress Consortium. \label{tabS1:at_data}}	
\end{table}

\begin{figure}[htbp]
\includegraphics[page=2, width=\textwidth]{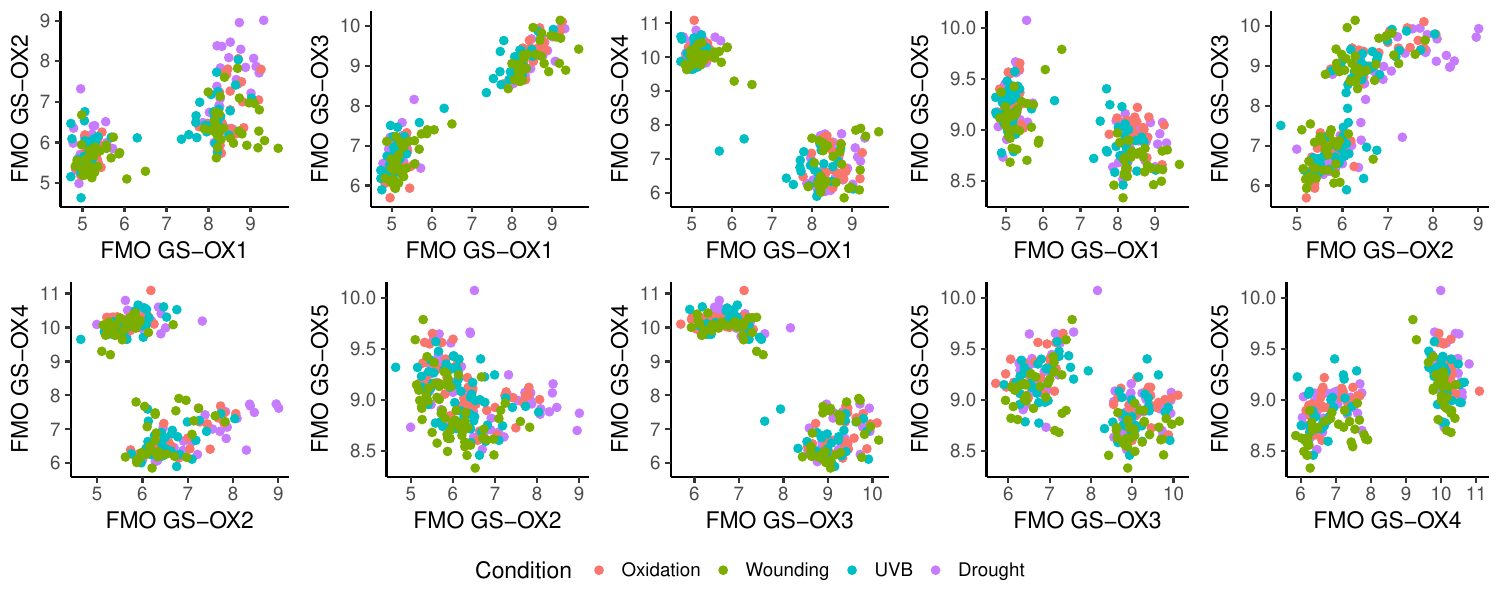}
\caption{Expression levels of the five Arabidopsis thaliana genes in Figure \ref{fig:motiv}A, with colors marked by the \texttt{condition} variable. \label{fig:motiv2}}	
\end{figure}

\begin{figure}[htbp]
\includegraphics[page=3, width=\textwidth]{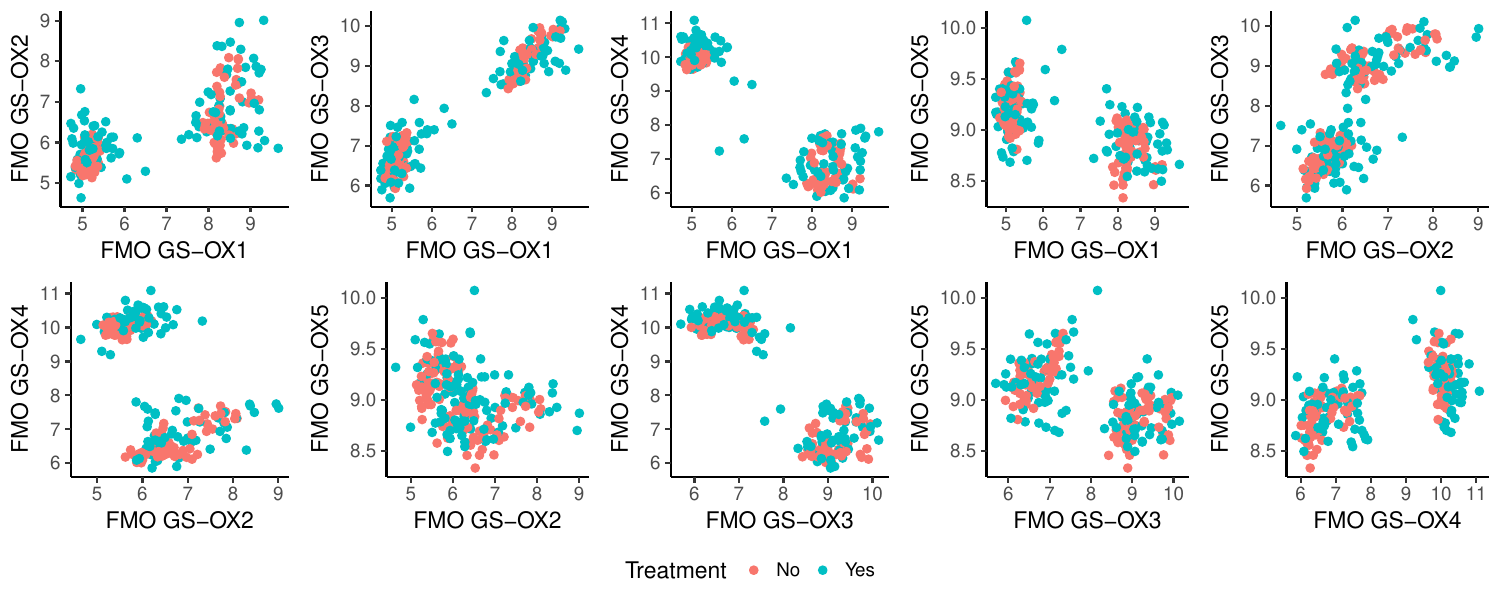}
\caption{Expression levels of the five Arabidopsis thaliana genes in Figure \ref{fig:motiv}A, with colors marked by the \texttt{treatment} variable. \label{fig:motiv3}}	
\end{figure}

\begin{figure}[htbp]
\includegraphics[page=4, width=\textwidth]{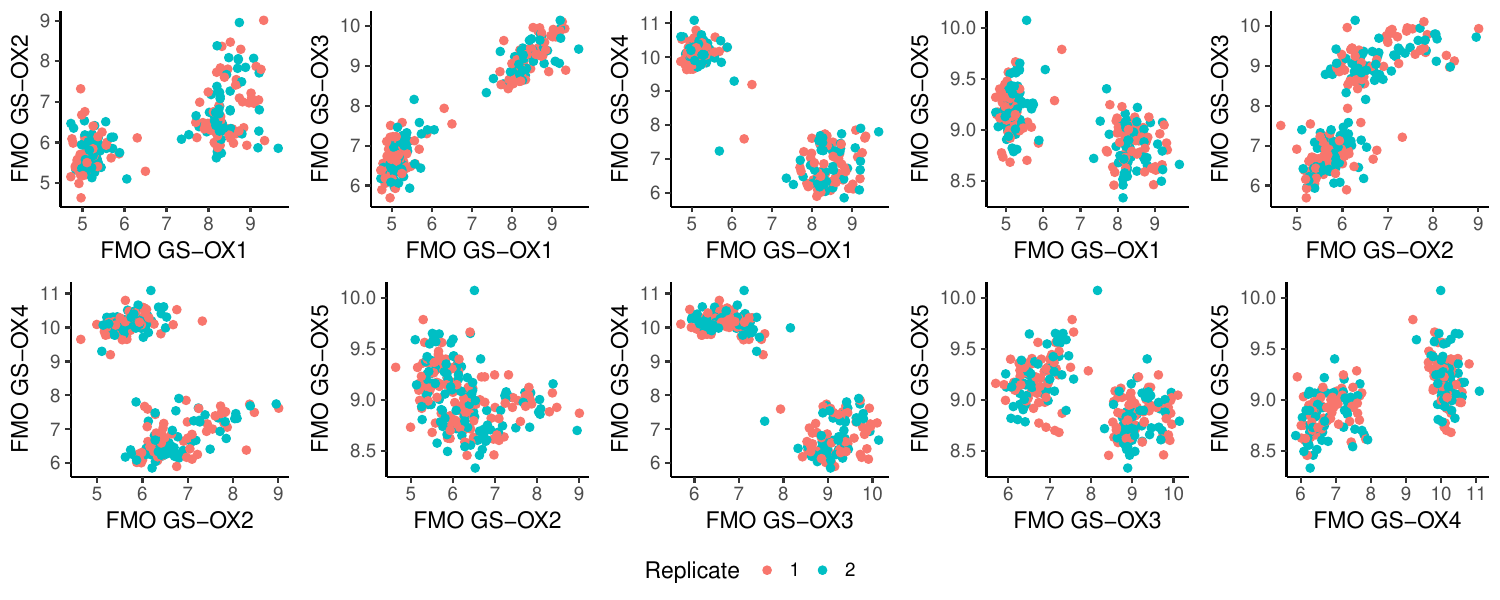}
\caption{Expression levels of the five Arabidopsis thaliana genes in Figure \ref{fig:motiv}A, with colors marked by the \texttt{replicate} variable. \label{fig:motiv4}}	
\end{figure}

Figures \ref{fig:motiv2}-\ref{fig:motiv4} shows pairwise scatterplots of the five FMO genes in Fig. \ref{fig:motiv}A colored by each of the four index variables: \texttt{condition} (oxidation, wounding, UV-B light, and drought), \texttt{treatment} (yes and no), \texttt{replicate} ($1$ and $2$), and \texttt{tissue} (root and shoot). It is clear that only the \texttt{tissue} variable serves as a good index variable for linear dependences. 

The processed data and the analysis codes are provided in the ``Supplementary Data and Codes" file available at \texttt{https://tinyurl.com/y4ypx6v2}

%

\clearpage
\bibliographystyle{chicago}
\bibliography{gR2_main.bib}

\begin{thebibliography}{}

\bibitem[\protect\citeauthoryear{Akaike}{Akaike}{1998}]{akaike1998information}
Akaike, H. (1998).
\newblock Information theory and an extension of the maximum likelihood
  principle.
\newblock In {\em Selected papers of hirotugu akaike}, pp.\  199--213.
  Springer.

\bibitem[\protect\citeauthoryear{Benaglia, Chauveau, Hunter, and
  Young}{Benaglia et~al.}{2009}]{benaglia2009mixtools}
Benaglia, T., D.~Chauveau, D.~Hunter, and D.~Young (2009).
\newblock mixtools: An r package for analyzing finite mixture models.
\newblock {\em Journal of Statistical Software\/}~{\em 32\/}(6), 1--29.

\bibitem[\protect\citeauthoryear{Bjerve and Doksum}{Bjerve and
  Doksum}{1993}]{bjerve1993correlation}
Bjerve, S. and K.~Doksum (1993).
\newblock Correlation curves: measures of association as functions of covariate
  values.
\newblock {\em The Annals of Statistics\/}, 890--902.

\bibitem[\protect\citeauthoryear{Blyth}{Blyth}{1972}]{blyth1972simpson}
Blyth, C.~R. (1972).
\newblock On simpson's paradox and the sure-thing principle.
\newblock {\em Journal of the American Statistical Association\/}~{\em
  67\/}(338), 364--366.

\bibitem[\protect\citeauthoryear{Bottou and Bengio}{Bottou and
  Bengio}{1995}]{bottou1995convergence}
Bottou, L. and Y.~Bengio (1995).
\newblock Convergence properties of the k-means algorithms.
\newblock In {\em Advances in neural information processing systems}, pp.\
  585--592.

\bibitem[\protect\citeauthoryear{Breiman and Friedman}{Breiman and
  Friedman}{1985}]{breiman1985estimating}
Breiman, L. and J.~H. Friedman (1985).
\newblock Estimating optimal transformations for multiple regression and
  correlation.
\newblock {\em Journal of the American statistical Association\/}~{\em
  80\/}(391), 580--598.

\bibitem[\protect\citeauthoryear{Buettner, Natarajan, Casale, Proserpio,
  Scialdone, Theis, Teichmann, Marioni, and Stegle}{Buettner
  et~al.}{2015}]{buettner2015computational}
Buettner, F., K.~N. Natarajan, F.~P. Casale, V.~Proserpio, A.~Scialdone, F.~J.
  Theis, S.~A. Teichmann, J.~C. Marioni, and O.~Stegle (2015).
\newblock Computational analysis of cell-to-cell heterogeneity in single-cell
  rna-sequencing data reveals hidden subpopulations of cells.
\newblock {\em Nature biotechnology\/}~{\em 33\/}(2), 155.

\bibitem[\protect\citeauthoryear{Cover and Thomas}{Cover and
  Thomas}{2012}]{cover2012elements}
Cover, T.~M. and J.~A. Thomas (2012).
\newblock {\em Elements of information theory}.
\newblock John Wiley \& Sons.

\bibitem[\protect\citeauthoryear{De~Veaux}{De~Veaux}{1989}]{de1989mixtures}
De~Veaux, R.~D. (1989).
\newblock Mixtures of linear regressions.
\newblock {\em Computational Statistics \& Data Analysis\/}~{\em 8\/}(3),
  227--245.

\bibitem[\protect\citeauthoryear{Delicado and Smrekar}{Delicado and
  Smrekar}{2009}]{delicado2009measuring}
Delicado, P. and M.~Smrekar (2009).
\newblock Measuring non-linear dependence for two random variables distributed
  along a curve.
\newblock {\em Statistics and Computing\/}~{\em 19\/}(3), 255.

\bibitem[\protect\citeauthoryear{Ferguson}{Ferguson}{1996}]{ferguson2017course}
Ferguson, T.~S. (1996).
\newblock {\em A course in large sample theory}.
\newblock Chapman \& Hall.

\bibitem[\protect\citeauthoryear{Gebelein}{Gebelein}{1941}]{gebelein1941statistische}
Gebelein, H. (1941).
\newblock Das statistische problem der korrelation als variations-und
  eigenwertproblem und sein zusammenhang mit der ausgleichsrechnung.
\newblock {\em ZAMM-Journal of Applied Mathematics and Mechanics/Zeitschrift
  f{\"u}r Angewandte Mathematik und Mechanik\/}~{\em 21\/}(6), 364--379.

\bibitem[\protect\citeauthoryear{Gretton, Bousquet, Smola, and
  Sch{\"o}lkopf}{Gretton et~al.}{2005}]{gretton2005measuring}
Gretton, A., O.~Bousquet, A.~Smola, and B.~Sch{\"o}lkopf (2005).
\newblock Measuring statistical dependence with hilbert-schmidt norms.
\newblock In {\em International conference on algorithmic learning theory},
  pp.\  63--77. Springer.

\bibitem[\protect\citeauthoryear{Haiman, Patterson, Freedman, Myers, Pike,
  Waliszewska, Neubauer, Tandon, Schirmer, McDonald, et~al.}{Haiman
  et~al.}{2007}]{haiman2007multiple}
Haiman, C.~A., N.~Patterson, M.~L. Freedman, S.~R. Myers, M.~C. Pike,
  A.~Waliszewska, J.~Neubauer, A.~Tandon, C.~Schirmer, G.~J. McDonald, et~al.
  (2007).
\newblock Multiple regions within 8q24 independently affect risk for prostate
  cancer.
\newblock {\em Nature genetics\/}~{\em 39\/}(5), 638.

\bibitem[\protect\citeauthoryear{Hawkins, Allen, and Stromberg}{Hawkins
  et~al.}{2001}]{hawkins2001determining}
Hawkins, D.~S., D.~M. Allen, and A.~J. Stromberg (2001).
\newblock Determining the number of components in mixtures of linear models.
\newblock {\em Computational Statistics \& Data Analysis\/}~{\em 38\/}(1),
  15--48.

\bibitem[\protect\citeauthoryear{Heller, Heller, and Gorfine}{Heller
  et~al.}{2012}]{heller2012consistent}
Heller, R., Y.~Heller, and M.~Gorfine (2012).
\newblock A consistent multivariate test of association based on ranks of
  distances.
\newblock {\em Biometrika\/}~{\em 100\/}(2), 503--510.

\bibitem[\protect\citeauthoryear{Hirschfeld}{Hirschfeld}{1935}]{hirschfeld1935connection}
Hirschfeld, H.~O. (1935).
\newblock A connection between correlation and contingency.
\newblock In {\em Mathematical Proceedings of the Cambridge Philosophical
  Society}, Volume~31, pp.\  520--524. Cambridge University Press.

\bibitem[\protect\citeauthoryear{Hoeffding}{Hoeffding}{1948}]{hoeffding1948non}
Hoeffding, W. (1948).
\newblock A non-parametric test of independence.
\newblock {\em The annals of mathematical statistics\/}, 546--557.

\bibitem[\protect\citeauthoryear{Hurn, Justel, and Robert}{Hurn
  et~al.}{2003}]{hurn2003estimating}
Hurn, M., A.~Justel, and C.~P. Robert (2003).
\newblock Estimating mixtures of regressions.
\newblock {\em Journal of computational and graphical statistics\/}~{\em
  12\/}(1), 55--79.

\bibitem[\protect\citeauthoryear{Jacobs, Jordan, Nowlan, and Hinton}{Jacobs
  et~al.}{1991}]{jacobs1991adaptive}
Jacobs, R.~A., M.~I. Jordan, S.~J. Nowlan, and G.~E. Hinton (1991).
\newblock Adaptive mixtures of local experts.
\newblock {\em Neural computation\/}~{\em 3\/}(1), 79--87.

\bibitem[\protect\citeauthoryear{Jolliffe}{Jolliffe}{1982}]{jolliffe1982note}
Jolliffe, I.~T. (1982).
\newblock A note on the use of principal components in regression.
\newblock {\em Applied Statistics\/}, 300--303.

\bibitem[\protect\citeauthoryear{Jones and McLachlan}{Jones and
  McLachlan}{1992}]{jones1992fitting}
Jones, P. and G.~McLachlan (1992).
\newblock Fitting finite mixture models in a regression context.
\newblock {\em Australian \& New Zealand Journal of Statistics\/}~{\em
  34\/}(2), 233--240.

\bibitem[\protect\citeauthoryear{Kendall}{Kendall}{1938}]{kendall1938new}
Kendall, M.~G. (1938).
\newblock A new measure of rank correlation.
\newblock {\em Biometrika\/}~{\em 30\/}(1/2), 81--93.

\bibitem[\protect\citeauthoryear{Kim, Jiang, Teng, Feldman, and Huang}{Kim
  et~al.}{2012}]{kim2012using}
Kim, K., K.~Jiang, S.~L. Teng, L.~J. Feldman, and H.~Huang (2012).
\newblock Using biologically interrelated experiments to identify pathway genes
  in arabidopsis.
\newblock {\em Bioinformatics\/}~{\em 28\/}(6), 815--822.

\bibitem[\protect\citeauthoryear{Kohonen}{Kohonen}{1989}]{kohonen2012self}
Kohonen, T. (1989).
\newblock {\em Self-organization and associative memory}, Volume 3rd edition.
\newblock Springer-Verlag, Berlin.

\bibitem[\protect\citeauthoryear{Kraskov, St{\"o}gbauer, and
  Grassberger}{Kraskov et~al.}{2004}]{kraskov2004estimating}
Kraskov, A., H.~St{\"o}gbauer, and P.~Grassberger (2004).
\newblock Estimating mutual information.
\newblock {\em Physical review E\/}~{\em 69\/}(6), 066138.

\bibitem[\protect\citeauthoryear{Leisch}{Leisch}{2008}]{leisch2008modelling}
Leisch, F. (2008).
\newblock Modelling background noise in finite mixtures of generalized linear
  regression models.
\newblock In {\em COMPSTAT 2008}, pp.\  385--396. Springer.

\bibitem[\protect\citeauthoryear{Li, Hansen, Ober, Kliebenstein, and
  Halkier}{Li et~al.}{2008}]{li2008subclade}
Li, J., B.~G. Hansen, J.~A. Ober, D.~J. Kliebenstein, and B.~A. Halkier (2008).
\newblock Subclade of flavin-monooxygenases involved in aliphatic glucosinolate
  biosynthesis.
\newblock {\em Plant Physiology\/}~{\em 148\/}(3), 1721--1733.

\bibitem[\protect\citeauthoryear{Li}{Li}{2002}]{li2002genome}
Li, K.-C. (2002).
\newblock Genome-wide coexpression dynamics: theory and application.
\newblock {\em Proceedings of the National Academy of Sciences\/}~{\em
  99\/}(26), 16875--16880.

\bibitem[\protect\citeauthoryear{Lloyd}{Lloyd}{1982}]{lloyd1982least}
Lloyd, S. (1982).
\newblock Least squares quantization in pcm.
\newblock {\em IEEE transactions on information theory\/}~{\em 28\/}(2),
  129--137.

\bibitem[\protect\citeauthoryear{Morrison}{Morrison}{1967}]{morrison1967multivariate}
Morrison, D.~F. (1967).
\newblock {\em Multivariate statistical methods}.
\newblock McGraw Hill, New York.

\bibitem[\protect\citeauthoryear{Mukai, Yabuta, Yoshida, Okamoto, Miura,
  Furuta, Abe, and Nojima}{Mukai et~al.}{2015}]{mukai2015lats1}
Mukai, S., N.~Yabuta, K.~Yoshida, A.~Okamoto, D.~Miura, Y.~Furuta, T.~Abe, and
  H.~Nojima (2015).
\newblock Lats1 suppresses centrosome overduplication by modulating the
  stability of cdc25b.
\newblock {\em Scientific reports\/}~{\em 5}, 16173.

\bibitem[\protect\citeauthoryear{Murtaph and Raftery}{Murtaph and
  Raftery}{1984}]{Murtaph.Raftery.1984}
Murtaph, F. and A.~Raftery (1984).
\newblock Fitting straight lines to point patterns.
\newblock {\em Pattern recognition\/}~{\em 17}, 479--483.

\bibitem[\protect\citeauthoryear{Pearson, Lee, and Bramley-Moore}{Pearson
  et~al.}{1899}]{pearson1899mathematical}
Pearson, K., A.~Lee, and L.~Bramley-Moore (1899).
\newblock Mathematical contributions to the theory of evolution. vi. genetic
  (reproductive) selection: Inheritance of fertility in man, and of fecundity
  in thoroughbred racehorses.
\newblock {\em Philosophical Transactions of the Royal Society of London.
  Series A, Containing Papers of a Mathematical or Physical Character\/}~{\em
  192}, 257--330.

\bibitem[\protect\citeauthoryear{Pollard}{Pollard}{1981}]{pollard1981strong}
Pollard, D. (1981).
\newblock Strong consistency of $ k $-means clustering.
\newblock {\em The Annals of Statistics\/}~{\em 9\/}(1), 135--140.

\bibitem[\protect\citeauthoryear{Quandt and Ramsey}{Quandt and
  Ramsey}{1978}]{quandt1978estimating}
Quandt, R.~E. and J.~B. Ramsey (1978).
\newblock Estimating mixtures of normal distributions and switching
  regressions.
\newblock {\em Journal of the American statistical Association\/}~{\em
  73\/}(364), 730--738.

\bibitem[\protect\citeauthoryear{R{\'e}nyi}{R{\'e}nyi}{1959}]{renyi1959measures}
R{\'e}nyi, A. (1959).
\newblock On measures of dependence.
\newblock {\em Acta mathematica hungarica\/}~{\em 10\/}(3-4), 441--451.

\bibitem[\protect\citeauthoryear{Reshef, Reshef, Finucane, Grossman, McVean,
  Turnbaugh, Lander, Mitzenmacher, and Sabeti}{Reshef
  et~al.}{2011}]{reshef2011detecting}
Reshef, D.~N., Y.~A. Reshef, H.~K. Finucane, S.~R. Grossman, G.~McVean, P.~J.
  Turnbaugh, E.~S. Lander, M.~Mitzenmacher, and P.~C. Sabeti (2011).
\newblock Detecting novel associations in large data sets.
\newblock {\em science\/}~{\em 334\/}(6062), 1518--1524.

\bibitem[\protect\citeauthoryear{Scharl, Gr{\"u}n, and Leisch}{Scharl
  et~al.}{2009}]{scharl2009mixtures}
Scharl, T., B.~Gr{\"u}n, and F.~Leisch (2009).
\newblock Mixtures of regression models for time course gene expression data:
  evaluation of initialization and random effects.
\newblock {\em Bioinformatics\/}~{\em 26\/}(3), 370--377.

\bibitem[\protect\citeauthoryear{Shannon, Weaver, and Burks}{Shannon
  et~al.}{1951}]{shannon1951mathematical}
Shannon, C.~E., W.~Weaver, and A.~W. Burks (1951).
\newblock The mathematical theory of communication.

\bibitem[\protect\citeauthoryear{Simon and Tibshirani}{Simon and
  Tibshirani}{2014}]{simon2014comment}
Simon, N. and R.~Tibshirani (2014).
\newblock Comment on ``detecting novel associations in large data sets" by
  reshef et al, science dec 16, 2011.
\newblock {\em arXiv preprint arXiv:1401.7645\/}.

\bibitem[\protect\citeauthoryear{Simpson}{Simpson}{1951}]{simpson1951interpretation}
Simpson, E.~H. (1951).
\newblock The interpretation of interaction in contingency tables.
\newblock {\em Journal of the Royal Statistical Society. Series B
  (Methodological)\/}, 238--241.

\bibitem[\protect\citeauthoryear{Smith}{Smith}{2009}]{smith2009use}
Smith, R.~J. (2009).
\newblock Use and misuse of the reduced major axis for line-fitting.
\newblock {\em American Journal of Physical Anthropology: The Official
  Publication of the American Association of Physical Anthropologists\/}~{\em
  140\/}(3), 476--486.

\bibitem[\protect\citeauthoryear{Spearman}{Spearman}{1904}]{spearman1904proof}
Spearman, C. (1904).
\newblock The proof and measurement of association between two things.
\newblock {\em The American journal of psychology\/}~{\em 15\/}(1), 72--101.

\bibitem[\protect\citeauthoryear{Sz{\'e}kely, Rizzo, et~al.}{Sz{\'e}kely
  et~al.}{2009}]{szekely2009brownian}
Sz{\'e}kely, G.~J., M.~L. Rizzo, et~al. (2009).
\newblock Brownian distance covariance.
\newblock {\em The annals of applied statistics\/}~{\em 3\/}(4), 1236--1265.

\bibitem[\protect\citeauthoryear{Sz{\'e}kely, Rizzo, and Bakirov}{Sz{\'e}kely
  et~al.}{2007}]{szekely2007measuring}
Sz{\'e}kely, G.~J., M.~L. Rizzo, and N.~K. Bakirov (2007).
\newblock Measuring and testing dependence by correlation of distances.
\newblock {\em The annals of statistics\/}, 2769--2794.

\bibitem[\protect\citeauthoryear{Turner}{Turner}{2000}]{turner2000estimating}
Turner, T.~R. (2000).
\newblock Estimating the propagation rate of a viral infection of potato plants
  via mixtures of regressions.
\newblock {\em Journal of the Royal Statistical Society: Series C (Applied
  Statistics)\/}~{\em 49\/}(3), 371--384.

\bibitem[\protect\citeauthoryear{Wang, Jiang, and Liu}{Wang
  et~al.}{2017}]{wang2017generalized}
Wang, X., B.~Jiang, and J.~S. Liu (2017).
\newblock Generalized r-squared for detecting dependence.
\newblock {\em Biometrika\/}~{\em 104\/}(1), 129--139.

\bibitem[\protect\citeauthoryear{Wang, Waterman, and Huang}{Wang
  et~al.}{2014}]{wang2014gene}
Wang, Y.~R., M.~S. Waterman, and H.~Huang (2014).
\newblock Gene coexpression measures in large heterogeneous samples using count
  statistics.
\newblock {\em Proceedings of the National Academy of Sciences\/}~{\em
  111\/}(46), 16371--16376.

\bibitem[\protect\citeauthoryear{Wedel and DeSarbo}{Wedel and
  DeSarbo}{1994}]{wedel1994review}
Wedel, M. and W.~S. DeSarbo (1994).
\newblock A review of recent developments in latent class regression models.

\bibitem[\protect\citeauthoryear{Wheeler, Leong, Liu, Hivert, Strawbridge,
  Podmore, Li, Yao, Sim, Hong, et~al.}{Wheeler
  et~al.}{2017}]{wheeler2017impact}
Wheeler, E., A.~Leong, C.-T. Liu, M.-F. Hivert, R.~J. Strawbridge, C.~Podmore,
  M.~Li, J.~Yao, X.~Sim, J.~Hong, et~al. (2017).
\newblock Impact of common genetic determinants of hemoglobin a1c on type 2
  diabetes risk and diagnosis in ancestrally diverse populations: A transethnic
  genome-wide meta-analysis.
\newblock {\em PLoS medicine\/}~{\em 14\/}(9), e1002383.

\bibitem[\protect\citeauthoryear{Xu and Jordan}{Xu and
  Jordan}{1996}]{xu1996convergence}
Xu, L. and M.~I. Jordan (1996).
\newblock On convergence properties of the em algorithm for gaussian mixtures.
\newblock {\em Neural computation\/}~{\em 8\/}(1), 129--151.

\bibitem[\protect\citeauthoryear{Yabuta, Okada, Ito, Hosomi, Nishihara,
  Sasayama, Fujimori, Okuzaki, Zhao, Ikawa, et~al.}{Yabuta
  et~al.}{2007}]{yabuta2007lats2}
Yabuta, N., N.~Okada, A.~Ito, T.~Hosomi, S.~Nishihara, Y.~Sasayama,
  A.~Fujimori, D.~Okuzaki, H.~Zhao, M.~Ikawa, et~al. (2007).
\newblock Lats2 is an essential mitotic regulator required for the coordination
  of cell division.
\newblock {\em Journal of Biological Chemistry\/}~{\em 282\/}(26),
  19259--19271.

\bibitem[\protect\citeauthoryear{Yule}{Yule}{1903}]{yule1903notes}
Yule, G.~U. (1903).
\newblock Notes on the theory of association of attributes in statistics.
\newblock {\em Biometrika\/}~{\em 2\/}(2), 121--134.

\bibitem[\protect\citeauthoryear{Zheng, Shi, and Zhang}{Zheng
  et~al.}{2012}]{zheng2012generalized}
Zheng, S., N.-Z. Shi, and Z.~Zhang (2012).
\newblock Generalized measures of correlation for asymmetry, nonlinearity, and
  beyond.
\newblock {\em Journal of the American Statistical Association\/}~{\em
  107\/}(499), 1239--1252.

\end{thebibliography}

\end{document}